\documentclass[namedreferences]{SolarPhysics}

\usepackage[optionalrh]{spr-sola-addons} 
\pdfoutput=1
\usepackage{graphicx}                    
\usepackage{amssymb}                    
\usepackage{color}                       
\usepackage{url}                         
\usepackage{hyperref}
\usepackage{multirow}

\hypersetup{
    pdftitle={Stokes Diagnostics of Magneto-Acoustic Wave Propagation in the Magnetic Network on the Sun},
    pdfauthor={G. Vigeesh}, 
    colorlinks=true,       
    linkcolor=black,          
    citecolor=blue,        
}

\begin{document}

\begin{article}

\begin{opening}

\title{Stokes Diagnostics of Magneto-Acoustic Wave Propagation in the
Magnetic Network on the Sun}

%
\author{G.~\surname{Vigeesh}$^{1}$\sep
        O.~\surname{Steiner}$^{2}$\sep
        S.~S.~\surname{Hasan}$^{1}$
       }

%
\runningauthor{Vigeesh et al.}
\runningtitle{Stokes diagnostics of wave propagation}

%
  \institute{$^{1}$ Indian Institute of Astrophysics, Block II Koramangala,\\
                    Bangalore-560034, India\\
                    email: \url{vigeesh@iiap.res.in} email: \url{hasan@iiap.res.in}\\
             $^{2}$ Kiepenheuer-Institut f\"ur Sonnenphysik, Sch\"oneckstrasse 6,\\
                    79104 Freiburg, Germany\\
                    email: \url{steiner@kis.uni-freiburg.de} \\
             }

\begin{abstract}
The solar atmosphere is magnetically structured and highly dynamic. Owing
to the dynamic nature of the regions in which the magnetic structures exist,
waves can be excited in them. Numerical investigations of wave propagation in
small-scale magnetic flux concentrations in the magnetic network on the Sun
have shown that the nature of the excited modes depends on the value of plasma
$\beta$ (the ratio of gas to magnetic pressure) where the driving motion occurs.
Considering that these waves should give rise to observable characteristic signatures,
we have attempted a study of synthesized emergent spectra from numerical simulations
of magneto-acoustic wave propagation. We find that the signatures of wave propagation in a magnetic element can be detected when the spatial resolution is sufficiently
high to clearly resolve it, enabling observations in
different regions within the flux concentration. The possibility to probe
various lines of sight around the flux concentration bears the potential to reveal
different modes of the magnetohydrodynamic waves and mode conversion. We highlight the
feasibility of using the Stokes-$V$ asymmetries as a diagnostic tool to study the wave
propagation within magnetic flux concentrations. These quantities can possibly
be compared with existing and new observations in order to place constraints on different
wave excitation mechanisms.

\end{abstract}

%
\keywords{Magnetic fields, Photosphere; Magnetic fields, Models; Magnetohydrodynamics; Spectral Line, Intensity and Diagnostics; Polarization, Optical; Waves, Magnetohydrodynamic; Waves, Modes }

\end{opening}

%
\section{Introduction}\label{s:introduction}
Spectral lines inform us on the properties of the atmosphere in which they form.
In addition, any presence of magnetic fields in the atmosphere modifies the polarization state of the light emerging from the surface.
A wealth of information about the structure and dynamics of the magnetized regions of the Sun is hidden in the polarization state of spectral lines.
In this paper, we will study the polarization signatures of magneto-acoustic wave propagation in a photospheric magnetic flux concentration using  magnetohydrodynamic numerical simulations thereof. The initial and simulated models are thought to represent a magnetic flux concentration in the magnetic network of the Sun.

The two-dimensional simulations by
\inlinecite{steiner1998} and
\inlinecite{grossmann1998}
showed that the Stokes profiles vary strong\-ly in response to a dynamic magnetic atmosphere.
Using a similar forward modelling, the effect of waves on Stokes-$V$ profiles of Ca \textsc{ii} infrared lines were studied by
\inlinecite{pietarila2006}.
They saw a clear time-dependent behaviour of the Stokes-$V$ profiles as a result of wave propagation and shock formation occurring in the numerically simulated atmosphere.
Even though they were able to reproduce the atmospheric dynamics in the form of observational signatures in the Stokes profile, their work was limited to the weak field case. Several authors
\cite{rosenthal2002,bogdan2003,cranmer2005,hasan2005,hasan2008,khomenko2008,vigeesh2009,fedun2009,kato+al2011}
have investigated wave phenomena in magnetically structured atmospheres.
\inlinecite{shelyag2010} constructed a photospheric bright point model and studied the observational signatures of wave propagation in them including the response in Stokes $V$. The effect of a direct excitation of a magnetic flux concentration by granular buffeting and the corresponding spectral signatures of wave propagation and mode coupling and mode conversion in these structures have not been studied so far.
\inlinecite{fujimura2009} report on the observation of magnetohydrodynamic wave propagating along magnetic flux tubes in the solar atmosphere. Their work is based on observations with the SOT/SP instrument onboard the {\it Hinode} spacecraft and demonstrates the feasibility of such measurements.

Inspired by these observations and by numerical simulations, and driven by the desire to find possible polarimetric signatures of wave propagation in magnetic network elements, we have attempted to study the feasibility of using the Stokes-$V$ spectra obtained from our simulation as a diagnostic for magnetohydrodynamic wave propagation.

The paper is structured as follows: Section~\ref{s:initial_eq_model} will give the construction of the initial equilibrium model. In Section~\ref{s:simulation}, we will explain our numerical method and boundary conditions. In Section~\ref{s:dynamics}, we will present three different experiments of wave propagation in magnetic elements and in Section~\ref{s:stokes_diagnostics} we will discuss the properties of the Stokes profiles emerging from the simulation box. The summary and conclusions will be given in Section~\ref{s:summary_conclusion}.

%
\section{Initial Equilibrium Model}\label{s:initial_eq_model}
We construct a two-dimensional initial atmosphere in Cartesian coordinates containing a magnetic flux sheet. For the construction we use the numerical methods described in
\inlinecite{steiner1986}.
The magnetic field configuration and the pressure distribution in the physical domain is specified as in
\inlinecite{vigeesh2009}.

The magnetic field can be written
in terms of the flux function $\psi(x, z)$ as
\begin{equation}
 B_x = -\frac{\partial \psi}{\partial z},\quad B_z = \frac{\partial \psi}{\partial x}\,.
\label{eq:mag_component}
\end{equation}
Contours of constant flux value, $\psi$, correspond then to magnetic field lines.
We identify $\psi=0$ with the symmetry axis in the centre of the flux sheet and
$\psi = \pm\psi_{\rm max}$ with the side boundaries, which also defines the total
magnetic flux.
The gas pressure is prescribed as a function of height and field line,
$p(\psi,z)$, in the following way,
\begin{eqnarray}
  p(\psi,z) = \left\{ \begin{array}{l l}
           \displaystyle \frac{p(0,z)}{p_{0}}(p_{0}
                               + p_{2}\psi^{2})
         &\quad\mbox{if}\quad  0 \le  \psi \le \psi_{1}, \\[3ex]
            \multicolumn{2}{l}{\displaystyle  \frac{p(0,z)}{p_{0}}
            (a(\psi - \psi_{1})^{n} + b(\psi - \psi_{1})^2 +}\\[1.0ex]
            \hspace{3.0cm}+ c(\psi - \psi_{1}) + d)
         &\quad\mbox{if}\quad \psi_{1} < \psi < \psi_{2}, \\[2ex]
            \displaystyle \frac{p(0,z)}{p_{0}} (p_{0}
           + \frac{B_{0}^{2}}{8\pi})
         &\quad\mbox{if}\quad \psi_2 \le \psi \le \psi_{\rm max},
\end{array} \right.
\label{eq:pressure_height}
\end{eqnarray}
where the constants $a$, $b$, $c$, and $d$ are chosen such that the pressure and
its first derivative with respect to $\psi$ are continuous functions of
$\psi$ and where we choose $n=8$. $B_0$ and $p_0$ are the magnetic field
strength and the gas pressure, respectively, on the axis of the flux sheet at the reference height $z=0$.
$p_2$, $\psi_1$, and $\psi_2$ are chosen conveniently so as to obtain the desired cross-sectional
profile for the magnetic field, \textit{viz.}, $B_{z}(x)$ at $z=0$, as shown further below in
Figure~\ref{fig:photo_field_component}.

\begin{figure}[]
  \centering
\includegraphics[width=0.6\textwidth]{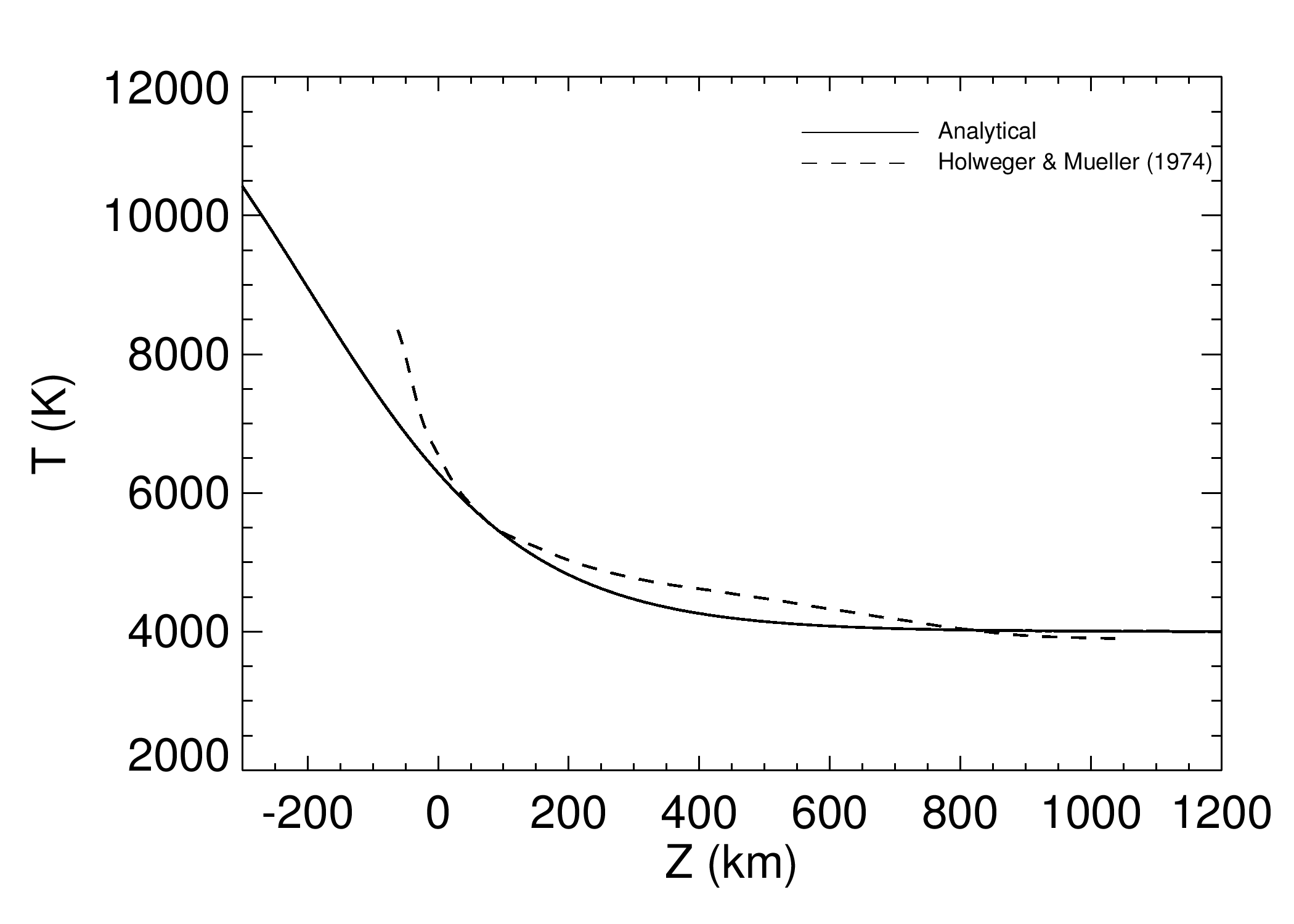}\\
\caption{Temperature as a function of height of the model atmosphere, according to Equation~(\ref{eq:temp_scale}) using $T_{0}=9000$~K, $\alpha=-5000$~K, $\gamma=3\times10^{-8}$~cm$^{-1}$, and $c=0.6$. $z=0$ corresponds
approximately to continuum optical depth unity, $\tau_c = 1$, for $\psi = \psi_{\rm max}$. The dashed curve shows the photospheric reference model of \protect\inlinecite{holweger1974}.}
\label{fig:photo_temp}
\end{figure}

The gas pressure along the axis is defined as,
\begin{equation}
\displaystyle p(0,z) = p_{0} \exp\left\{-\int_0^z \frac{\mu g}{R T(z)} {\rm d}z\right\}.\\
\label{eq:pressure_scale}
\end{equation}
Here $\mu$ is the mean molecular weight, taken to be $\mu=1.297$, $R$ the universal gas constant, and $g$ the gravitational acceleration at the solar surface. $T(z)$, the temperature as a function of height, is modelled by an analytical function of the form
\begin{equation}
\displaystyle T(z) = T_{0} + \alpha \tanh(\gamma z + c).
\label{eq:temp_scale}
\end{equation}
It is constant on levels of constant $z$, assuming that an efficient radiative exchange between flux sheet and the ambient medium levels the temperature to a horizontally isothermal state. With an appropriate choice of $T_{0}$, $\alpha$, $\gamma$, and $c$, we construct a photospheric temperature run as shown in Figure~\ref{fig:photo_temp}. For comparison, Figure~\ref{fig:photo_temp} also shows the temperature as a function of height of the solar atmospheric model of \inlinecite{holweger1974}, which was derived from observations under the assumption of local thermodynamic equilibrium (LTE). Initially, the temperature drops rapidly from 10\,500~K at the bottom boundary to 6\,300~K at $z=0$~km, then asymptotically decreases to 4\,000~K.
This temperature profile approximately reflects the temperature profile of the Holweger and M\"uller model in the photospheric part and enables us to compute spectral lines in absorption.
We do not include a chromospheric temperature rise because we intend to compute the spectral lines in LTE only. With the choice of an analytical expression for the temperature and an ideal gas equation of state with constant molecular weight, the initial atmosphere quickly relaxes to the static initial solution of the discrete numerical scheme, although at the expense of a more realistic atmosphere.

Having defined the gas pressure and the temperature distribution through
Equations~(\ref{eq:pressure_height})-(\ref{eq:temp_scale}),
we obtain the density distribution.
From the force balance perpendicular to the direction of the field lines, one obtains the electric current density,
\begin{equation}
  \displaystyle
j_{y} = \left. \frac{\partial p}{\partial \Psi}\right|_z\,.
  \label{eq:eqn_volume_current}
\end{equation}
The new magnetic field configuration can be calculated from the current density
using the Grad-Shafranov equation,
\begin{equation}
\displaystyle \frac{\partial^{2} \psi}{\partial x^{2}} + \frac{\partial^{2} \psi}{\partial z^{2}} = 4 \pi j_{y}.
\label{eq:grad_shafranov}
\end{equation}
A detailed derivation of these equations is given in \inlinecite{steiner2007}.

\begin{figure}[]
  \centering
  \includegraphics[width=.6\textwidth]{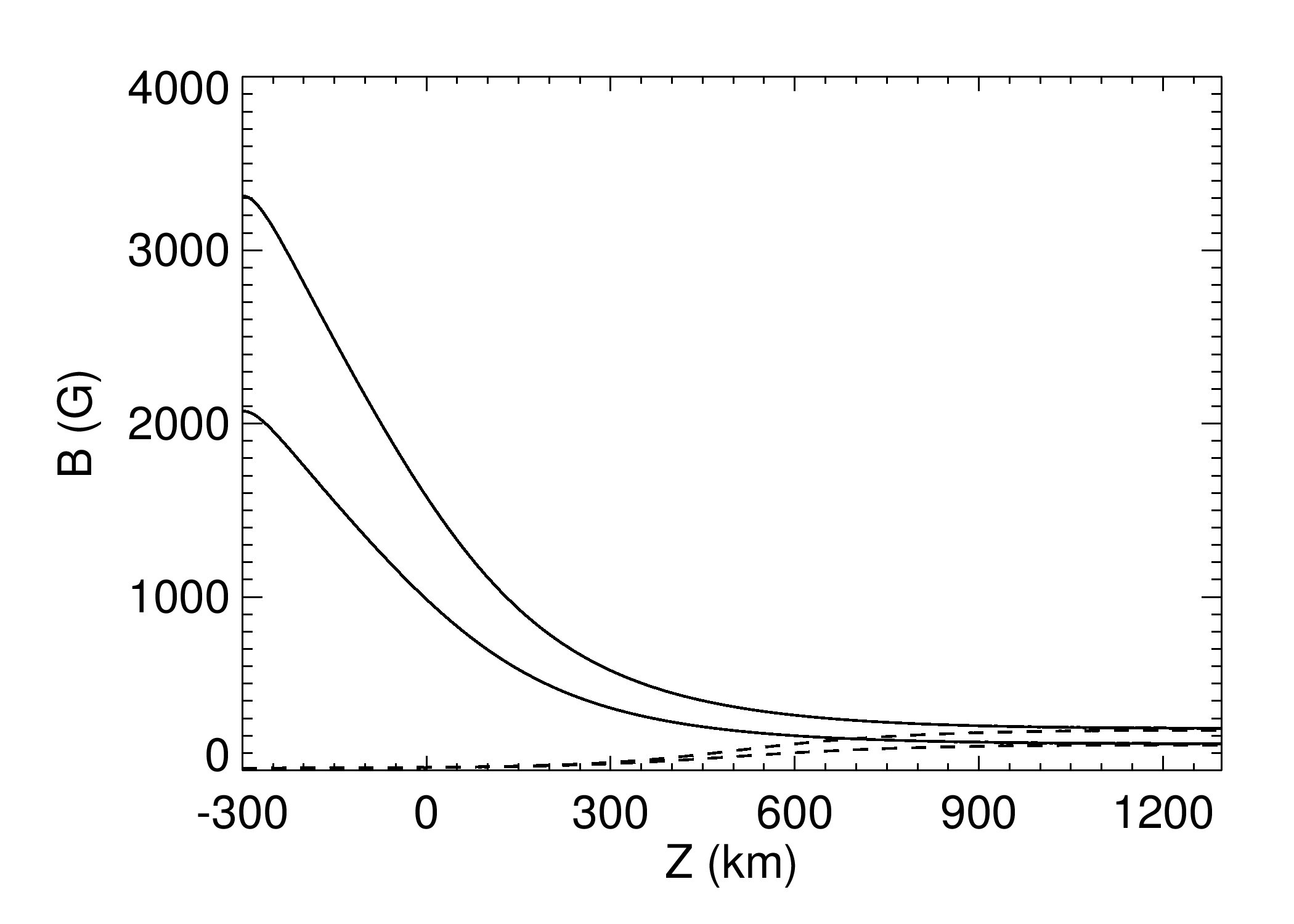}\\
  \caption{Magnetic field strength as a function of height on the axis (solid curve) and in the ambient medium (dashed curve) of photospheric magnetic flux sheets with field strengths of 1000~G and 1600~G at the axis at $z=0$.}
  \label{fig:photo_field_strength}
\end{figure}
%
\begin{figure}[]
   \centering
   \setlength{\tabcolsep}{0pt}
   \begin{tabular}{cc}
     \includegraphics[width=.49\textwidth]{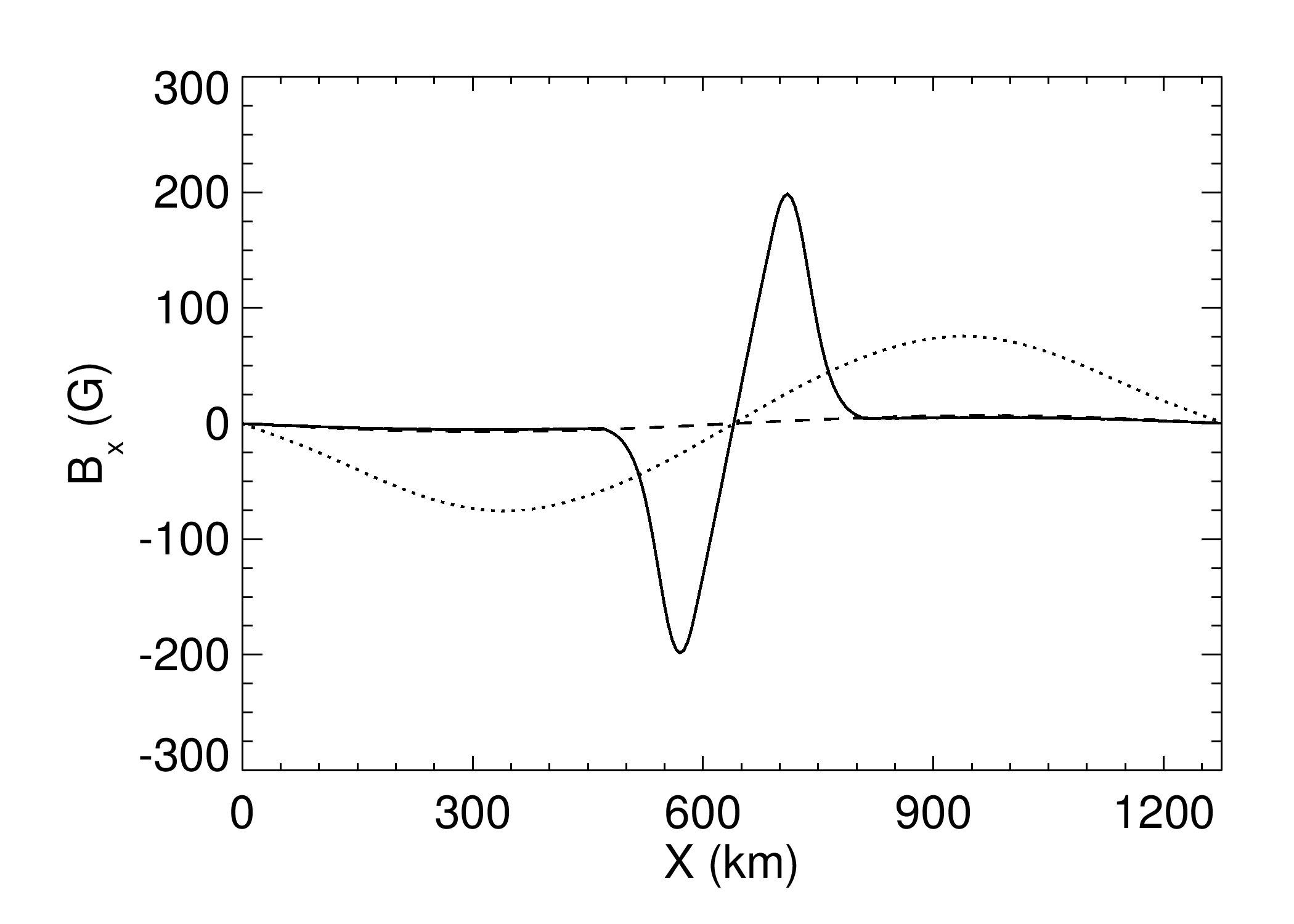} &
     \includegraphics[width=.49\textwidth]{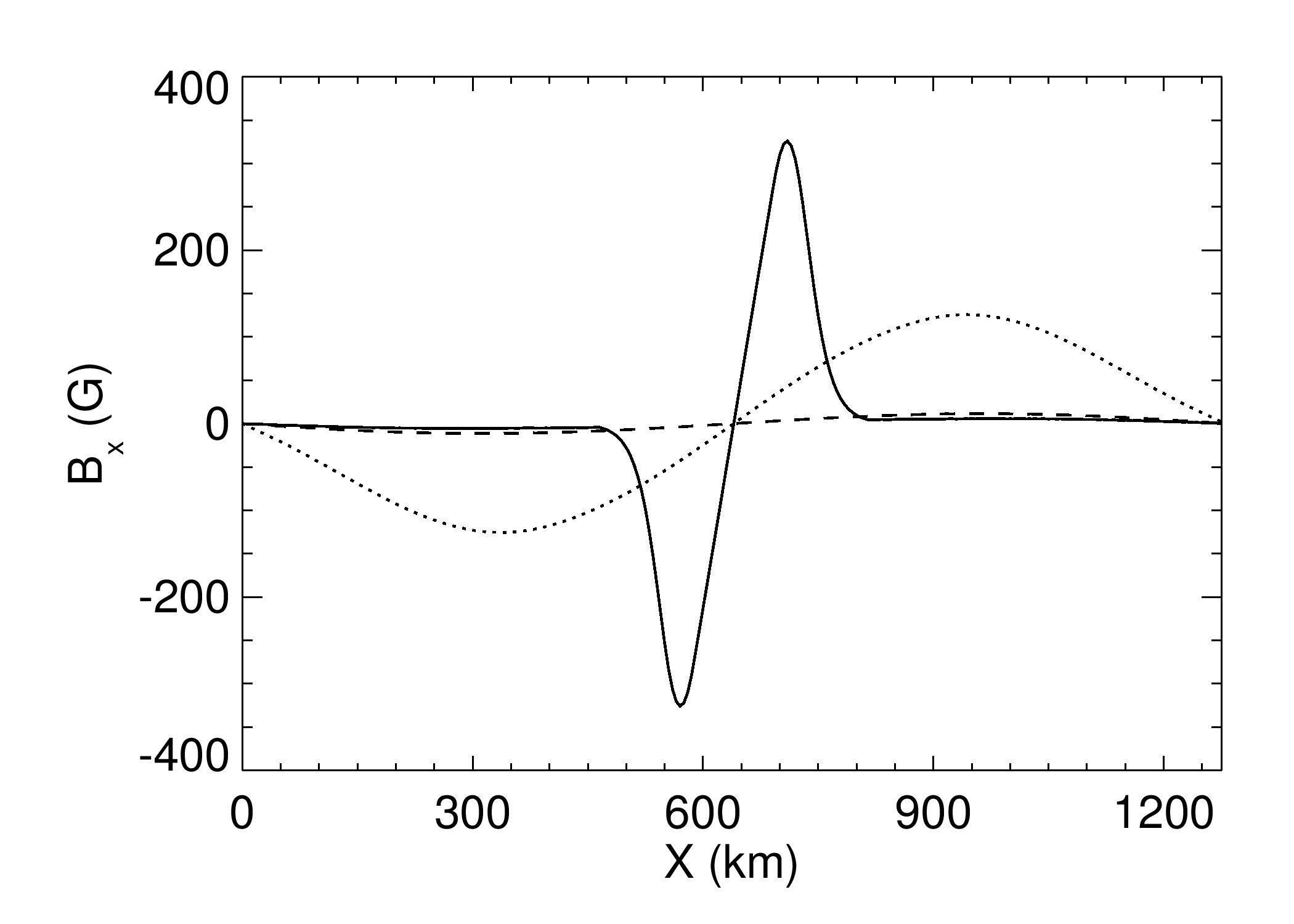}\\
     \includegraphics[width=.49\textwidth]{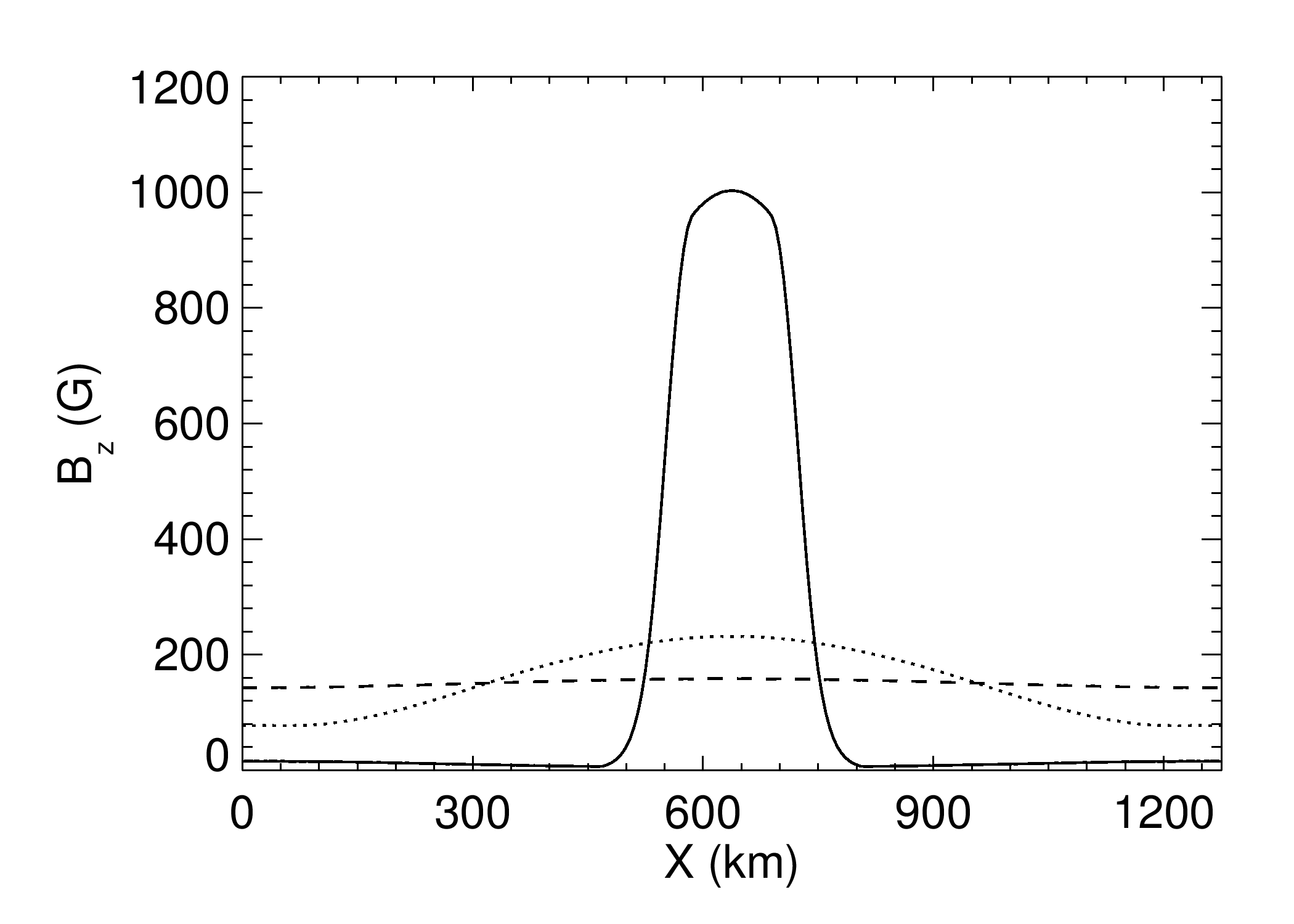} &
     \includegraphics[width=.49\textwidth]{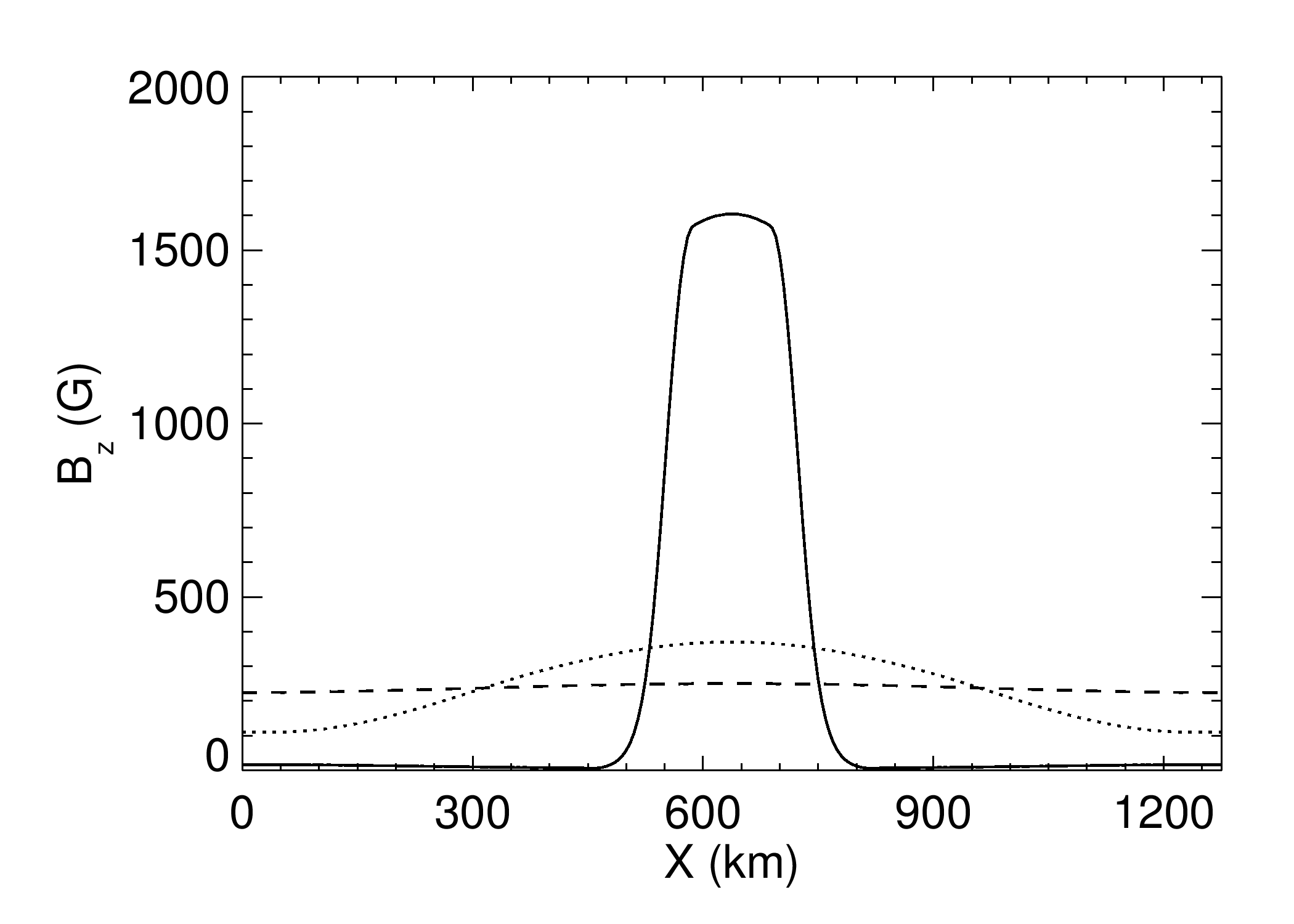}\\
     \mbox{\footnotesize a) 1000~G} & \mbox{\footnotesize b) 1600~G}
   \end{tabular}
   \caption{$B_{x}$ (top row) and $B_{z}$ (bottom row) components of the magnetic field as functions of horizontal distance at the following heights: $z=0$~km (solid curve), $z=500$~km (dotted curve), and $z=1000$~km (dashed curve). The plots refer to photospheric magnetic flux sheets with field strengths of (a) 1000~G (gauss) and (b) 1600~G at the axis at $z=0$.}
   \label{fig:photo_field_component}
\end{figure}

Due to the symmetry of the problem, we solve Equation~(\ref{eq:grad_shafranov}) in a computational domain that consists of only half of the flux sheet of horizontal and vertical extensions of 640~km and 1\,600~km, respectively. The bottom boundary is at a depth of $z = -300$~km. The domain is discretized on an equidistant rectangular mesh with a spacing of 5~km. The left side of the domain corresponds to the axis of the flux sheet. The value of $\psi$ is prescribed at the left and the at right side boundaries. At the top and bottom boundaries, we use the Neumann condition $\partial\psi/\partial z = 0$, assuming that the horizontal field component vanishes at these two boundaries. Starting from a reasonable initial field configurations, \textit{e.g.}, as derived from the thin flux-tube approximation, one obtains by iteration of Equations~(\ref{eq:mag_component})-(\ref{eq:grad_shafranov}), a final, self-consistent, magnetohydrostatic model.

\begin{table}
\caption{Equilibrium model characteristics for the 1000~G and 1600~G flux sheets. The numbers in the first row of each physical quantity correspond to the top boundary ($z=1300$~km) and the numbers in the second row correspond to the height $z=0$~km.}
\label{tab:photo_equilibrium_table}
\setlength{\tabcolsep}{3pt}
\begin{tabular}{llcccc}
\hline
\multirow{2}{*}{Physical quantity} & & \multicolumn{2}{c}{1000~G} & \multicolumn{2}{c}{1600~G} \\
\cline{3-6}
& & Sheet axis & Ambient medium & Sheet axis & Ambient medium\\
\cline{1-6}
\multirow{2}{*}{Temperature [K]}
& & 4001 & 4001 & 4001 & 4001 \\
& & 6342 & 6342 & 6342 & 6342\\[0pt]
\multirow{2}{*}{Density [g\,cm$^{-3}$]}  &  \rule{0pt}{10pt}
& $7.4 \times 10^{-13}$ &  $1.1 \times 10^{-12}$  & $1.7 \times 10^{-13}$ &  $1.1 \times 10^{-12}$ \\[0pt]
&  & $2.0 \times 10^{-7}$ & $3.0 \times 10^{-7}$ & $4.5 \times 10^{-8}$  & $3.0 \times 10^{-7}$\\
\multirow{2}{*}{Pressure [dyn\,cm$^{-2}$]}
&  & 0.2 & 0.3 & 0.04 & 0.29 \\
&  & $8.2 \times 10^{4}$& $1.2 \times 10^{5}$ & $ 1.8 \times 10^{4}$ & $12.2 \times10^{4}$ \\
\multirow{2}{*}{Sound speed [km\,s$^{-1}$]}
&  & 6.5 & 6.5 & 6.5 & 6.5 \\
 &  & 8.2 & 8.2 & 8.2 & 8.2\\
\multirow{2}{*}{Alfv\'{e}n speed [km\,s$^{-1}$]}
& & 504 & 394 & 1672 & 620 \\
 & & 6.3 & 0.1 & 21  & 0.08 \\
\multirow{2}{*}{Magnetic field [G]}
& & 154 & 147 & 243 & 232 \\
& & 1003 & 16 & 1604 & 16\\
\multirow{2}{*}{Plasma $\beta$ [--]} & \rule{0pt}{10pt}
& $2.0 \times 10^{-4}$  & $3.4 \times 10^{-4}$ & $1.9 \times 10^{-5}$  & $1.4 \times 10^{-4}$\\
&  & 2.1 & $1.3 \times 10^{4}$ & 0.2 & $1.2 \times 10^{4}$\\
\cline{1-6}
\end{tabular}
\end{table}

We calculate cases corresponding to different field strengths (at  $z = 0$) ranging between 1000~G (gauss) and 1600~G, on the axis of the sheet. For the case with weaker field strength, the $\beta=1$ layer lies well above the bottom boundary dropping to a minimum height of $370$~km only. Here $\beta$ is the ratio of the gas pressure to the magnetic pressure. Any field line originating from the bottom eventually crosses this layer, dividing the flux sheet into two regions: a lower region with $\beta >1$ and the upper region with $\beta <1$. We say that the flux sheet is rooted in a high $\beta$ region. In the case of a stronger field, \textit{e.g.}, when $B_{0}=1600$~G, the $\beta=1$ layer traces the boundary of the flux sheet. The entire flux sheet is in a region of $\beta<1$, and only the ambient medium has a high $\beta$ plasma value.

The variation of the magnetic field strength with height on the axis and in the ambient medium is shown in Figure~\ref{fig:photo_field_strength}. In both cases, the magnetic field drops to a uniform value within the flux sheet. In the ambient medium, the field strength in the lower part is negligible, but with increasing height, it settles down to the same uniform value as that of the flux sheet. Figure~\ref{fig:photo_field_component} shows the horizontal variation of the horizontal and vertical components of the magnetic field at three different heights, $z=0$~km (solid curve), $z=500$~km (dotted curve), and $z=1000$~km (dashed curve). The horizontal component of the field at $z=0$~km is close to zero, hence the field is almost vertical at this level. The flux sheet at this height has a vertical component of the magnetic field that drops sharply to the ambient value in the horizontal direction, confining it to a narrow region with a width of about 320~km. The flux sheet expands with height to cover the entire horizontal extent with a homogeneous vertical field, starting at a height of below $z=1000$~km.

The equilibrium characteristics of the two models are summarized in Table~\ref{tab:photo_equilibrium_table}. The values of the physical quantities are given at the sheet axis and in the ambient medium for the top boundary ($z=1300$~km) and for the height $z=0$~km in the first and second rows of each entry, respectively. Note that the plasma $\beta$ at the sheet axis at the base is 2.1 for the 1000~G case and 0.2 for the 1600~G case. The sound speed ($c_{s}$) and the Alfv\'{e}n speed ($v_{A}$) are defined as,
\begin{equation}
\label{eq:sound_speed}
\displaystyle c_{s} = \sqrt{\frac{\gamma p}{\rho}},
\end{equation}

\begin{equation}
\label{eq:alfven_speed}
\displaystyle v_{A} = \frac{B}{\sqrt{4\pi \rho}},
\end{equation}
where $\gamma$ is the ratio of specific heats taken to be 5/3 and $p$, $\rho$, and $B$ are the equilibrium values of gas pressure, density and magnetic field strength, respectively.
%

%
\section{Numerical Simulation: Methods and Boundary Condition}\label{s:simulation}
Wave propagation is studied by an impulsive transverse excitation of the lower boundary in the equilibrium model (similar to
\inlinecite{hasan2005}
and
\inlinecite{vigeesh2009}).
The system of MHD equations, given in conservation-law form for an inviscid adiabatic
fluid, is solved according to the method described in
\inlinecite{steiner1994}.

The side boundaries are open due to a constant extrapolation of the
variables from the physical domain to the boundary cells. The horizontal component of the momentum at the top and bottom boundaries and the vertical component at the top boundary are also set by a constant extrapolation. The density in the top and bottom boundary cells is
determined using a linear-log extrapolation.
For the temperature, constant extrapolation is used at the top boundary. The temperature in the
bottom boundary cells is determined using Equation~(\ref{eq:temp_scale}). The horizontal component
of the magnetic field at the top and bottom boundaries are set equal to the corresponding values
at the preceding interior point so that ${\rm d}B_{x}/{\rm d}z=0$. The vertical component of the magnetic field is determined by the
condition $\nabla \cdot \mathbf{B}= 0$.
With these boundary conditions, acoustic waves (slow modes) can propagate across the top boundary with little reflection.
They also work for the fast, magnetically dominated mode as it typically gets refracted and converted to an acoustic mode before leaving the computational domain across the side boundaries as will be shown in Section~\ref{s:dynamics}. Thus, only a small fraction of the fast-mode wave fronts reaches the top boundary: the bulk of it leaves the computational domain through the side boundaries.

Similar to
\inlinecite{vigeesh2009}, the transverse velocity $V_{x}$ at $z = 0$ is specified as follows:
\begin{eqnarray}
  V_{x}(x,0,t) = \left\{ \begin{array}{l l}
                 \displaystyle V_{0} \sin (2 \pi t / P)
                 &\mbox{for}\quad 0 \le t \le P/2\,, \\[1ex]
                 \displaystyle 0
                 &\mbox{for}\quad 0 > t > P/2\,,
\end{array} \right.
\label{eq:velocity}
\end{eqnarray}
where $V_{0}$ denotes the amplitude of the horizontal motion and $P$ is the
wave period. This form simulates an \textit{impulsive} transverse excitation of the flux sheet at the lower boundary. For simplicity, we assume that all points of the lower boundary have this motion: this does not generate any waves in the ambient medium, other than at the interface with the flux sheet.
In order to achieve significant intensity signals, we use $V_{0} = 5$~km\,s$^{-1}$
and $P$ = 24~s.
Such short duration motions are expected to be generated by the turbulent motion
in the convectively unstable subsurface layers, where the flux sheet is rooted.
\inlinecite{cranmer2005}
studied the kinematics of G-band bright points and suggested that there are two components involved: a ``random walk phase'' and a ``jump phase''.
Our work considers the case with
higher velocities, which represents the ``jump phase'' component of
\inlinecite{cranmer2005}.
This motion generates
magnetoacoustic waves in the flux sheet.

%
\section{Dynamics}\label{s:dynamics}
We consider a unidirectional horizontal displacement of the entire bottom region below $z=0$~km
(wide excitation region), first for the case in which the field strength is
1000~G at $z=0$ (moderate field case) and second, for the case in which the field strength is
1600~G at $z=0$ (strong field case). For the strong field case, we also consider a narrow excitation
region with excitation below $z=-150$~km only. The excitations correspond to the \emph{impulsive} case discussed in Section~\ref{s:simulation} and given by Equation~(\ref{eq:velocity}).

%
\subsection{Moderate Field Case}\label{ss:moderate_case}
\begin{figure}[h]
  \begin{flushright}
  \includegraphics*[width=0.8\textwidth]{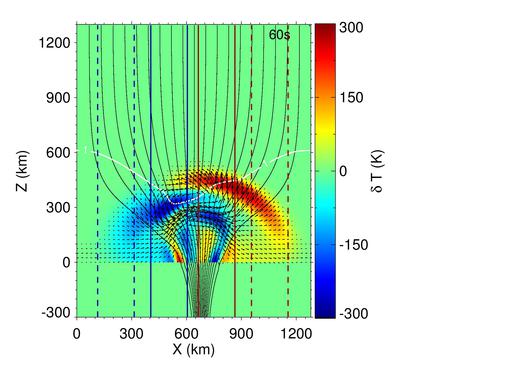}
  \end{flushright}
  \caption{Temperature perturbations and velocity field of a flux sheet in which the field strength at the axis at $z=0$ is 1000~G. The snapshot is taken at time 60~s after the initiation of an impulsive horizontal motion of the entire region below $z=0$~km according to Equation~(\ref{eq:velocity}). The amplitude of the motion is 5~km\,s$^{-1}$ and the period $P=24$~s. The black curves represent magnetic field lines and the white curve depicts the $\beta=1$ contour. Vertical lines indicate different bundles of lines of sight considered for the Stokes analysis.}
  \label{fig:1000_delta_t}
\end{figure}

The $\beta=1$ contour in this case is well above $z=0$ and hence all the magnetic field lines that emerge from the base cross this layer at some height as is visible in Figure~\ref{fig:1000_delta_t}.

The motion in the direction to the right hand side takes place in the region where $\beta>1$ (high-$\beta$ excitation) and results in the excitation of waves in the form of a fast (predominantly acoustic) wave and a slow (predominantly magnetic) wave, which propagate at the sound speed and the Alfv\'en speed, respectively. The fast wave manifests itself as a compression and rarefaction of the gas at the leading and trailing edges of the flux sheet, respectively. It can be clearly discerned in the snapshot of the temperature perturbation, $\delta T$ (the temperature  difference with respect to the initial value), shown in Figure~\ref{fig:1000_delta_t}, which is taken 60~s after the start of the perturbation. The black curves denote the magnetic field lines and the white curve depicts the $\beta=1$ contour. The perturbations are $180^\circ$ out of phase on the opposite sides of the sheet axis, leading to a quasi anti-symmetric wave pattern. As these fast waves travel upwards they eventually cross the layer of $\beta=1$, where they change from fast to slow, without changing their acoustic nature: this corresponds to  a ``mode transmission'' in the sense of
\inlinecite{cally2007}.
The transmission coefficient depends (among others) on the ``attack angle", \textit{i.e.},
the angle between the wave vector and the local direction of the magnetic
field
\cite{cally2007}.
On the $\beta=1$ layer, away from the sheet axis,
where the wave vector is not exactly parallel to the magnetic field, we do
not have complete transmission of the fast wave to a slow wave.  Rather,
there is a partial conversion  of the mode from fast acoustic to
fast magnetic, so that the energy in the acoustic mode is
reduced correspondingly. For more details, the reader is referred to
\inlinecite{vigeesh2009}.

%
\subsection{Strong Field Case}\label{ss:strong_case}
\begin{figure}[]
  \centering
  \setlength{\tabcolsep}{0pt}
  \begin{tabular}{cc}
    \includegraphics*[width=0.49\textwidth]{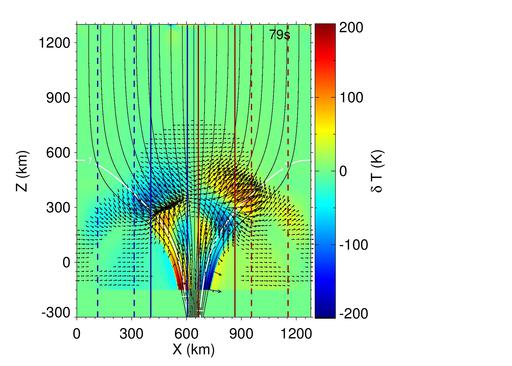} &   \includegraphics*[width=0.49\textwidth]{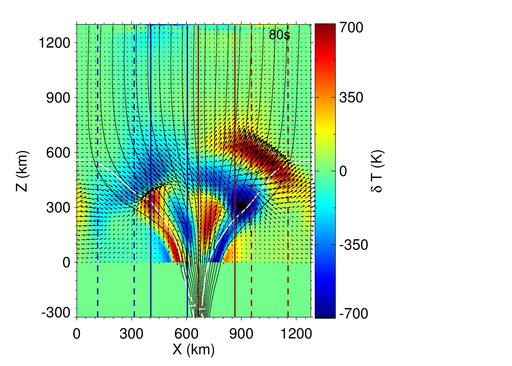}\\
    \includegraphics*[width=0.49\textwidth]{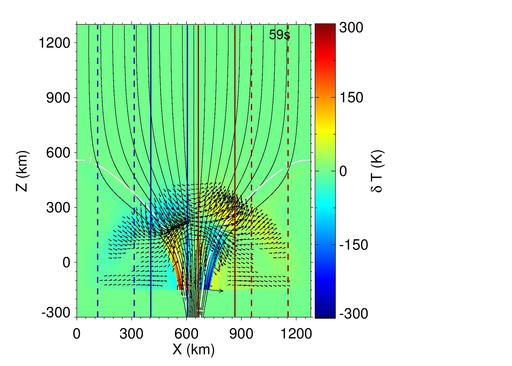} &   \includegraphics*[width=0.49\textwidth]{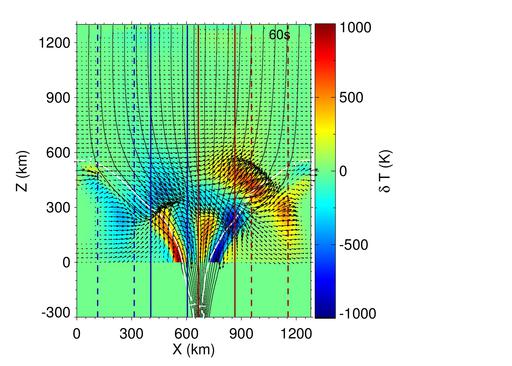}\\
    \includegraphics*[width=0.49\textwidth]{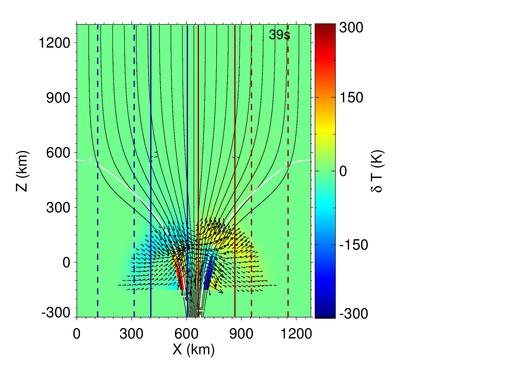} &   \includegraphics*[width=0.49\textwidth]{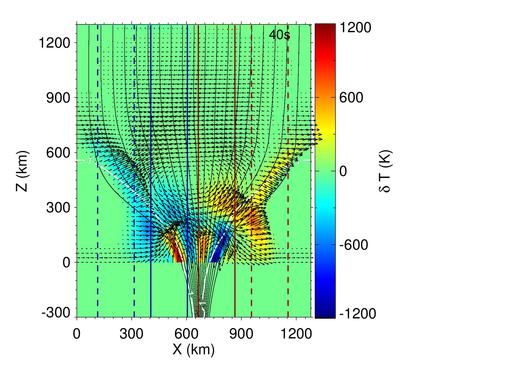} \\
    \mbox{\footnotesize (a) Narrow} & \mbox{\footnotesize (b) Wide} \\
  \end{tabular}
%
  \caption{Temperature perturbation for a narrow and a wide excitation depths for a flux sheet in which the field strength at the axis at $z=0$ is 1600~G. Time instances refer to 40, 60, and 80~s (from bottom to top) after the initiation of an impulsive
horizontal motion in the regions from (a) $z=-150$~km and (b) $z=0$~km to the bottom boundary. The duration of the motion is 12~s and has an amplitude of
750~m\,s$^{-1}$. The thin, black, vertically running curves represent field lines
and the white curve corresponds to the contour of $\beta=1$. The arrows indicate velocities of $50$ m\,s$^{-1}$ and more. Vertical lines indicate different bundles of lines of sight considered for the Stokes analysis.}
  \label{fig:1600_delta_t}
\end{figure}

We consider a uniform horizontal displacement of the entire bottom boundary region of a thickness of 150~km (which we term as narrow) and of a thickness of 300~km (wide) from the bottom boundary. This is to mimic the buffeting of the flux tubes by granular eddies of two different depths. The excitation corresponds to the impulsive case as given by Equation~(\ref{eq:velocity}). Figure~\ref{fig:1600_delta_t} shows the temperature perturbation $\delta T$ at 40, 60, and 80~s for the two cases of narrow and wide regions of excitation.

Here, the contour of $\beta=1$ approximately traces the boundary
of the flux sheet (see Figure~\ref{fig:1600_delta_t}). The transverse motion of
the lower boundary generates slow (predominantly acoustic) and fast (predominantly
magnetic) waves. Since the contour of $\beta=1$ runs along the boundary of the
flux sheet, the waves that travel within the flux sheet along the magnetic field lines upwards do not encounter this layer and hence do not undergo mode conversion. On the other hand, the fast wave, which can travel across the field, encounters the $\beta = 1$ contour at the boundary of the flux
sheet. As the fast wave crosses this boundary, it enters a region of negligible field and hence gets converted into a fast (acoustic) wave as can be seen in
the snapshot of temperature perturbations at an elapsed time of 40~s. At this time, the
fast wave in the low-$\beta$ region, which is essentially a magnetic wave,
undergoes mode conversion and becomes an acoustic wave, which creates
fluctuations in temperature visible as wing-like features at the periphery of
the flux sheet (approximately along the $\beta=1$ contour). The refraction of the fast wave, due to the gradients in the Alfv\'en speed, and the eventual mode conversion are prominent in the case of the wide excitation regime. It can be hardly seen in the case where the flux sheet is shaken over a narrow region. This is due to the fact that a wide excitation range in a flux tube embedded in a low-$\beta$ region creates more magnetic pressure fluctuations relative to gas pressure fluctuations and hence imparts more energy to the fast (magnetic) wave. Due to the gradients in the Alfv\'en speed, this mode gets refracted and returns back to a region with high-$\beta$, when this energy is transferred to the fast (acoustic) wave, producing larger temperature fluctuations. This has implications for a realistic atmosphere, where granular eddies of different sizes are likely to impact deep rooted flux tubes. An impact over a wide range on a flux tube with strong field will transfer more energy to the fast (magnetic) mode. The ambient atmosphere regains part of this energy in the form of a fast (acoustic) wave due to the refraction of the fast mode and eventual mode conversion. When the excitation range is narrow, there is relatively weak magnetic pressure fluctuation compared to gas pressure fluctuation and hence, most of the energy goes into the slow (acoustic) mode, which is channelled up along the flux tube and eventually dissipates by shock formation.

%
\section{Stokes Diagnostics}\label{s:stokes_diagnostics}
Even the largest modern solar telescopes are still not capable of resolving small-scale magnetic structures completely.
Observations of the Stokes parameter $I$ usually cannot reveal properties exclusive to the magnetic feature, since they suffer from mixing contributions from the magnetic field with contributions from the surrounding field-free or weak-field plasma.
Differently from that, Stokes $V$ gives us the properties intrinsic to the magnetic structure, because the circularly polarized light is formed only where the magnetic field is present.
Therefore, presently the most sensitive method to study the magnetic atmosphere is by analyzing the Stokes-$V$ spectra emerging from them
(see \inlinecite{sigwarth2000} for a review).

\begin{table}[]
\caption{Atomic parameters of the selected lines.}
\label{tab:atomic_parameters}
\begin{tabular}{ccccccc}
\hline
\multirow{2}{*}{Ion} & \multirow{2}{*}{Wavelength} & Excitation & \multirow{2}{*}{$\log({gf})$} & \multirow{2}{*}{$g_{\rm eff}$} & Lower & Upper\\
& & Potential & & & Level & Level \\
& (\AA) & (eV) & & & &\\
\hline
Fe \textsc{i} & 5250.21 & 0.121 & $-4.938$ & 3.0 & ${}^{5}D_{0}$ & ${}^{7}D_{1}$\\
Fe \textsc{i} & 5247.05 & 0.087 & $-4.946$ & 2.0 & ${}^{5}D_{2}$ & ${}^{7}D_{3}$\\
Fe \textsc{i} & 6301.50 & 3.654 & $-0.718$ & ~\,1.67 & ${}^{5}P_{2}$ & ${}^{5}D_{2}$\\
Fe \textsc{i} & 6302.49 & 3.686 & $-1.235$ & 2.5 & ${}^{5}P_{1}$ & ${}^{5}D_{0}$\\
\hline
\end{tabular}\\
{\footnotesize Values taken from \inlinecite{nave1994}.}
\end{table}

We have computed the emergent Stokes-$V$ profiles from the top of our simulation box for the weak field and the strong field cases, using the Stokes radiative transfer code DIAMAG
\cite{grossmann1994}.
This code calculates the normalized Stokes parameters by solving the Unno-Rachkovsky equations of radiative transfer. At the same time, it computes the line depression contribution function for each wavelength point. The program requires the temperature, gas pressure, magnetic field vector, velocity, and micro-turbulence to be specified on every grid point along the line of sight. The calculations were done for a set of four Fe \textsc{i} lines, \textit{viz.}, $\lambda\lambda$ 5250.2, 5247.05, 6301.5, and 6302.5~\AA. The atomic parameters of the selected lines are listed in Table~\ref{tab:atomic_parameters}. We set the microturbulence velocity to zero for all the line-transfer calculations.

Table~\ref{tab:atomic_parameters} lists the wavelengths of the lines, the excitation potentials of the lower level, oscillator strengths (log (\textit{gf})), and the effective Land\'{e} factors ($g_{\rm eff}$). The two pairs of lines were selected because each forms under similar conditions in the atmosphere, since the lines of each pair have similar excitation potentials and oscillator strengths, which means similar opacities. But the difference in Land\'{e} factor for these lines make them useful for measuring the magnetic field strength, in particular, deviations from the weak field regime. These lines are commonly used to study solar magnetic fields.
\inlinecite{socas2008}
have confirmed the reliability of using these four lines for the diagnostics of the quiet Sun magnetic field.

Here we study the spectral signature of wave propagation in magnetically structured atmospheres with dynamically varying magnetic field. The effects of wave propagation in the four Fe \textsc{i} lines, listed in Table~\ref{tab:atomic_parameters} are assessed, using the numerical simulations described in Section~\ref{s:dynamics}. The Stokes spectra for vertical lines of sight, separated by a horizontal distance of 10~km, were computed for each time step. These correspond to real observations at disk centre. Here, we present the analysis of the Stokes-$V$ spectra for the two cases of moderate and strong magnetic fields.

\begin{figure}[]
  \centering
\includegraphics[width=0.6\textwidth]{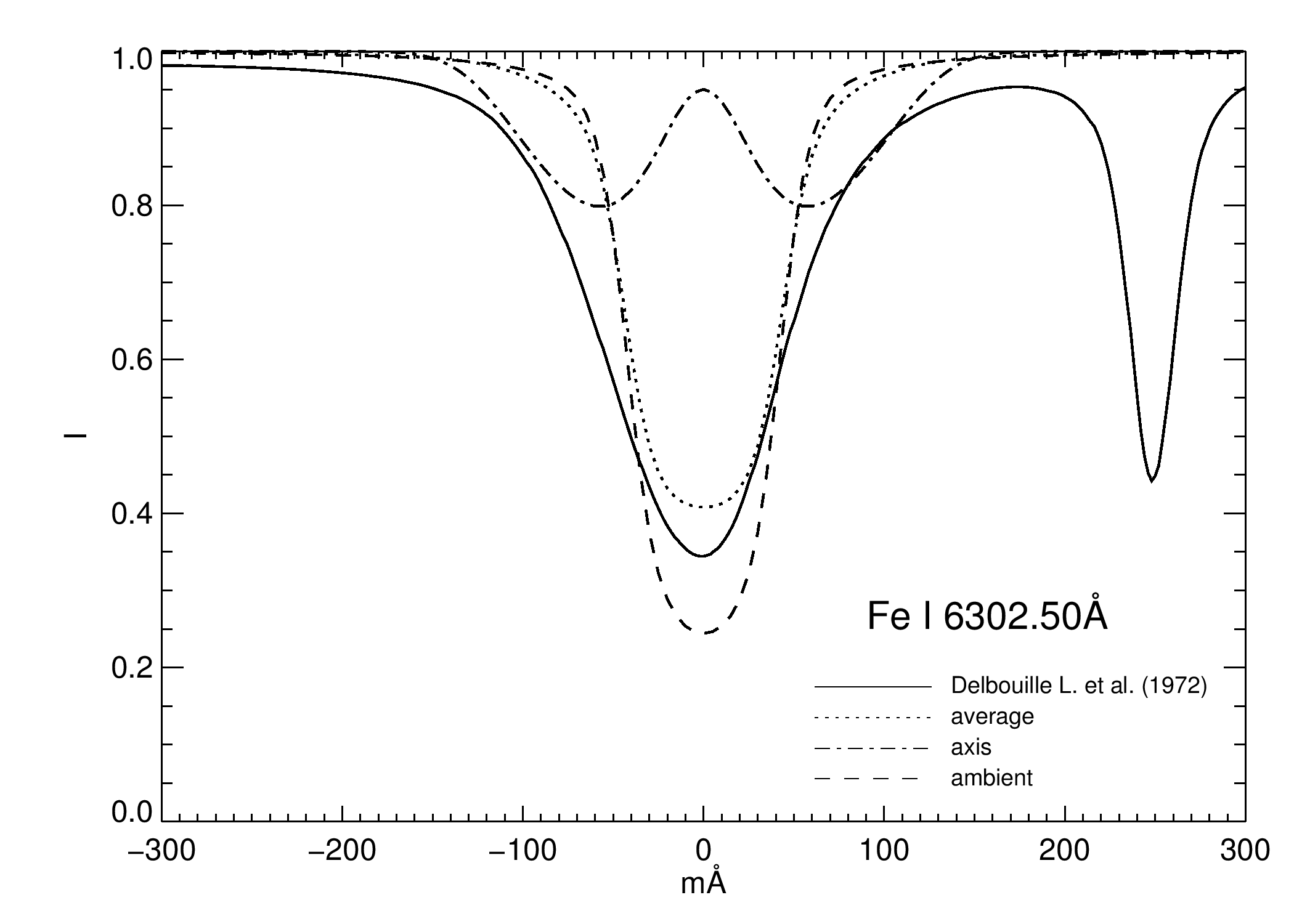}\\
\caption{Stokes-$I$ profiles for the spectral line Fe \textsc{i} 6302.5~\AA\ for a line of sight in the ambient medium (dashed), on the axis of the flux concentration (dot-dashed), and the profile averaged over the entire box (dotted) along with an observed profile (solid) from the atlas of \protect\inlinecite{delbouille1973}.}
\label{fig:stokes_i}
\end{figure}

Our analytical description of the atmosphere according to Equations~(\ref{eq:pressure_height})-(\ref{eq:temp_scale}) yields of course only a rough approximation to observed spectral lines.
In order to assess this approximation we compare the synthesized Stokes-$I$ line profiles of the Fe \textsc{i} 6302.5 line for a line of sight in the ambient medium and on the axis of the flux concentration, and the profile averaged over the entire box with the corresponding spectral line profile from the Jungfraujoch solar atlas
\cite{delbouille1973}
in Figure~\ref{fig:stokes_i}.
The synthetic spectral lines were calculated without including micro turbulence and hence they are narrower than the observed line.
The average profile is weaker than the profile emerging from the ambient medium in accordance with the observed ``line weakening'' or ``line gap'' in magnetic elements
\cite{sheeley1967}.
Although the Zeeman-broadening is strong, we found (by arbitrarily setting the magnetic field to zero for the radiation transfer) that the line weakening is mainly due to the strong difference in the temperature as a function of optical depth between the magnetic element and the ambient medium.
This is in agreement with the findings of
\inlinecite{shelyag2007}
from three-dimensional MHD simulations.
The central line of sight produces a Stokes $I$, which is much weaker than the ambient profile and is fully split.
Indeed, a similar behavior is seen in the observed Stokes-$I$ profile from the fully resolved magnetic element of
\inlinecite{lagg2010}.
Since our aim is to understand the basic principles governing the response of Stokes-$V$ profiles to the propagation of MHD waves and not to perform a detailed comparison with observed Stokes-$V$ profiles, we judge the chosen analytical form of the atmosphere as good enough and do not strive for a better fit of the synthetic with the observed Stokes $I$ by introducing microturbulent and mesoturbulent velocities.

The Stokes-$V$ profiles were computed along the vertical lines of sight by integration of the radiative transfer equation for polarized light. If the profiles emerging from the top boundary are spatially averaged over the entire width of the box, the profiles do not show significant variation with time revealing no sign of wave propagation inside the box. This is due to the fact that the anti-symmetric flow pattern within the flux sheet will average out to give zero net contribution to the Stokes-$V$ variation. This is different in the case when the horizontal integration is carried out over only a narrow spatial window on either side of the flux sheet axis. In this case, profiles show signatures of wave propagation. Therefore, it is necessary to observe at very high spatial resolution in order to study the effect of wave propagation in individual flux concentrations. The lines of sight that cover only one half of the flux sheet or a small part of it give more information about the wave activity in the domain. In order to quantitatively study the signatures of wave propagation, we study the evolution of the wavelength shift of the central zero crossing of Stokes $V$ ($\delta\lambda_{\rm zc}$) and the area and amplitude asymmetries, $\delta A$ and $\delta a$, respectively, according to Equations~(\ref{eq:zero_crossing})-(\ref{eq:ampli_asymmetry}). The zero-crossing shift of Stokes $V$ is defined as,
\begin{equation}
\delta\lambda_{\rm zc} = \lambda_{\rm zc} - \lambda_{0},
\label{eq:zero_crossing}
\end{equation}
where $\lambda_{\rm zc}$ is the wavelength of the central zero-crossing of the Stokes $V$ profile and $\lambda_{0}$ a reference wavelength, which for the present purpose is the rest wavelength of the spectral line.
The asymmetries between the blue and red lobe areas of the Stokes-$V$ profiles, $A_{\rm b}$ and $A_{\rm r}$, and the amplitudes of the blue and the red lobe, $a_{\rm b}$ and $a_{\rm r}$, are defined as,
\begin{equation}
\delta A = \frac{|A_{\rm b}| - |A_{\rm r}|}{|A_{\rm b}| + |A_{\rm r}|},
\label{eq:area_asymmetry}
\end{equation}
and
\begin{equation}
\delta a = \frac{|a_{\rm b}| - |a_{\rm r}|}{|a_{\rm b}| + |a_{\rm r}|}.
\label{eq:ampli_asymmetry}
\end{equation}
%

\begin{figure}
  \setlength{\tabcolsep}{0pt}
  \centering
  \begin{tabular}{cc}
    \includegraphics[width=0.49\textwidth]{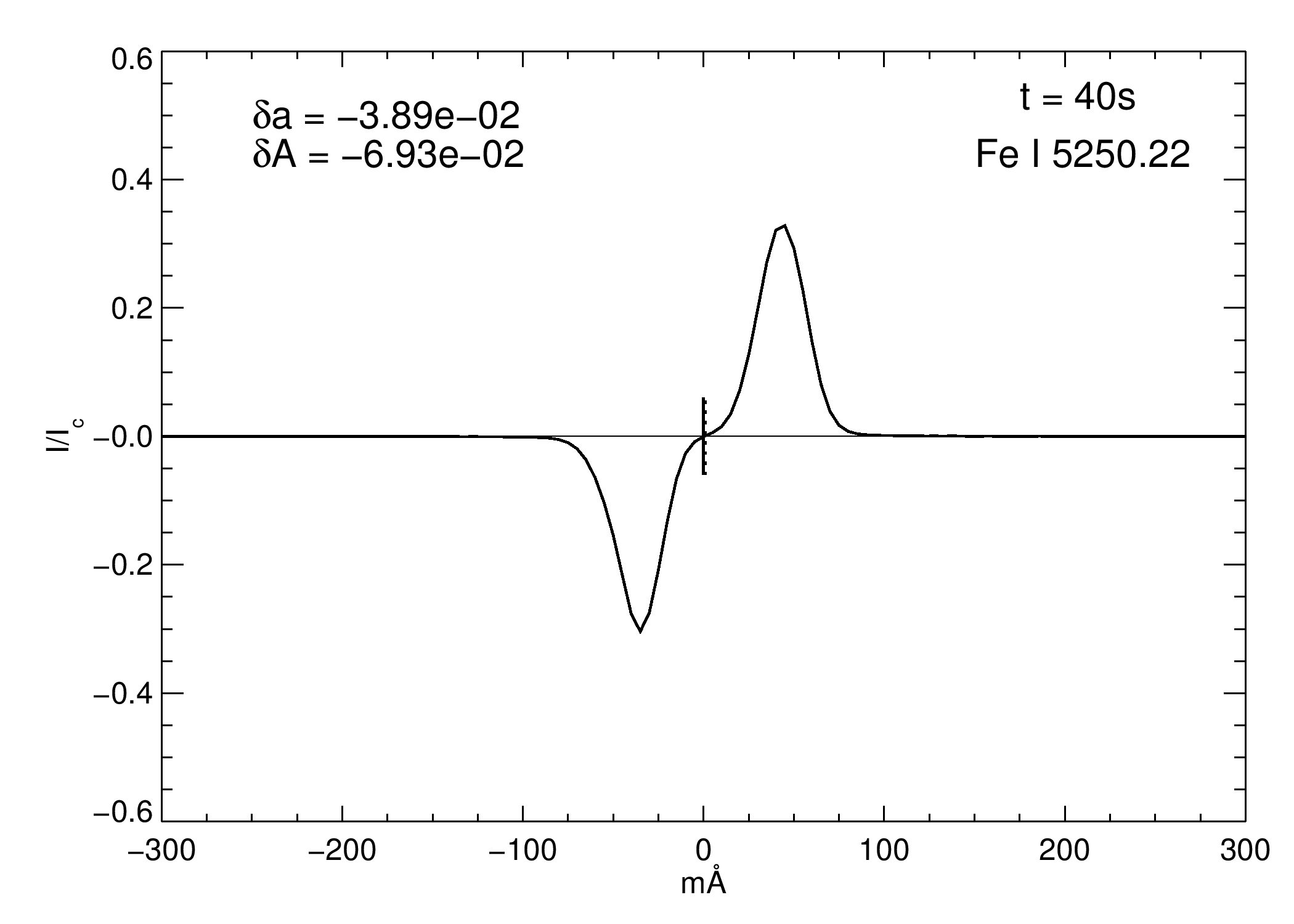} &
    \includegraphics[width=0.49\textwidth]{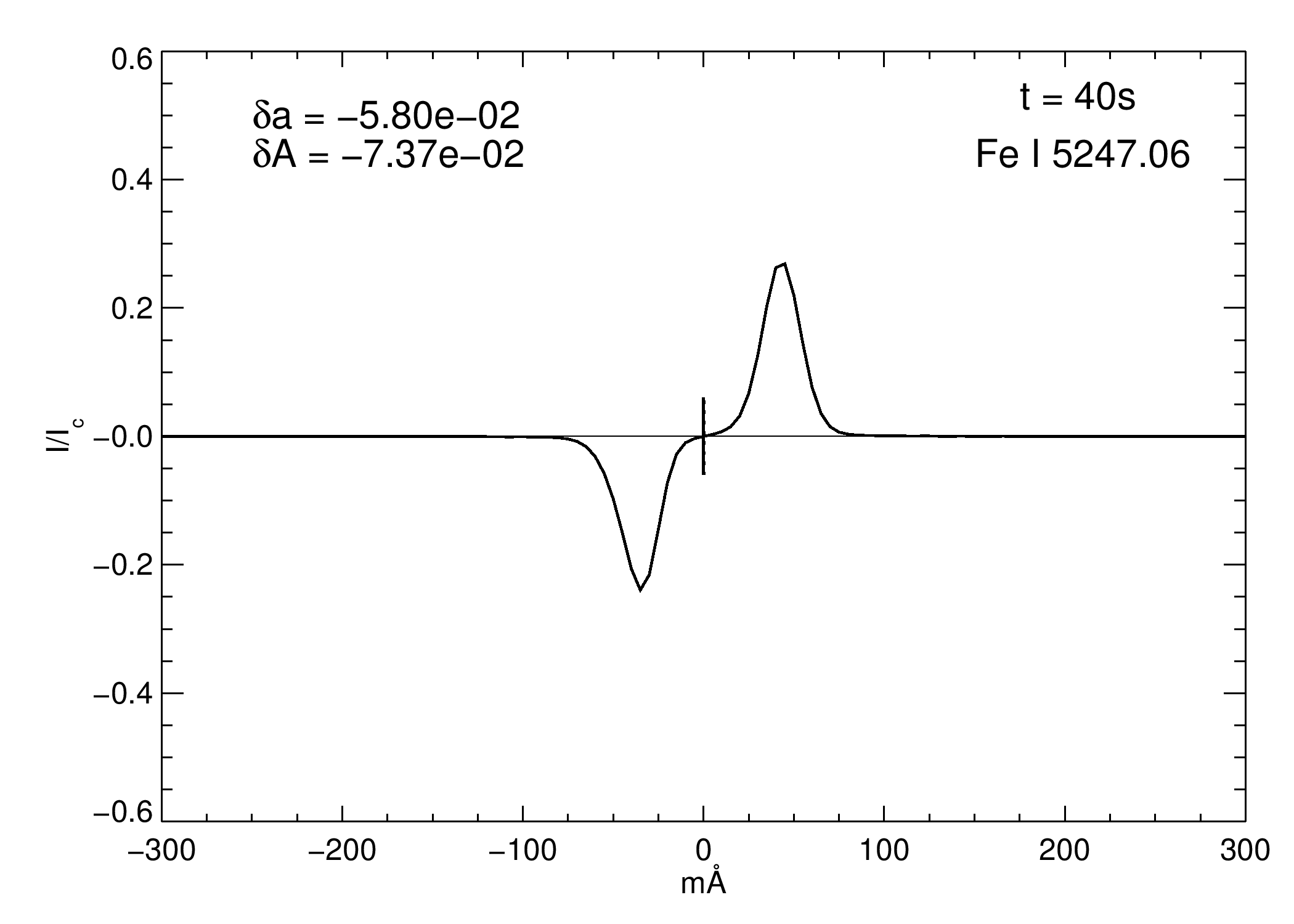}\\[-5pt]
{\small a)}&{\small b)}\\
    \includegraphics[width=0.49\textwidth]{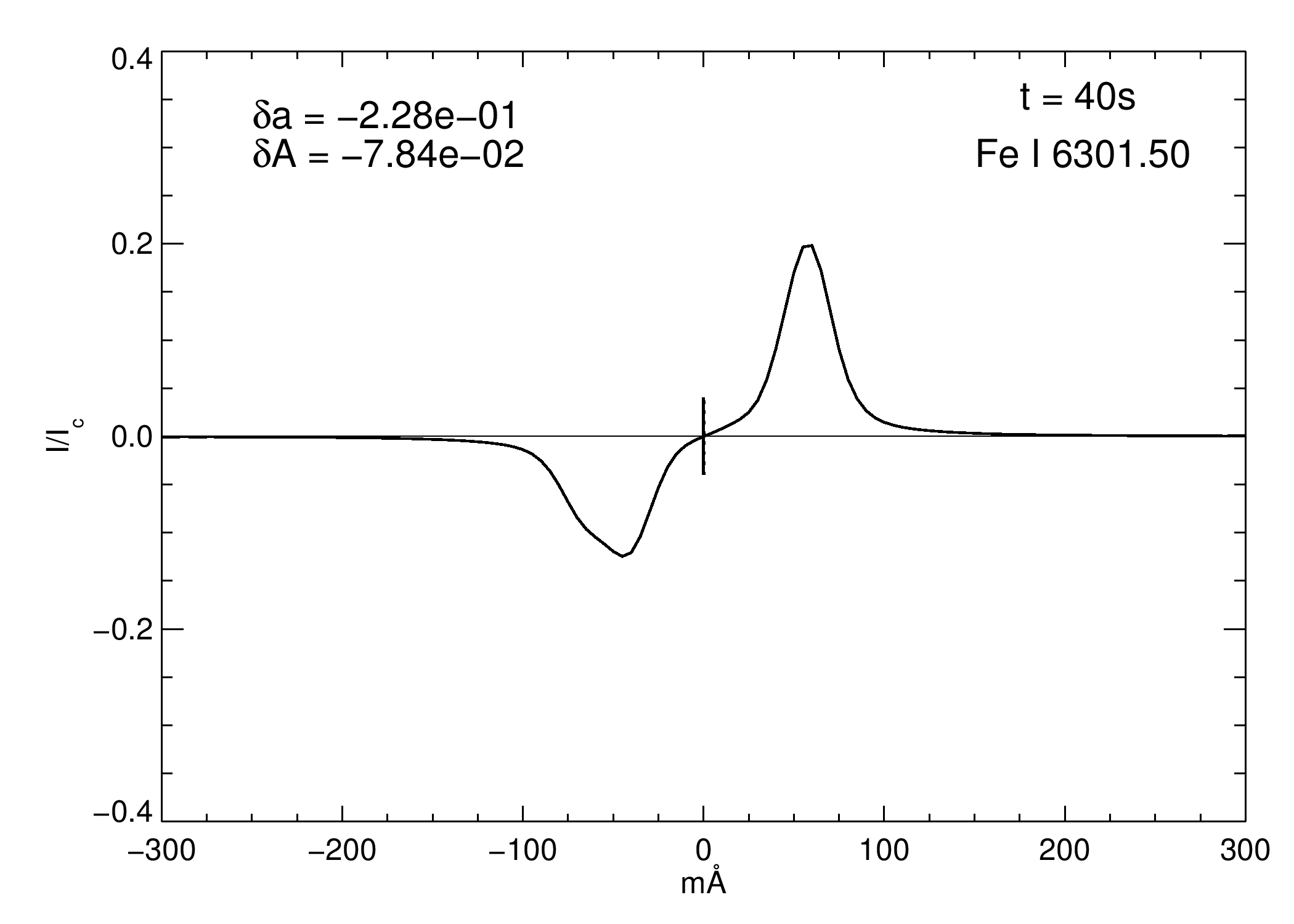} &
    \includegraphics[width=0.49\textwidth]{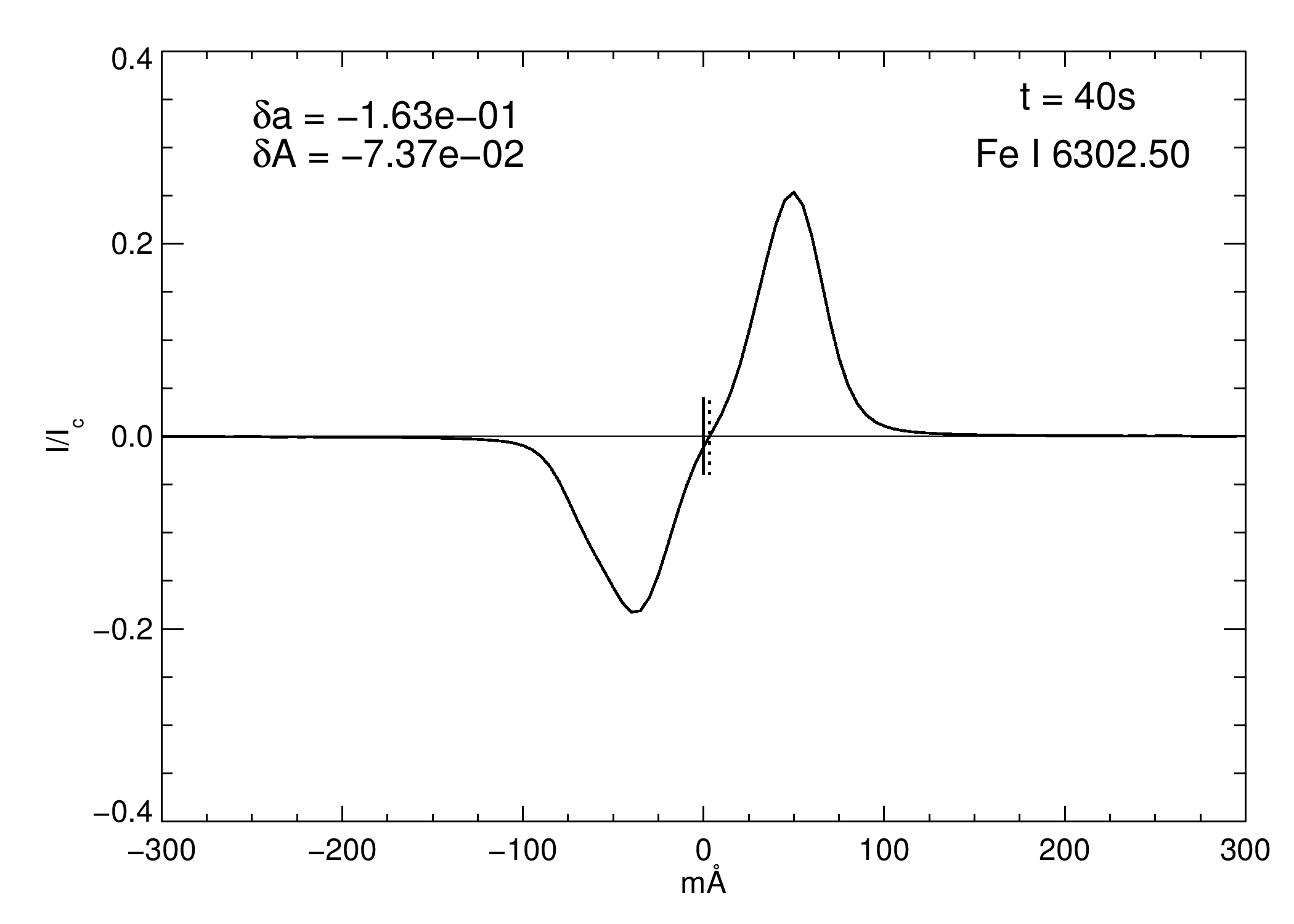}\\[-5pt]
{\small c)}&{\small d)}\\
  \end{tabular}
  \caption{Stokes-$V$ profiles of Fe~\textsc{i} (a) $\lambda$~5250.2~\AA, (b) $\lambda$~5247.06~\AA, (c) $\lambda$~6301.5~\AA, and (d) $\lambda$~6302.5~\AA\ at an elapsed time of 40~s from the vertical lines of sight in a slice ranging from $x=410$~km to $x=610$~km (left of the axis). The magnetic flux sheet has a field strength of 1000~G on the axis at $z=0$. The solid vertical line marks the unshifted central wave-length position, the dotted line the zero-crossing position of Stokes $V$.}
  \label{fig:Stokes_1000_left}
\end{figure}

%
\subsection{Moderate Field Case}
\begin{figure}
  \setlength{\tabcolsep}{0pt}
  \centering
  \begin{tabular}{cc}
    \includegraphics[width=0.49\textwidth]{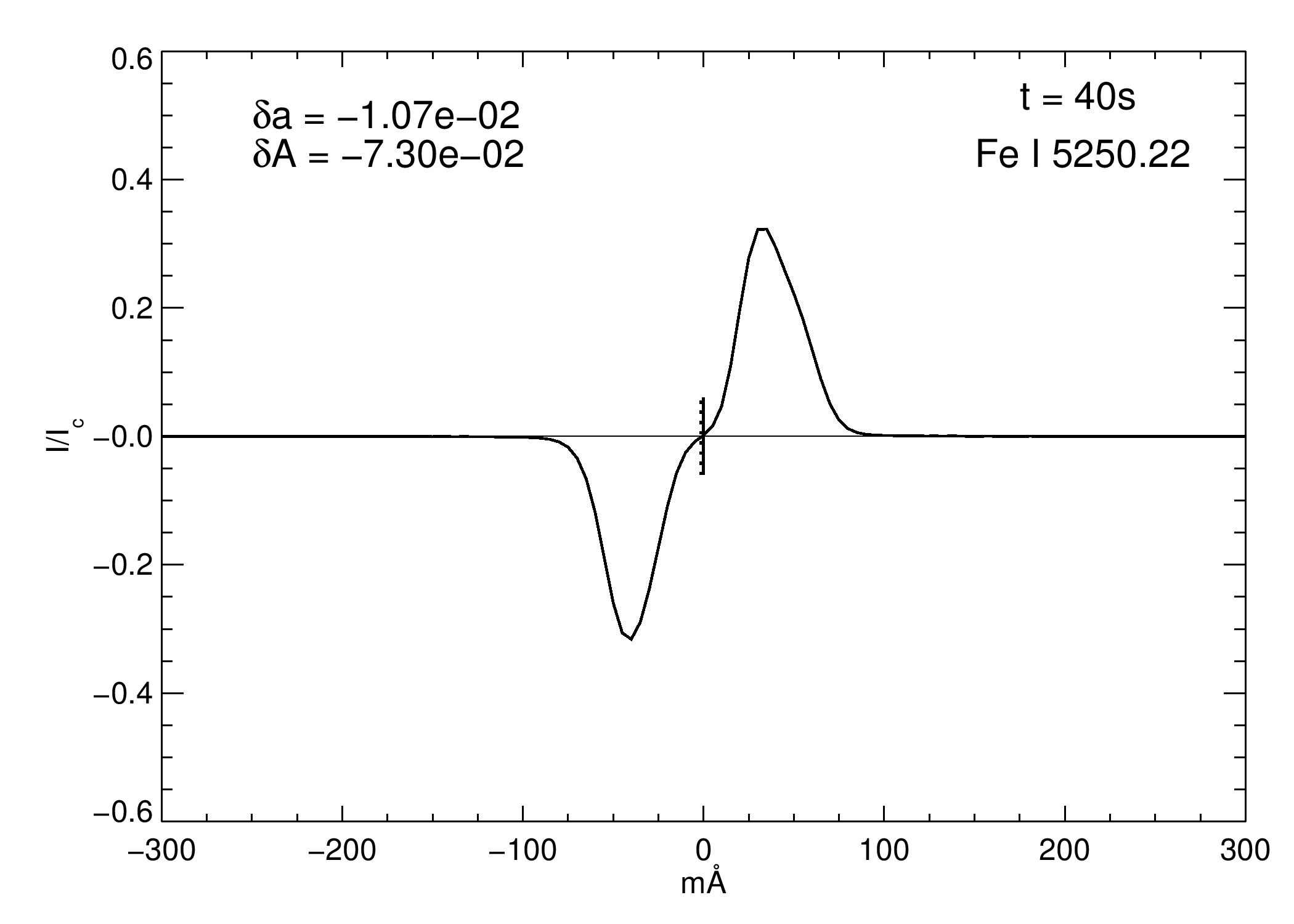} &
    \includegraphics[width=0.49\textwidth]{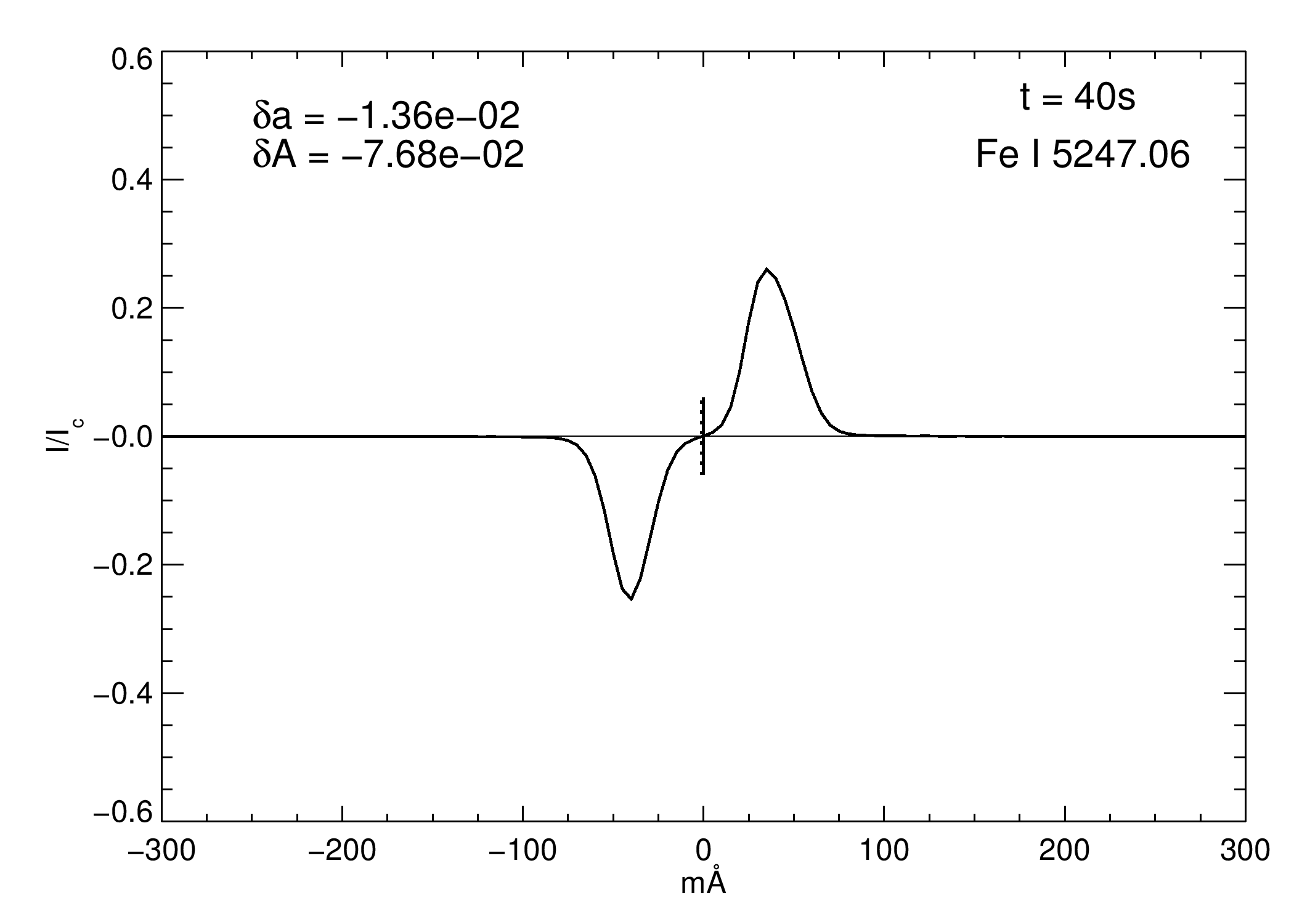}\\[-5pt]
    {\small a)} &{\small b)}\\
    \includegraphics[width=0.49\textwidth]{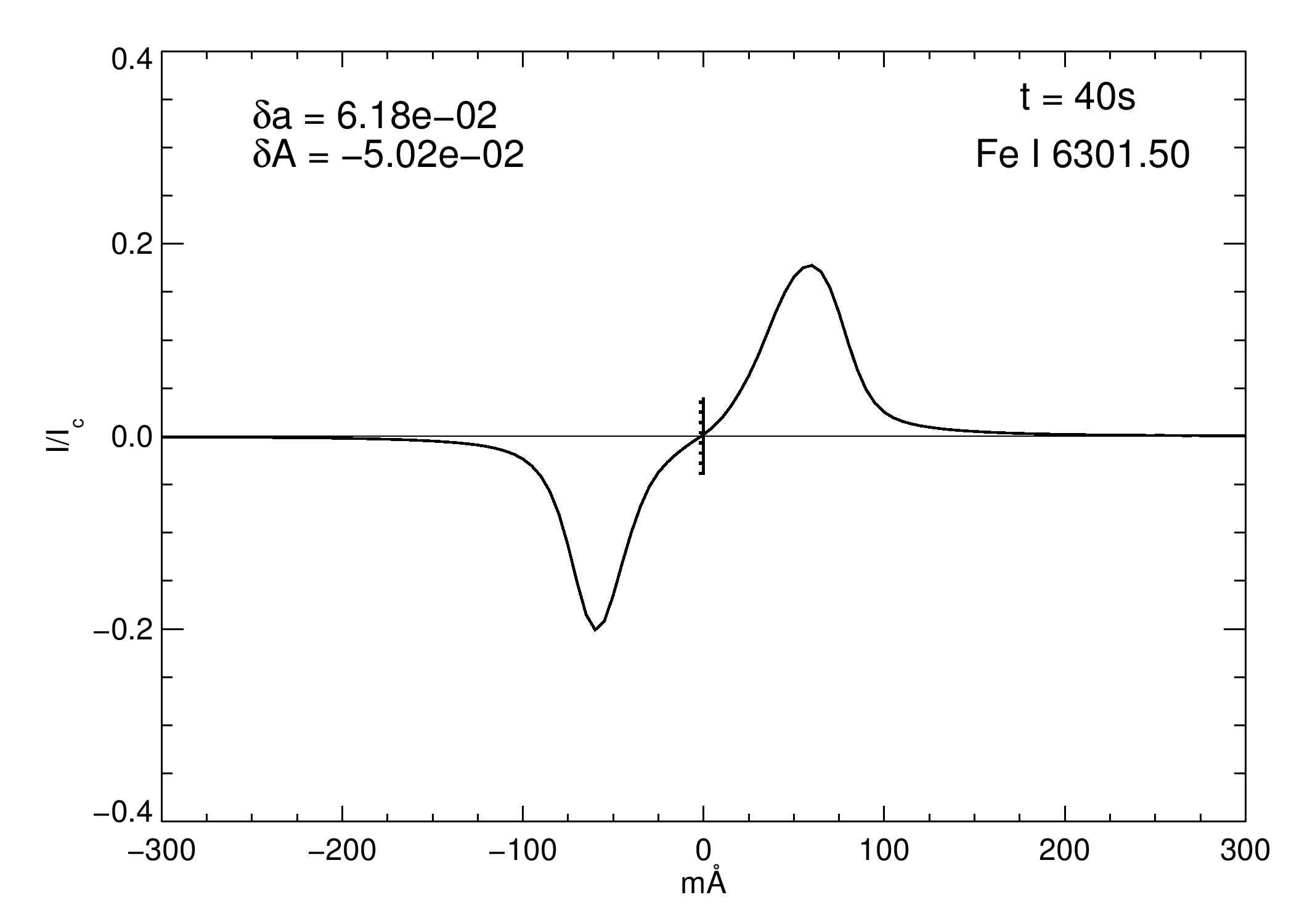} &
    \includegraphics[width=0.49\textwidth]{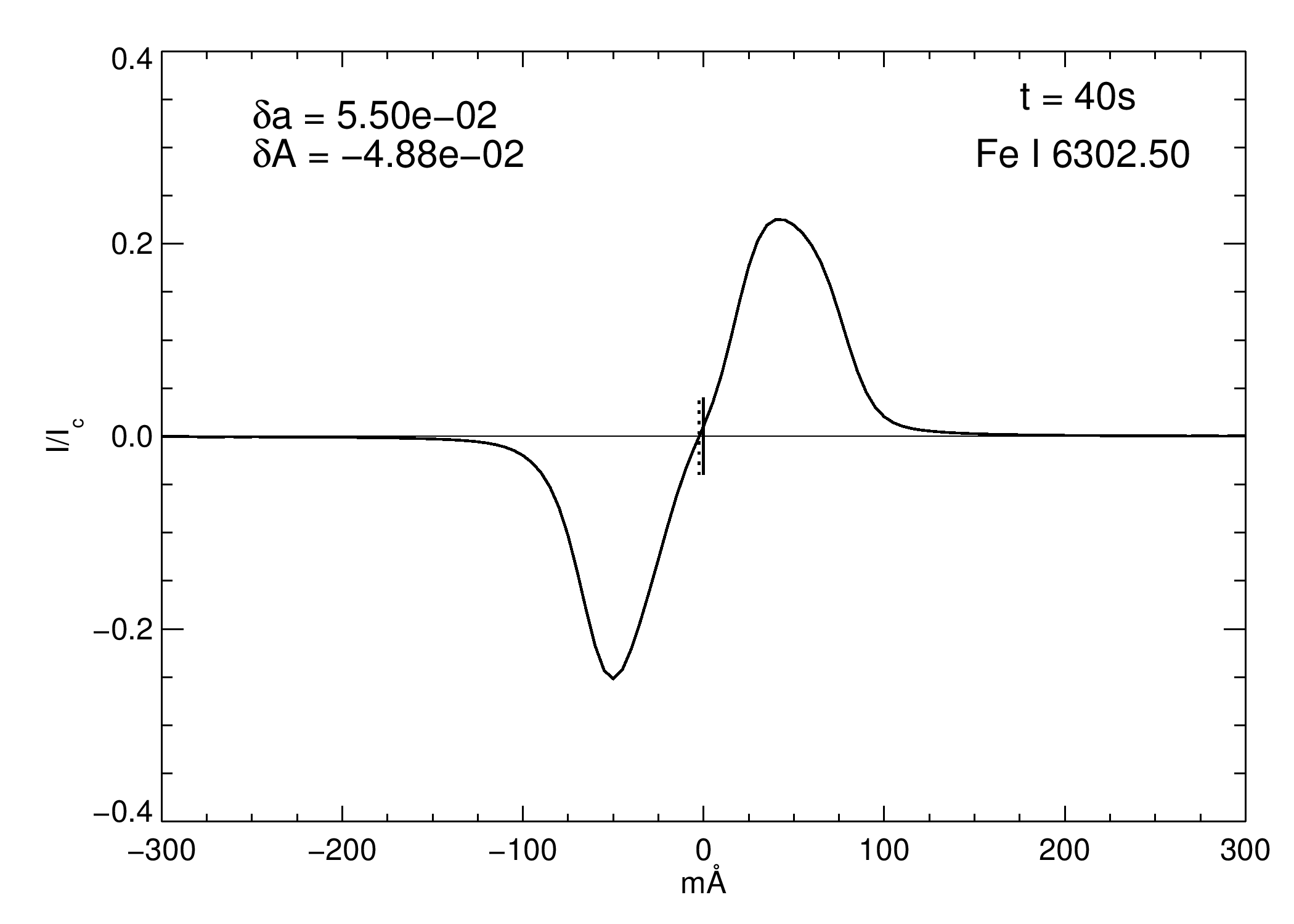}\\[-5pt]
    {\small c)}& {\small d)}\\
  \end{tabular}
  \caption{Stokes-$V$ profiles of Fe~\textsc{i} (a) $\lambda$~5250.2~\AA, (b) $\lambda$~5247.06~\AA, (c) $\lambda$~6301.5~\AA, and (d) $\lambda$~6302.5~\AA\ at an elapsed time of 40~s from the vertical lines of sight in a slice ranging from $x=670$~km to $x=870$~km (right of the axis). The magnetic flux sheet has a field strength of 1000~G on the axis at $z=0$. The solid and dotted vertical lines are as in Figure~\ref{fig:Stokes_1000_left}.}
  \label{fig:Stokes_1000_right}
\end{figure}

Figures~\ref{fig:Stokes_1000_left} and \ref{fig:Stokes_1000_right} show snapshots of Stokes-$V$ profiles of Fe~\textsc{i} $\lambda\lambda$~5250.2, 5247.06, 6301.5, and 6302.5~\AA, at time $t=40$~s after the start of the simulation of the moderate field case (see Figure~\ref{fig:1000_delta_t} for reference). Figure~\ref{fig:Stokes_1000_left} shows the Stokes-$V$ profiles averaged over a horizontal distance from $x=410$~km to $x=610$~km (left of the symmetry axis) and Figure~\ref{fig:Stokes_1000_right} shows the profiles averaged over $x=670$~km to $x=870$~km (right of the symmetry axis). The lines Fe~\textsc{i} $\lambda$~5250.2~\AA\ and Fe~\textsc{i} $\lambda$~5247.06~\AA\ belong to the same multiplet of iron, differing only in the effective Land\'{e} factor, which are 3 and 2, respectively. Hence, the Stokes-$V$ amplitudes are different for the two lines and scale approximately according to the ratio given by the Land\'{e} factor as 3:2. This can be seen in the plots shown in Figure~\ref{fig:Stokes_1000_left}, where the amplitude of Fe~\textsc{i} $\lambda$~5247.06~\AA\ is lower than that of Fe~\textsc{i} $\lambda$~5250.2~\AA. Similarly, the amplitudes of Fe~\textsc{i} $\lambda$~6301.5~\AA\ and Fe~\textsc{i} $\lambda$~6302.5~\AA\ scale according to the Land\'{e} factors of 1.67 and 2.5, respectively. However, the amplitude ratios are not strictly according to the Land\'{e} factor ratios in the strong-field regime because of saturation effects (\opencite{stenflo1994book}).

\par
\begin{figure}
  \setlength{\tabcolsep}{0pt}
  \centering
  \begin{tabular}{cc}
    \includegraphics[width=0.49\textwidth]{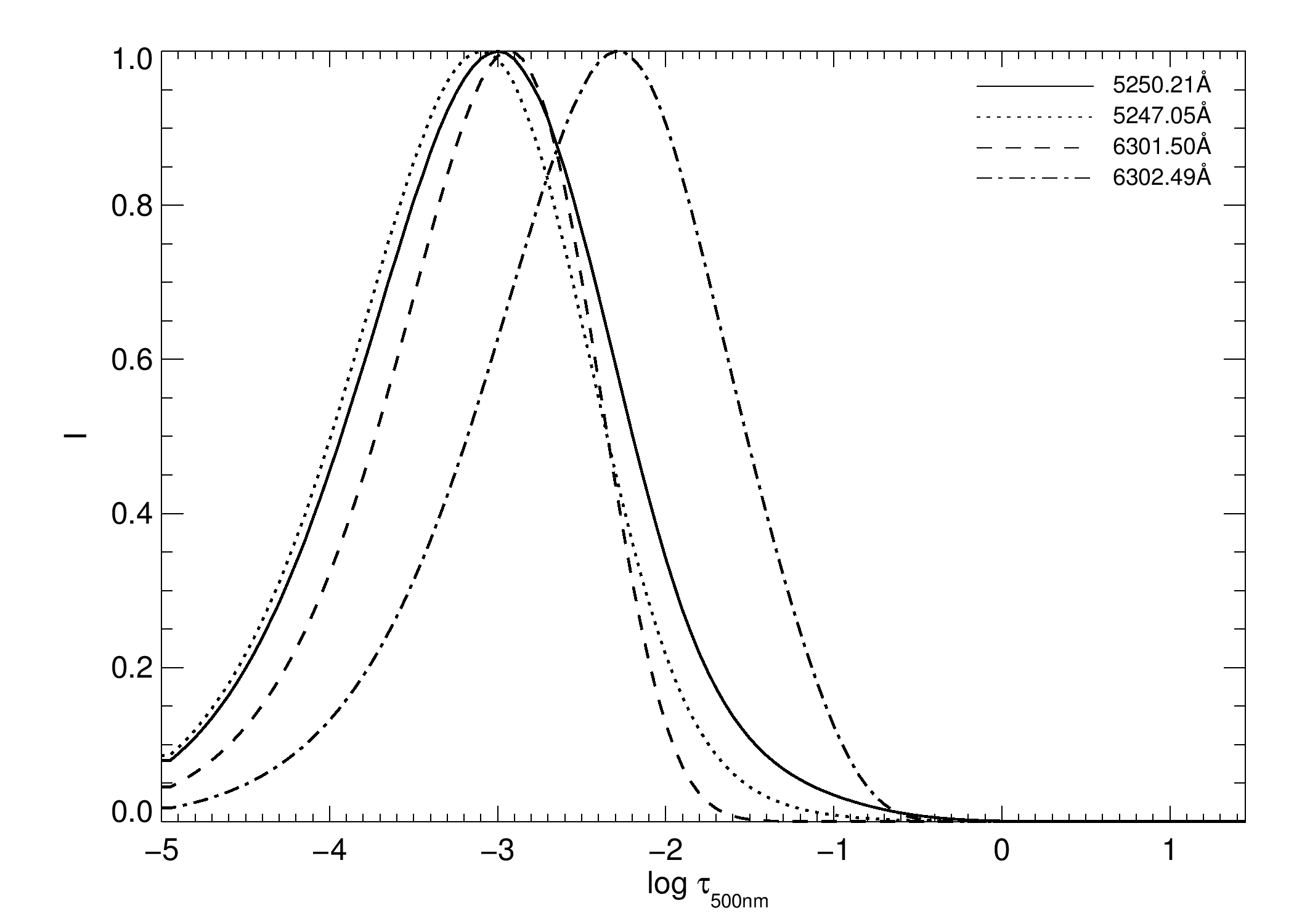}&
    \includegraphics[width=0.49\textwidth]{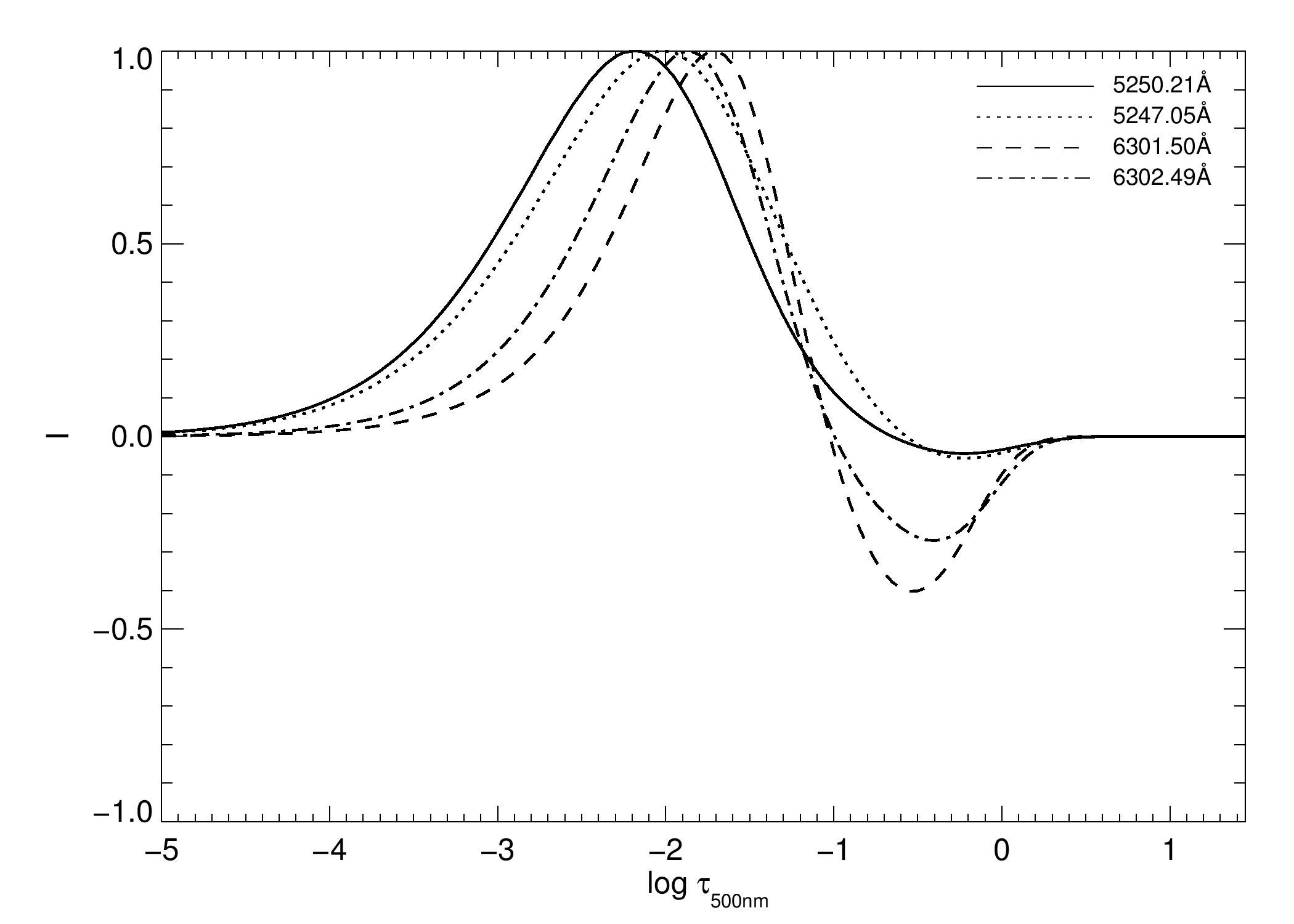}\\
{\footnotesize a)}&{\footnotesize b)}\\
  \end{tabular}
  \caption{The line depression contribution functions for the four Fe~\textsc{i} lines of Table~\ref{tab:atomic_parameters}, evaluated for a line of sight located at $x=520$~km, in the initial model with 1000~G field strength at $z=0$~km. (a) Line depression contribution functions for Stokes $I$ in the line core. (b) Corresponding contribution functions for Stokes $V$ at $\displaystyle\lambda=\lambda_{V_{\rm min}}$, \textit{i.e.}, the wavelength of the blue Stokes-$V$ peak.}
  \label{fig:stokes_contrib_function}
\end{figure}

 Apart from the thermodynamic properties of the plasma wherein the spectral lines are formed, the height of formation of the spectral lines also depends on the line strength, which in turn depends on the excitation potential and the log (\textit{gf}) value. The region of formation spans a wide range and is affected by the presence and the height variations of the magnetic field---the latter matters in particular for Stokes parameters $Q$, $U$, and $V$. The line depression contribution functions for the four Fe~\textsc{i} lines are plotted in Figure~\ref{fig:stokes_contrib_function} for the initial model with moderate field strength and for the line of sight at $x=520$~km, \textit{i.e.}, in the centre of the bundle of lines of sight that was considered in Figure~\ref{fig:Stokes_1000_left}. Figure~\ref{fig:stokes_contrib_function}a shows the Stokes-$I$ line depression contribution functions in the line cores and Figure~\ref{fig:stokes_contrib_function}b shows the Stokes-$V$ line depression contribution functions at the wavelength position of minimal $V$ signals at $\displaystyle\lambda_{V_{\rm min}}$. For the definition of the Stokes line depression contribution function we refer to
\inlinecite{grossmann1988}.
Clearly, the maximum contribution to both Stokes $I$ and $V$ comes for the doublet Fe~\textsc{i} 5250.22~\AA\ and Fe~\textsc{i} 5247.05~\AA\ from higher layers in the atmosphere compared to Fe~\textsc{i} 6301.50~\AA\ and Fe~\textsc{i} 6302.50~\AA.
This difference in the relative height of formation of these lines is consistent with the theoretical models of
\inlinecite{khomenko2007}
and multi-line spectropolarimetric observations of
\inlinecite{socas2008}.


The asymmetry in the Stokes-$V$ profiles after 40~s can be clearly seen in the Fe~\textsc{i} 6301.5 and 6302.5~\AA\ lines plotted in Figures~\ref{fig:Stokes_1000_left} and \ref{fig:Stokes_1000_right}. The effect of the wave propagation is first sensed by these lines as they are formed lower in the atmosphere than the other two lines. The Stokes-$V$ asymmetries as functions of time for the above four lines give a clearer picture. Figure~\ref{fig:Stokes_1000_evolution} shows (a) the Stokes-$V$  amplitude asymmetry and (b) the Stokes-$V$ area asymmetry for the four Fe~\textsc{i} lines as functions of time. The red and blue colours represent the narrow bundles of lines of sight on the two sides of the flux-sheet axis. The blue solid curves correspond to lines of sight spanning $x=410$~km to $x=610$~km
(left) and the blue dashed curves for $x=120$~km to $x=320$~km (far left). Similarly on the right side, the red solid curves are for $x=670$~km to $x=870$~km (right) and the red dashed curves are for $x=960$~km to $x=1160$~km (far right). In order to explain Figure~\ref{fig:Stokes_1000_evolution}, it is essential to look at the velocity field of the simulation, which is shown in Figure~\ref{fig:1000_delta_t}. There, the vertical lines mark the boundaries of the line-of-sight bundles.

%
\subsubsection{Amplitude Asymmetry}
\begin{figure}[]
\centering
\begin{tabular}{cc}
\includegraphics[scale=0.3]{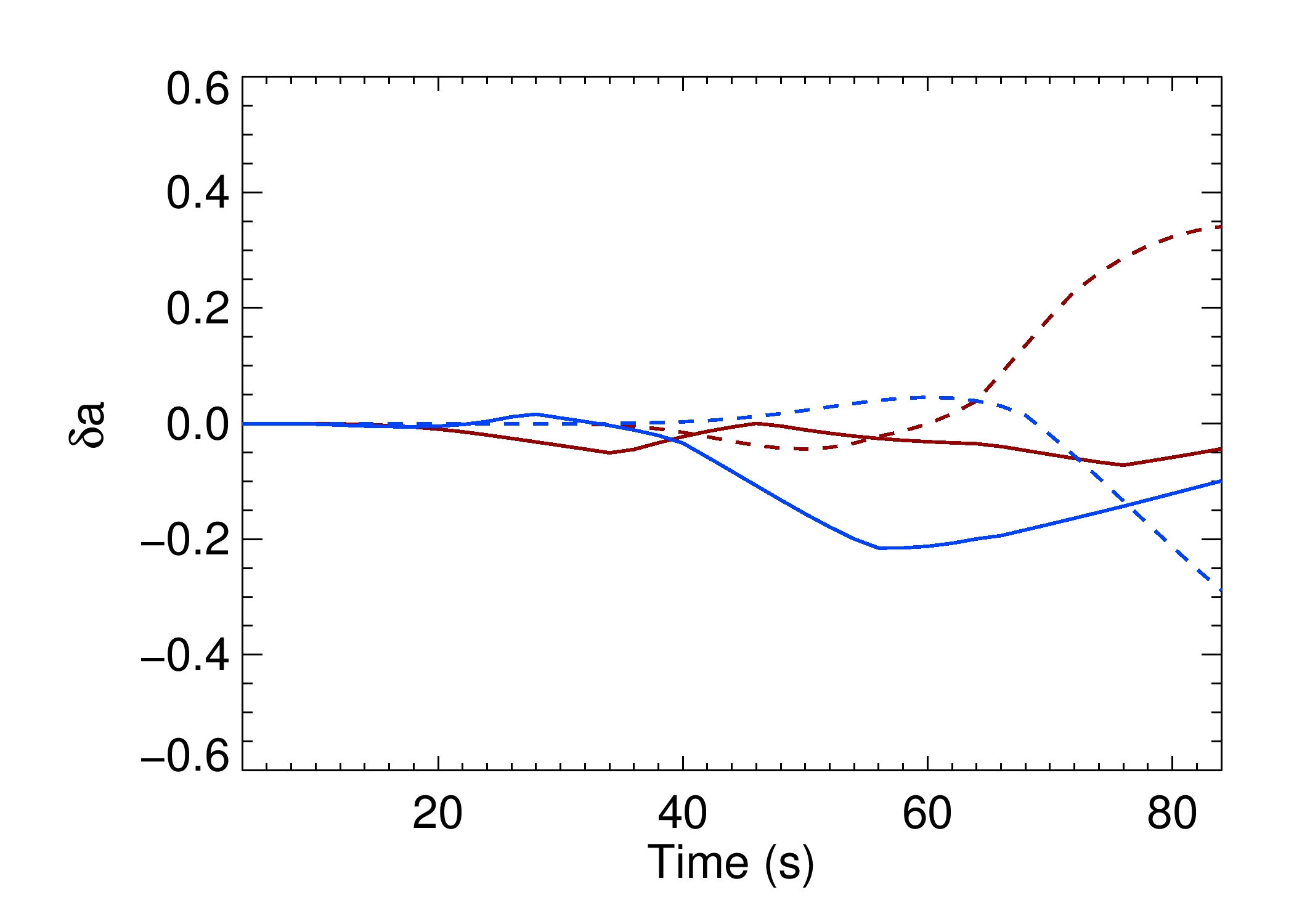}  &
\includegraphics[scale=0.3]{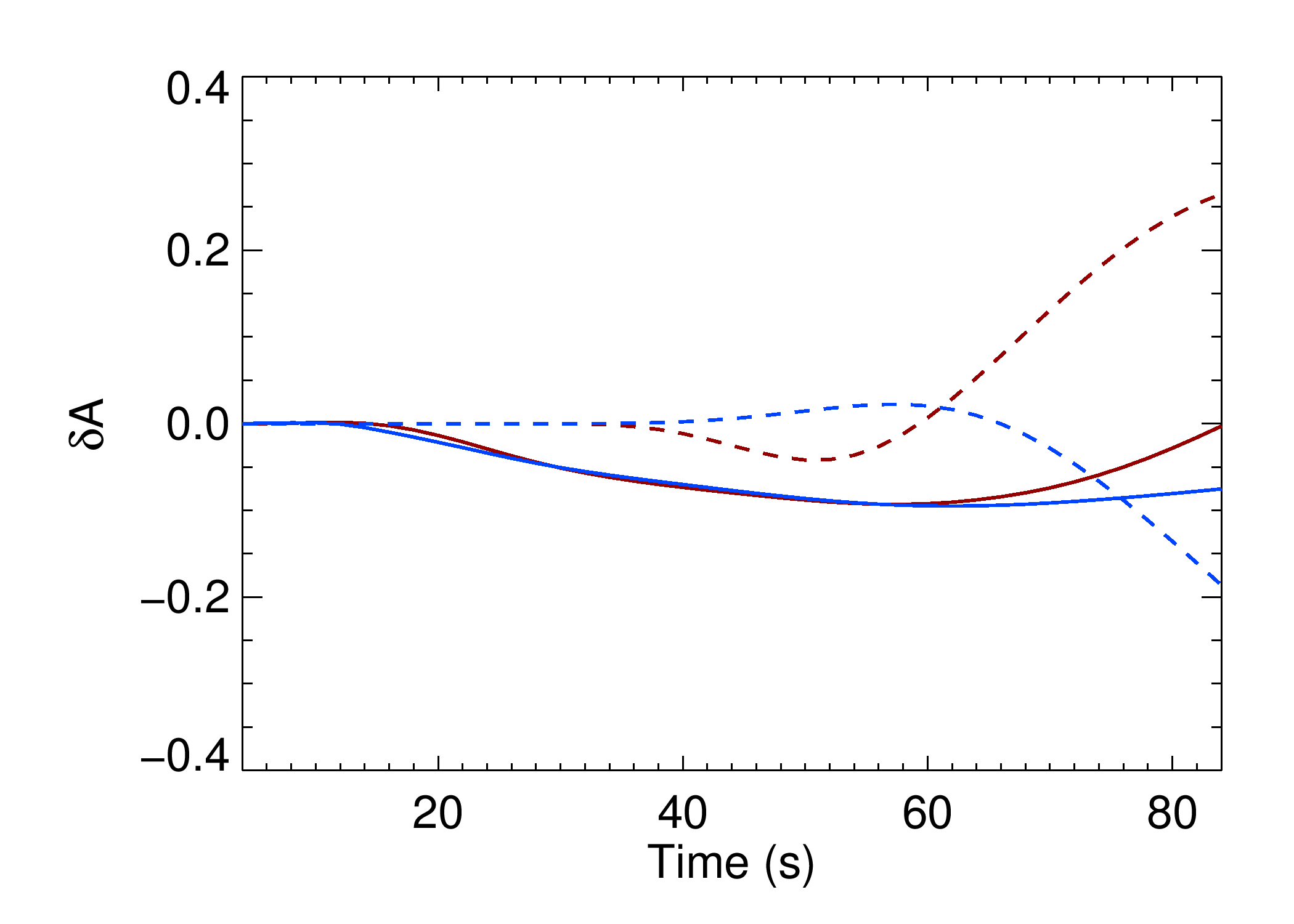} \\[-10pt]
\multicolumn{2}{c}{\footnotesize i) Fe~\textsc{i} $\lambda$ 5247.06~\AA}\\
\includegraphics[scale=0.3]{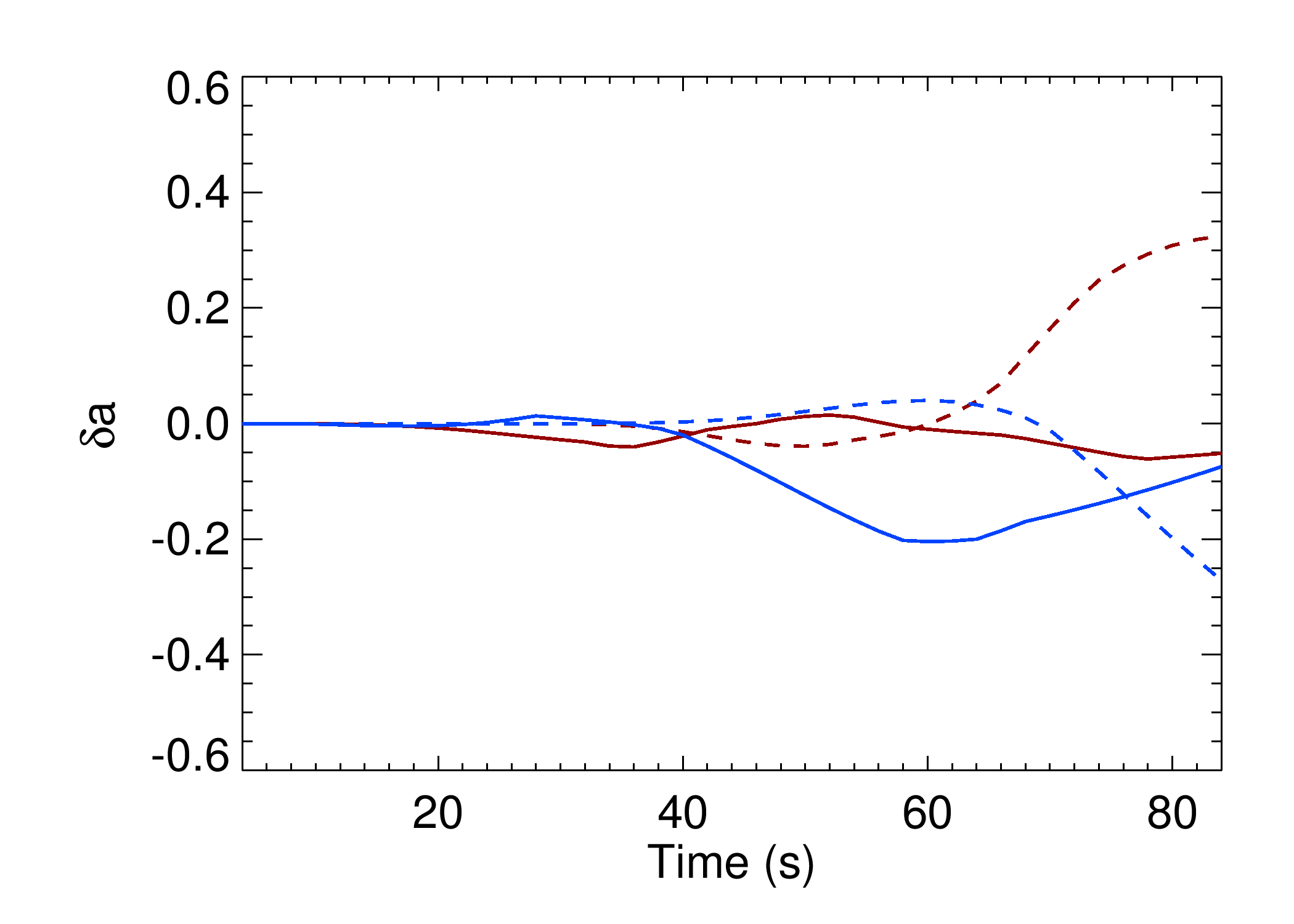}  &
\includegraphics[scale=0.3]{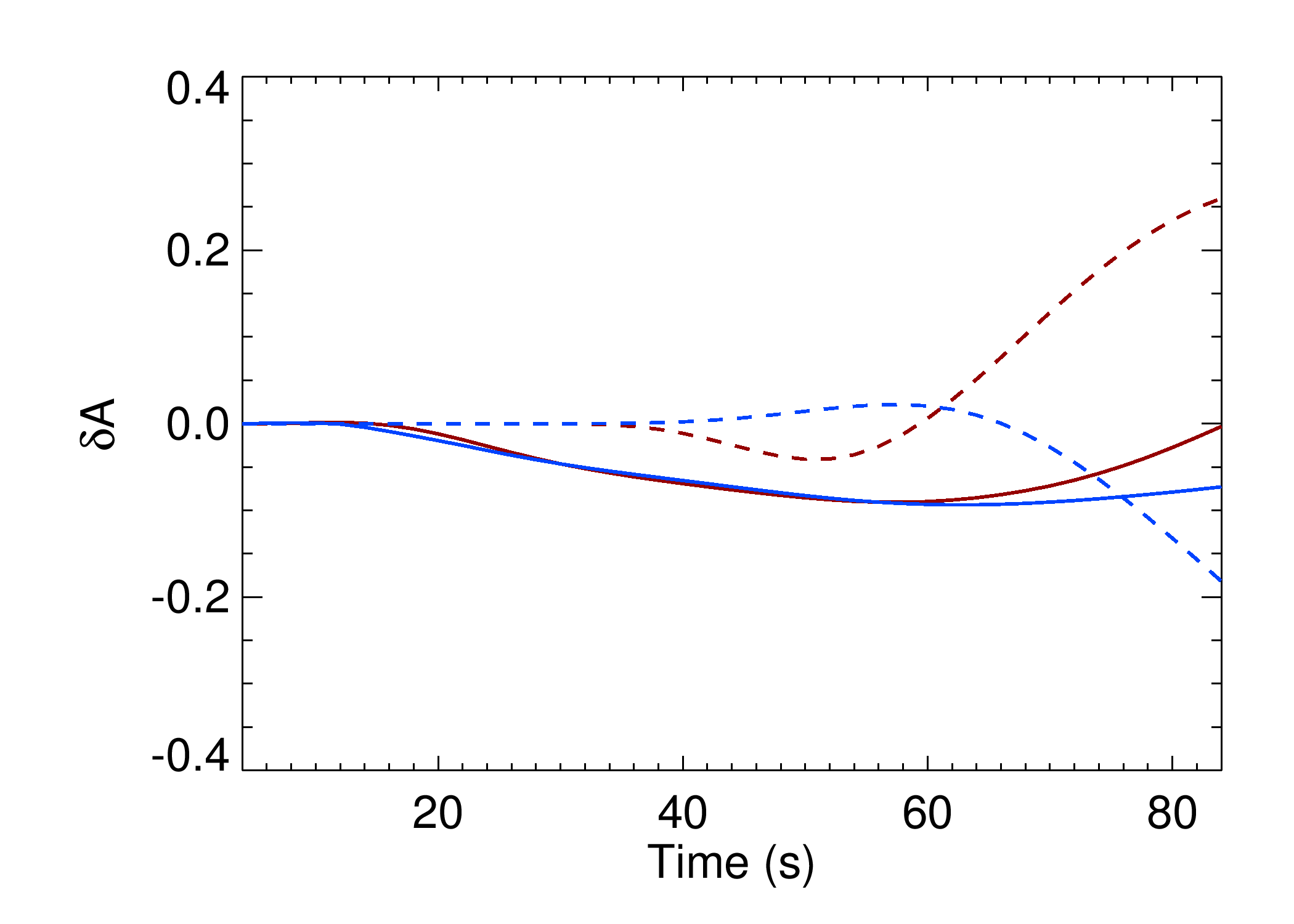} \\[-10pt]
\multicolumn{2}{c}{\footnotesize ii) Fe~\textsc{i} $\lambda$ 5250.2~\AA}\\
\includegraphics[scale=0.3]{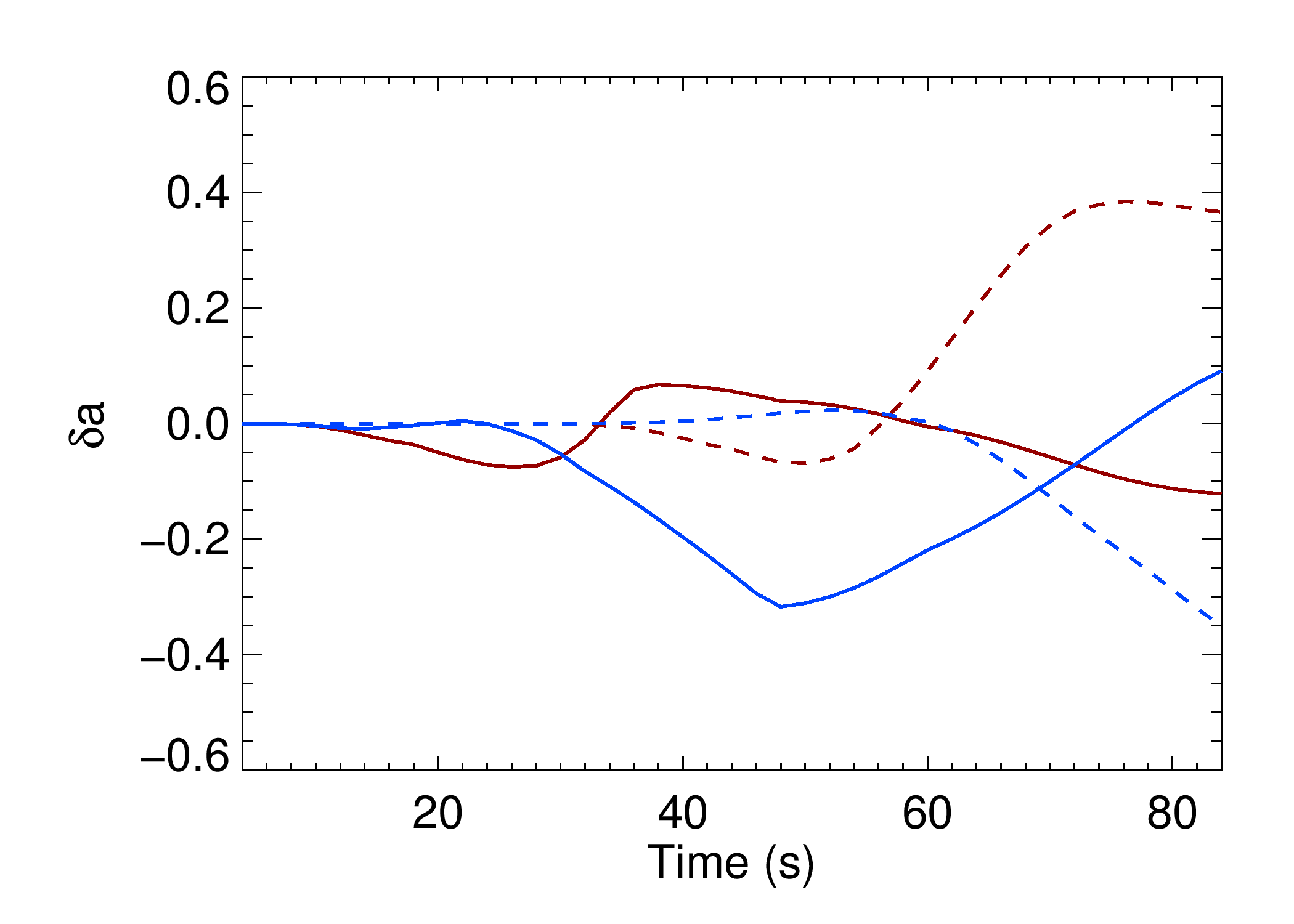}  &
\includegraphics[scale=0.3]{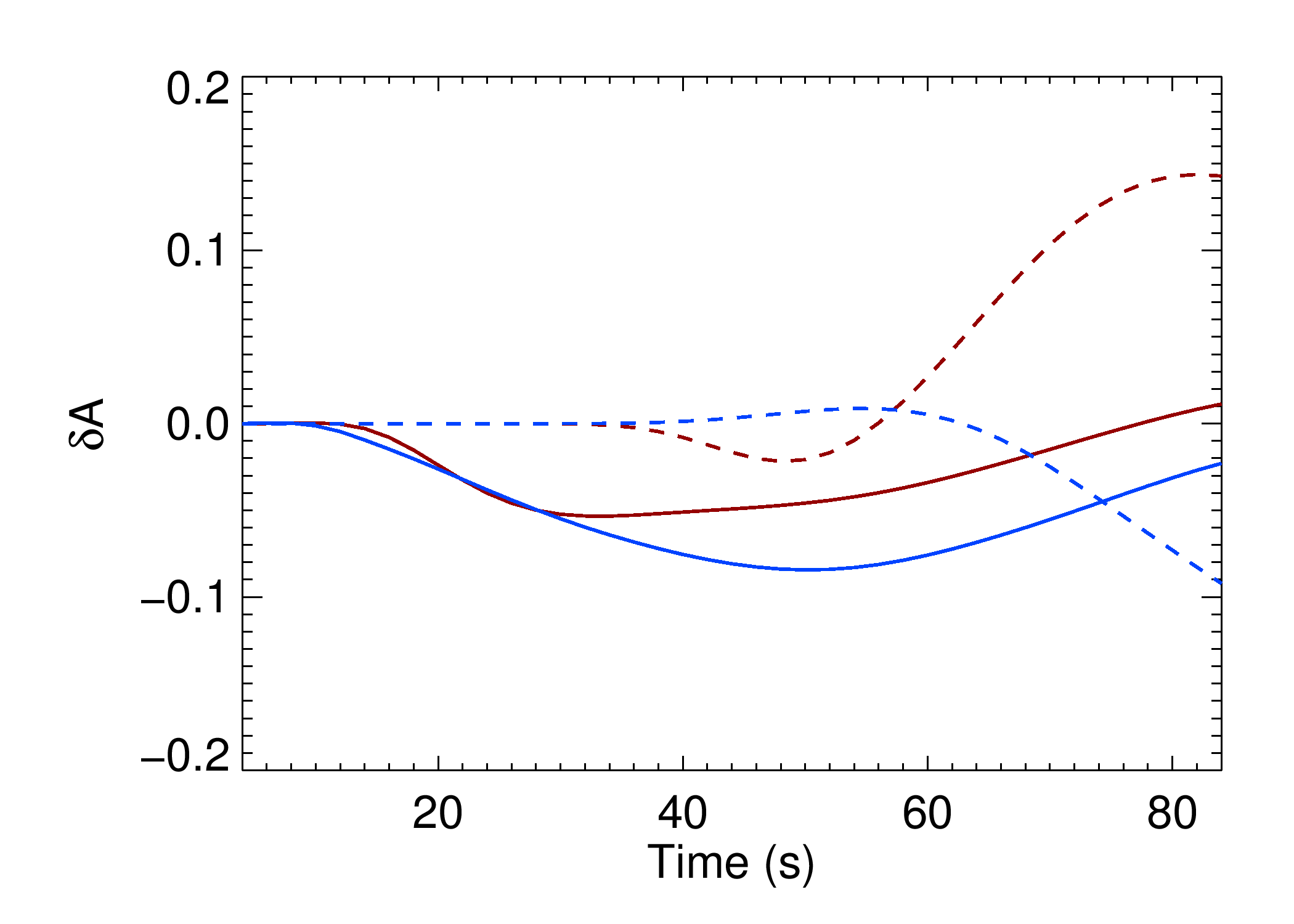} \\[-10pt]
\multicolumn{2}{c}{\footnotesize iii) Fe~\textsc{i} $\lambda$ 6301.5~\AA}\\
\includegraphics[scale=0.3]{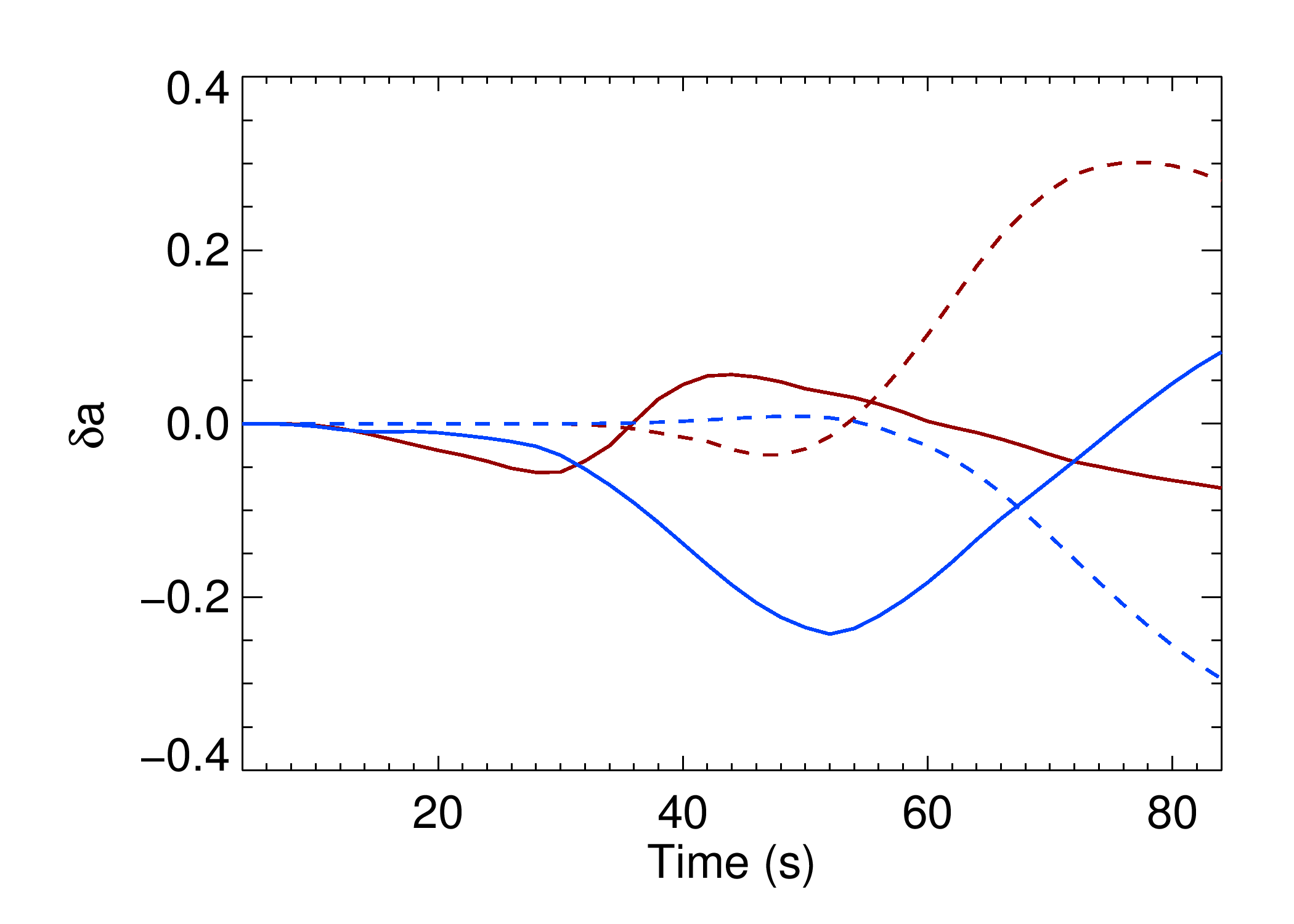}  &
\includegraphics[scale=0.3]{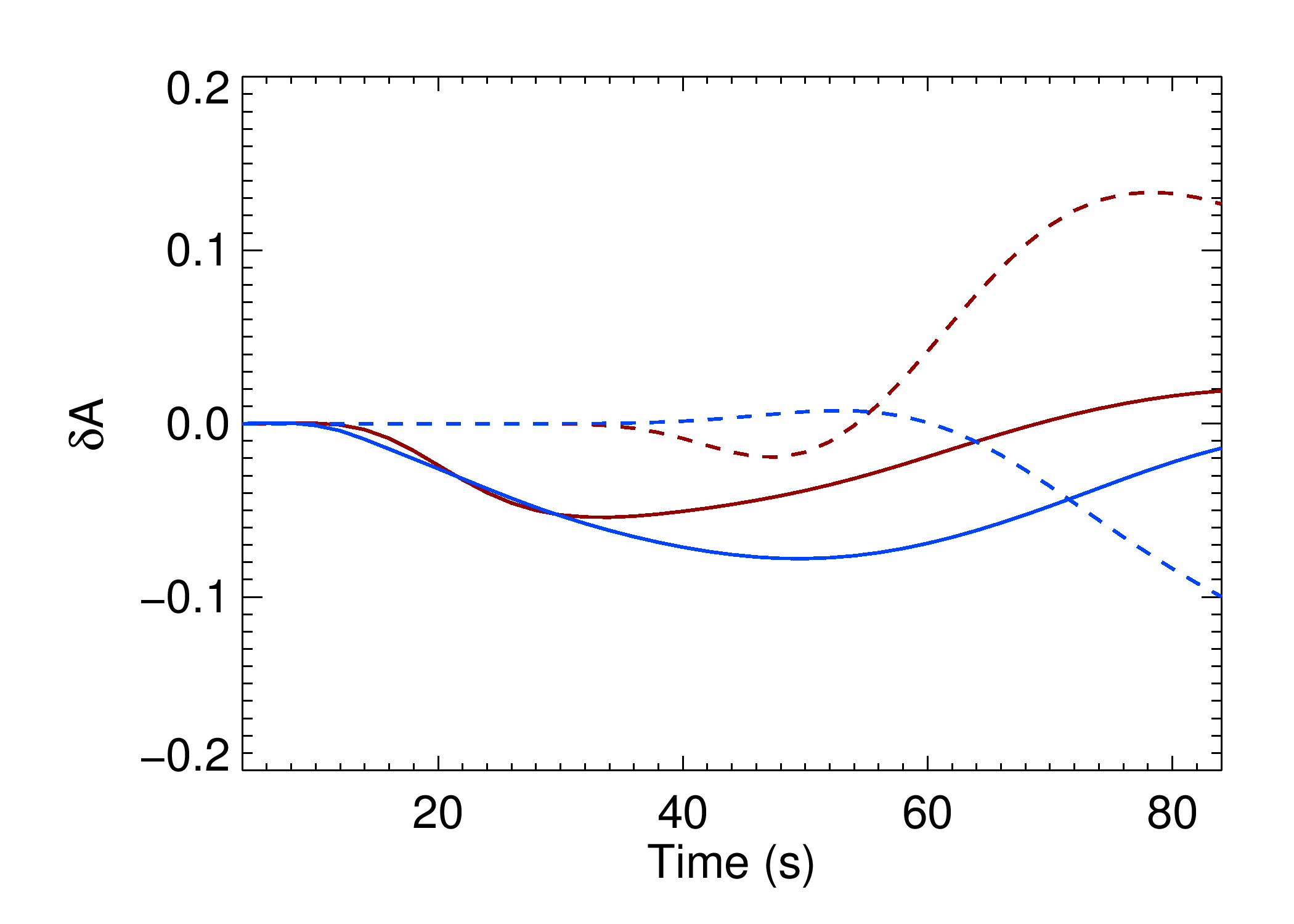} \\[-10pt]
\multicolumn{2}{c}{\footnotesize iv) Fe~\textsc{i} $\lambda$ 6302.5~\AA}\\
{\small a) Amplitude asymmetry} & b) {\small Area asymmetry} \\
\end{tabular}
\caption{The Stokes-$V$ (a) amplitude asymmetry and (b) area asymmetry for the four {Fe~\textsc{i}} lines listed in Table~\ref{tab:photo_equilibrium_table} as functions of time for the moderate field case with 1000~G. The red solid curves represent the bundle of lines of sight on the right side of the axis. The red dashed curves represent the bundle of lines of sight on the far right. The blue solid curves are for the left slice and blue dashed curves are for the far left bundle.
}
\label{fig:Stokes_1000_evolution}
\end{figure}

Initially, there are no velocities inside the domain and hence the Stokes profiles are anti-symmetric and therefore $\delta a$ is zero. After the start of the simulation, velocities start building up in the domain, consequently giving rise to asymmetries in the Stokes profiles. Let us now consider the bundle of lines of sight on the left side of the axis and the response in $\delta a$ for the lines Fe~\textsc{i} 6301.5 or 6302.5~\AA\ as shown in Figure~\ref{fig:Stokes_1000_evolution}a.
After 20~s, there develops a small downdraft within the (blue) bundle of lines of sight in the magnetic region. This leads to a red shift of the Stokes-$V$ profile that forms in this region, \textit{viz.}, in the lines Fe~\textsc{i} 6301.5 and 6302.5~\AA. The spectral line emerging from the quasi-static, partially field-free layer further below is less affected by Doppler shifts. It causes an asymmetric illumination of the two flanks of the red-shifted line contribution formed further above. This results in the blue lobe of the emerging Stokes-$V$ profile be suppressed and consequently the amplitude asymmetry tends to become negative. The amplitude asymmetry starts decreasing and reaches a minimum at around 50~s, after which time the downdraft moves out of the line formation region and a following updraft starts to dominate (see Figure~\ref{fig:1000_delta_t}), making the blue lobe strong again and the amplitude asymmetry to rise.

On the other hand, the line of sight on the right side of the axis shows an inverse time dependence, which comes about because the waves are 180$^{\circ}$ out of phase on the opposite sides of the flux-sheet axis (Section~\ref{ss:moderate_case}).
There develops a strong updraft within the line of sight after $\approx 30$\,s into the simulation.
The contribution to the line formation that stems from within this updraft is then shifted to the blue relative to the line emerging from the quasi-static layer below. This situation suppresses the red lobe of the emerging Stokes-$V$ profile, as long as the latter is mainly formed within the updraft, making the amplitude asymmetry to rise towards positive values.
This trend is seen until slightly after 40~s when the downdraft of a following wave phase replaces the updraft in these regions. The maximum value is reached before the time of minimum value of $\delta a(t)$ on the left hand side because the wave on the right hand sides of the flux-sheet axis is preceding the wave on the left hand side.
However, we notice that the two curves for $\delta a(t)$ are not symmetric (relative to the time axis)---the
(red) curve for the right side shows initially even a slight trend towards negative values like the (blue) curve for the left side and its amplitude is smaller than that for the left side. This asymmetry is due to the fact that the excitation of the flux sheet is asymmetric too. The flux sheet continuously moves to
the right side until it comes to a halt after 12~s and 38.2~km to the right of the initial symmetry axis
(from Equation~(\ref{eq:velocity})). This can be seen in Figure~\ref{fig:1000_delta_t}, where the blue solid
and the red solid lines of sight sample more peripheral and more central parts of the flux sheet, respectively. We found that the opposite temperature perturbations on the opposite sides of
the flux-sheet axis has only a minor effect on the behaviour of $\delta a(t)$. This is also true for
the transversal magnetic field component.

In the case of lines formed higher in the atmosphere, like the lines Fe~\textsc{i} 5250.2~\AA\ and 5247.06~\AA, we see a similar behaviour like described above but with a time lag, depending on the arrival time of the perturbations. We see a delay of 9~s between the two pairs of lines corresponding to roughly a distance of 58~km, since the sound speed at these heights is around 6.5~km\,s$^{-1}$ (see Table~\ref{tab:photo_equilibrium_table}). This distance corresponds to approximately the distance in $\log\tau$ from $-1.8$ to $-2.1$ of the maximum peaks of the Stokes-$V$ contribution functions for the two line pairs as can be seen from Figure~\ref{fig:stokes_contrib_function}b.

We now consider the lines of sight further away from the flux-sheet axis. These are depicted as dashed vertical lines in Figure~\ref{fig:1000_delta_t} and the dashed curves in Figure~\ref{fig:Stokes_1000_evolution}a show the corresponding asymmetry $\delta a$. The perturbations become significant in this region after 50~s only when the first front arrives. On the right side of the flux sheet, we see that the velocities are directed upward. This suppresses the red lobe making the amplitude asymmetry positive. On the left side of the flux sheet, the downward velocities shift the amplitude asymmetry towards negative values. We notice, however,
that initially there exists the opposite tendency: especially the right side showing negative $\delta a$.
This behaviour is even more pronounced for the inner bundles of lines of sight on the right hand side and
corresponding red solid curves of Figure~\ref{fig:Stokes_1000_evolution}a. It indicates that the explanation 
provided above for the origin of the asymmetries is not complete even though it can serve as a
rough guideline. The origin of this tendency to opposite asymmetries is that, prior to the wave affecting
the formation of Stokes $V$, it affects the field-free layers below, where Stokes $I$ is already forming.
In particular, the updraft in the leading wave on the right hand side first causes Stokes $I$ to be
blue-shifted, which suppresses the blue lobe of Stokes $V$ formed higher up in the magnetic region,
which is not yet affected by the wave. This causes $\delta a$ to become negative. Since the velocities
are still moderate at this stage, the asymmetry remains moderate as well. As the wave moves further
up it grows in amplitude and enters the magnetic region in the bundles of lines of sight on the far right
side. This leads to a blue shift of the Stokes-$V$ contribution, while the velocities in the field-free
layer below are decreasing as the wave moves out of this region. Hence,
the red lobe of Stokes $V$ gets suppressed. Not until then, $\delta a$ rises to the expected positive
values and these values become substantial because the velocities are growing rapidly.

\begin{figure}[]
\setlength{\tabcolsep}{0pt}
\centering
\begin{tabular}{cc}
\includegraphics[width=0.49\textwidth]{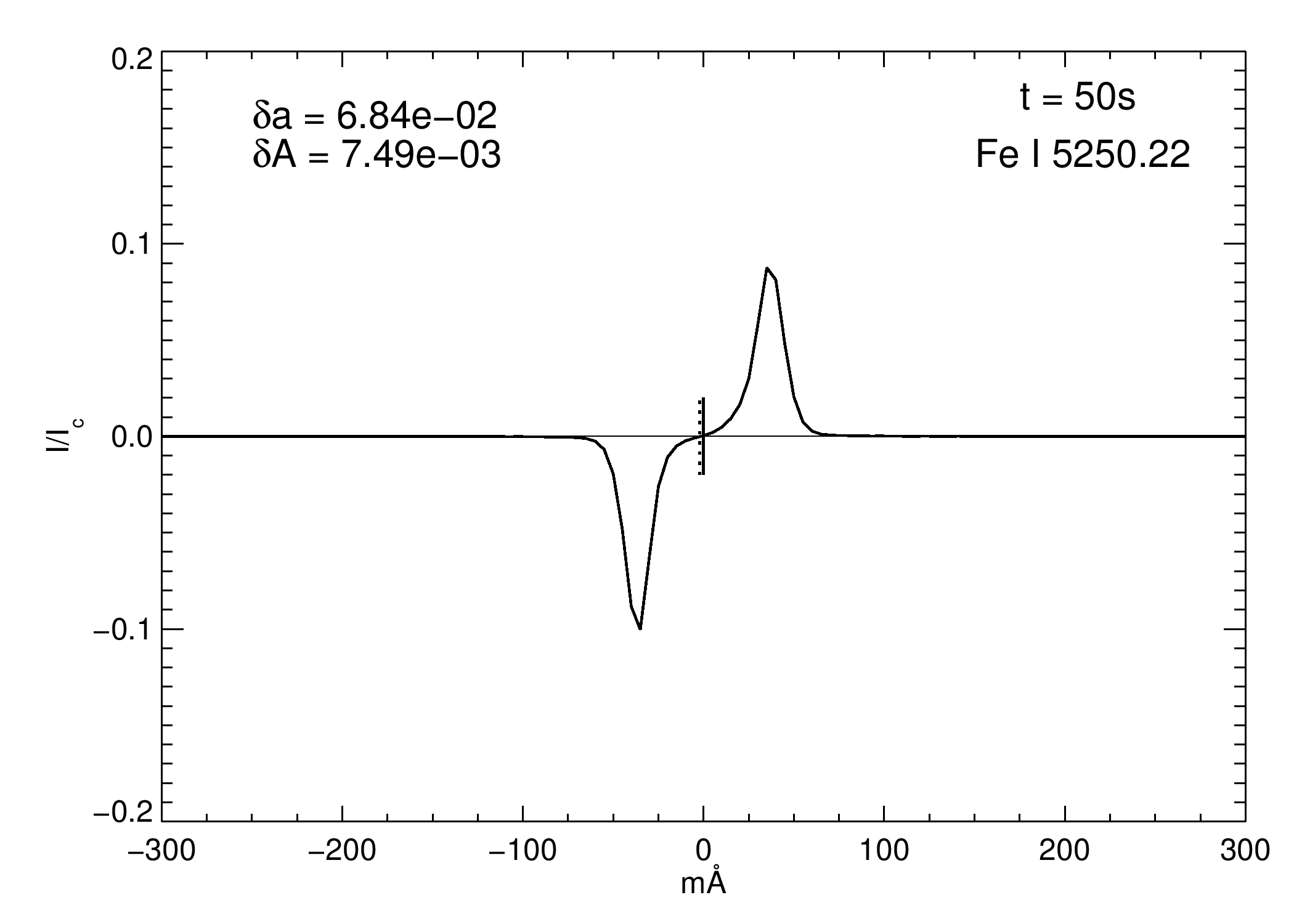} &
\includegraphics[width=0.49\textwidth]{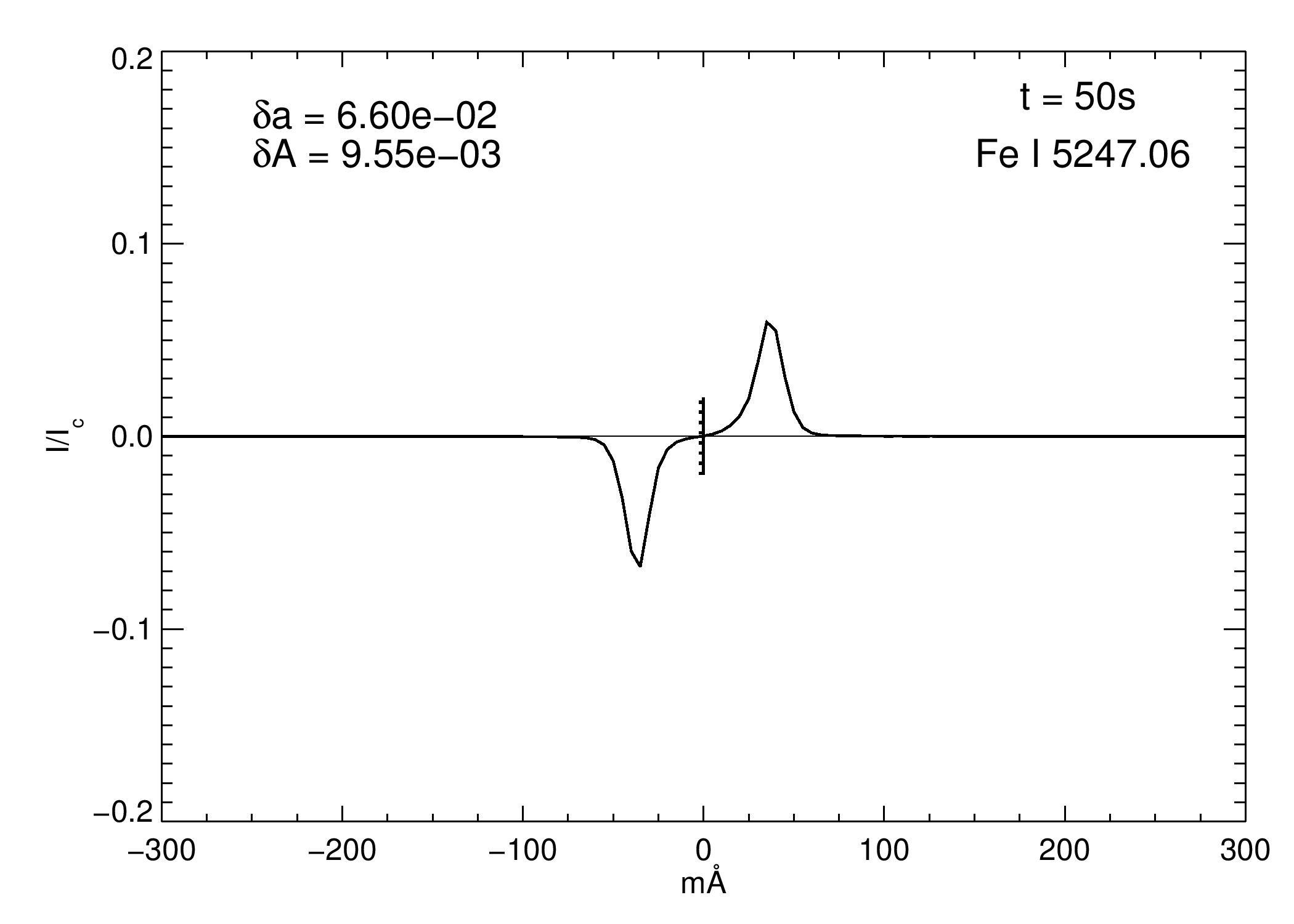}\\[-5pt]
{\small a)}& {\small b)}\\
\includegraphics[width=0.49\textwidth]{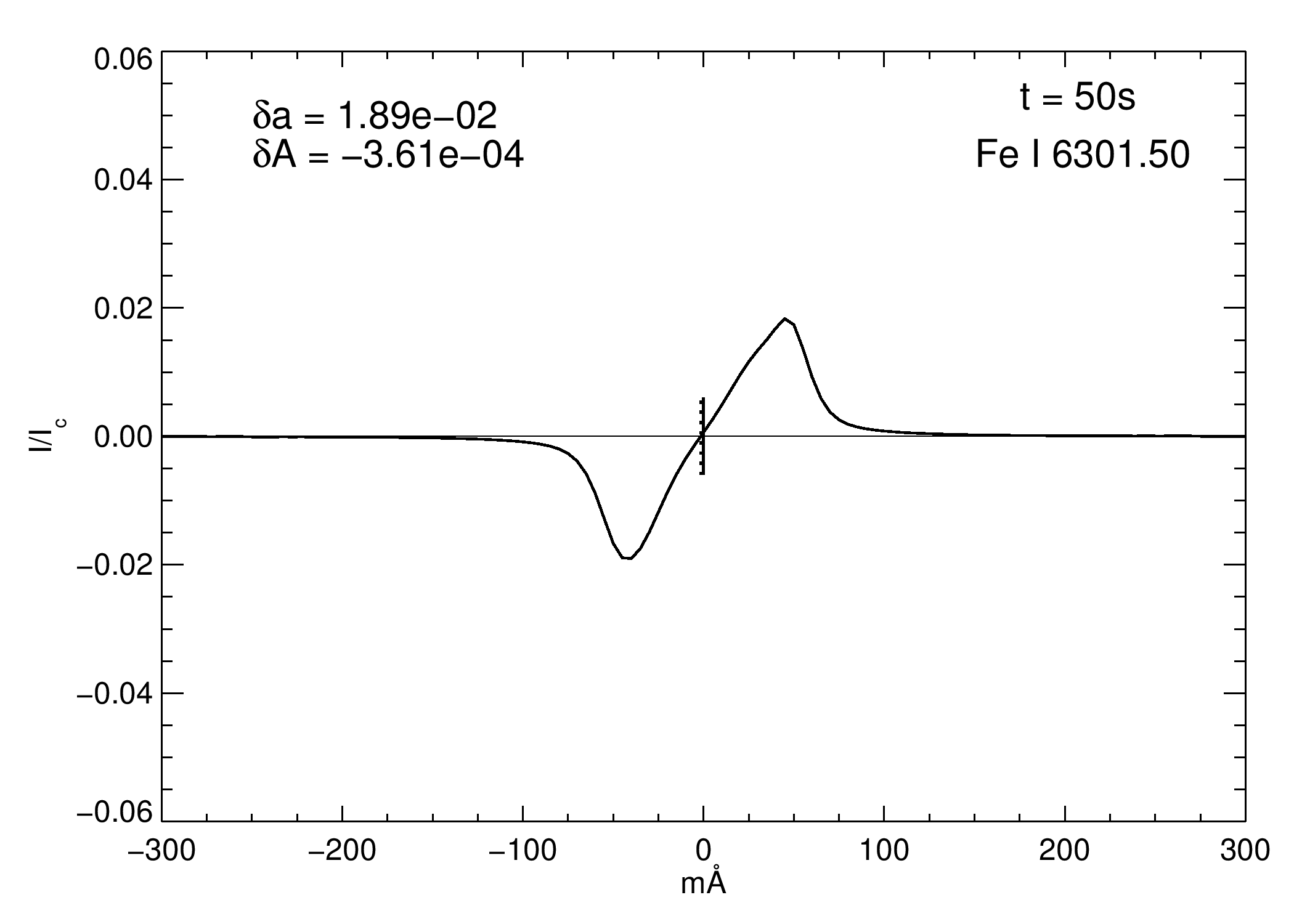} &
\includegraphics[width=0.49\textwidth]{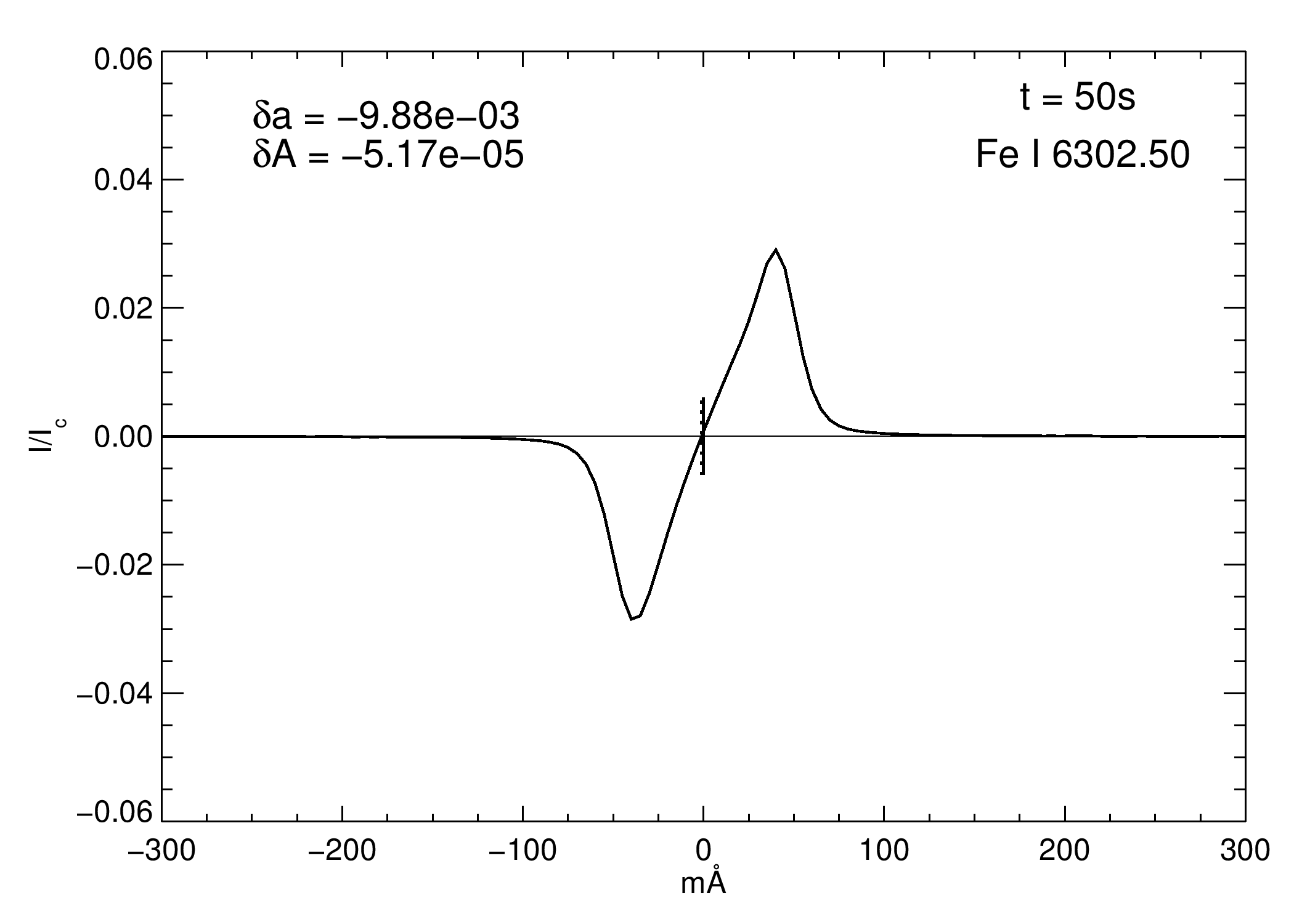}\\[-5pt]
{\small c)}& {\small d)}\\
\end{tabular}
\caption{Stokes-$V$ profiles of Fe~\textsc{i} (a) $\lambda$~5250.2~\AA, (b) $\lambda$~5247.06~\AA, (c) $\lambda$~6301.5~\AA, and (d) $\lambda$~6302.5~\AA\ at an elapsed time of 50~s from the vertical lines of sight in a slice ranging from $x=120$~km to $x=320$~km (far left of the axis). The magnetic flux sheet has a field strength of 1600~G on the axis at $z=0$. The solid and dotted vertical lines are as in Figure~\ref{fig:Stokes_1000_left}.}
\label{fig:Stokes_1600_left}
\end{figure}

%
\subsubsection{Area Asymmetry}
The gradients in velocity and magnetic field cause the asymmetry. A rule to calculate the sign of the area asymmetry, $\delta A$, for a purely longitudinal component has been provided by
\inlinecite{solanki1988}
(see also \opencite{steiner1999}):
\begin{equation}
\frac{{\rm d}|B(\tau)|}{{\rm d}\tau}\cdot\frac{{\rm d}v(\tau)}{{\rm d}\tau}
\left\{\begin{array}{lll}
< 0 &\;\Rightarrow\; & \delta A > 0, \\
> 0 &\;\Rightarrow\; & \delta A < 0.
\end{array}\right.
\label{eq:solanki_equation}
\end{equation}
By convention, $v(\tau)$ is taken to be positive for flows in the direction of increasing optical depth and \textit{vice versa}, where $v$ is the line-of-sight velocity. In case of a flux tube expanding with height, a line of sight along the tube axis will have ${\rm d}|B(\tau)|/{\rm d}\tau > 0$. Given this information, together with Equation~(\ref{eq:solanki_equation}), the correct interpretation of the time dependence of $\delta A$ would be relatively straightforward, except that for a propagating wave the term ${\rm d}v(\tau)/{\rm d}\tau$ changes
sign for each half wave. If the line of sight is eccentric, it may traverse the flux-sheet boundary, where the field drops suddenly with increasing optical depth, so that ${\rm d}|B(\tau)|/{\rm d}\tau < 0$. In this case, a positive $\delta A$ is realized if ${\rm d}v(\tau)/{\rm d}\tau>0$, \textit{i.e.}, if there is an accelerating downflow. This is for instance the case when there is no velocity inside the flux sheet but a downflow in the field-free surrounding region, or in case with an upflow within the flux sheet and no flow in the outside field-free region
\cite{grossmann1988b}.
Thus, $\delta A(t)$ in Figure~\ref{fig:Stokes_1000_evolution}b has contributions from gradients in magnetic field and velocity that stem from inside the magnetic flux sheet as well as from the boundary of the flux sheet. These contributions may have opposite signs. This is also true for the contributions from different phases of the wave. In addition, $\delta A$ is an integral quantity with contributions over all wavelength of the spectral line. Hence, $\delta A$ forms over a wide height range and therefore includes more than a single wave crest or wave trough. This renders the correct interpretation of $\delta A(t)$ in this case more intricate than it is for $\delta a(t)$. This is in particular true for the inner bundles of lines of sight. The outer bundles show a behaviour very similar to that of  $\delta a(t)$.

\begin{figure}[]
\setlength{\tabcolsep}{0pt}
\centering
\begin{tabular}{cc}
\includegraphics[width=0.49\textwidth]{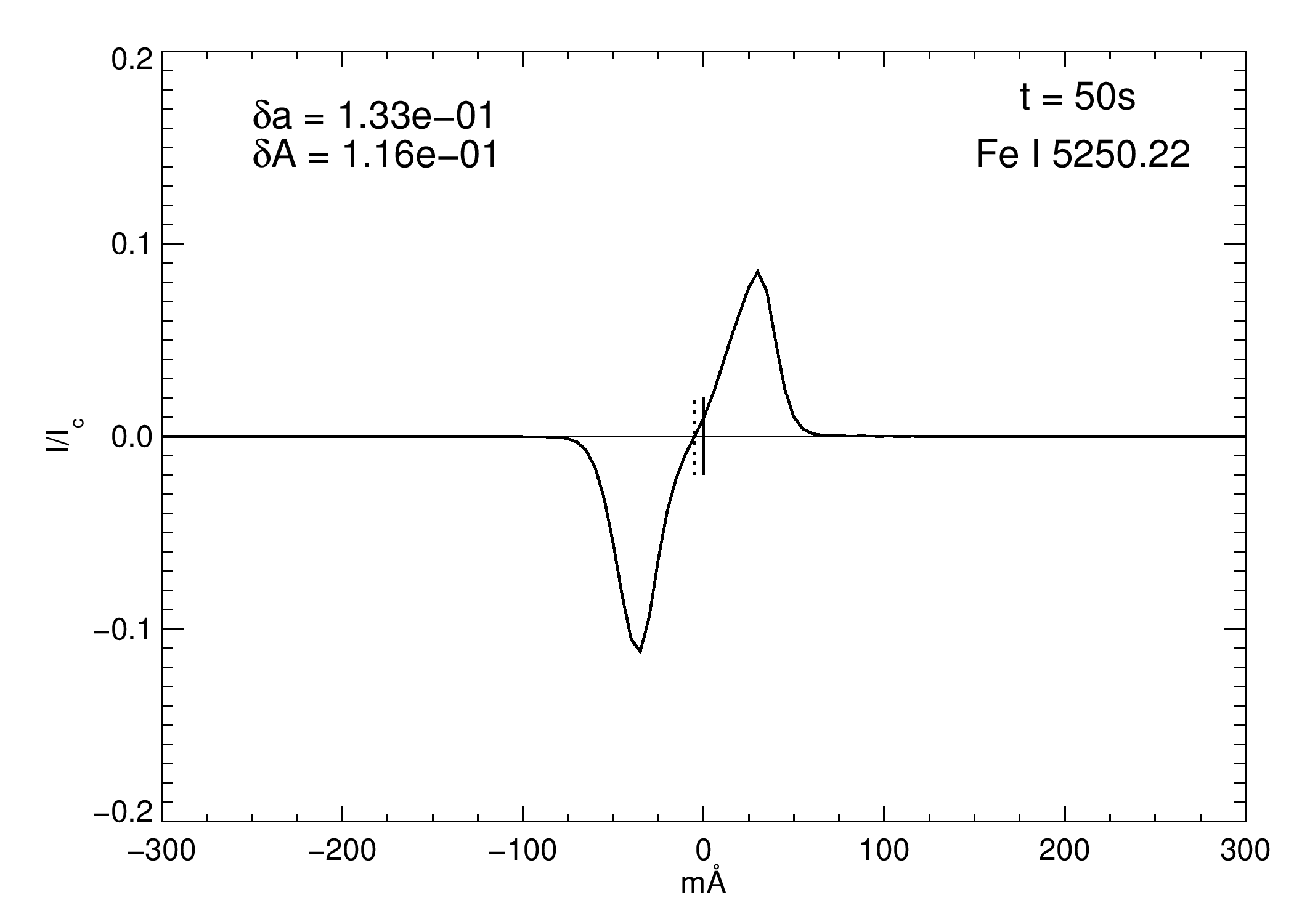} &
\includegraphics[width=0.49\textwidth]{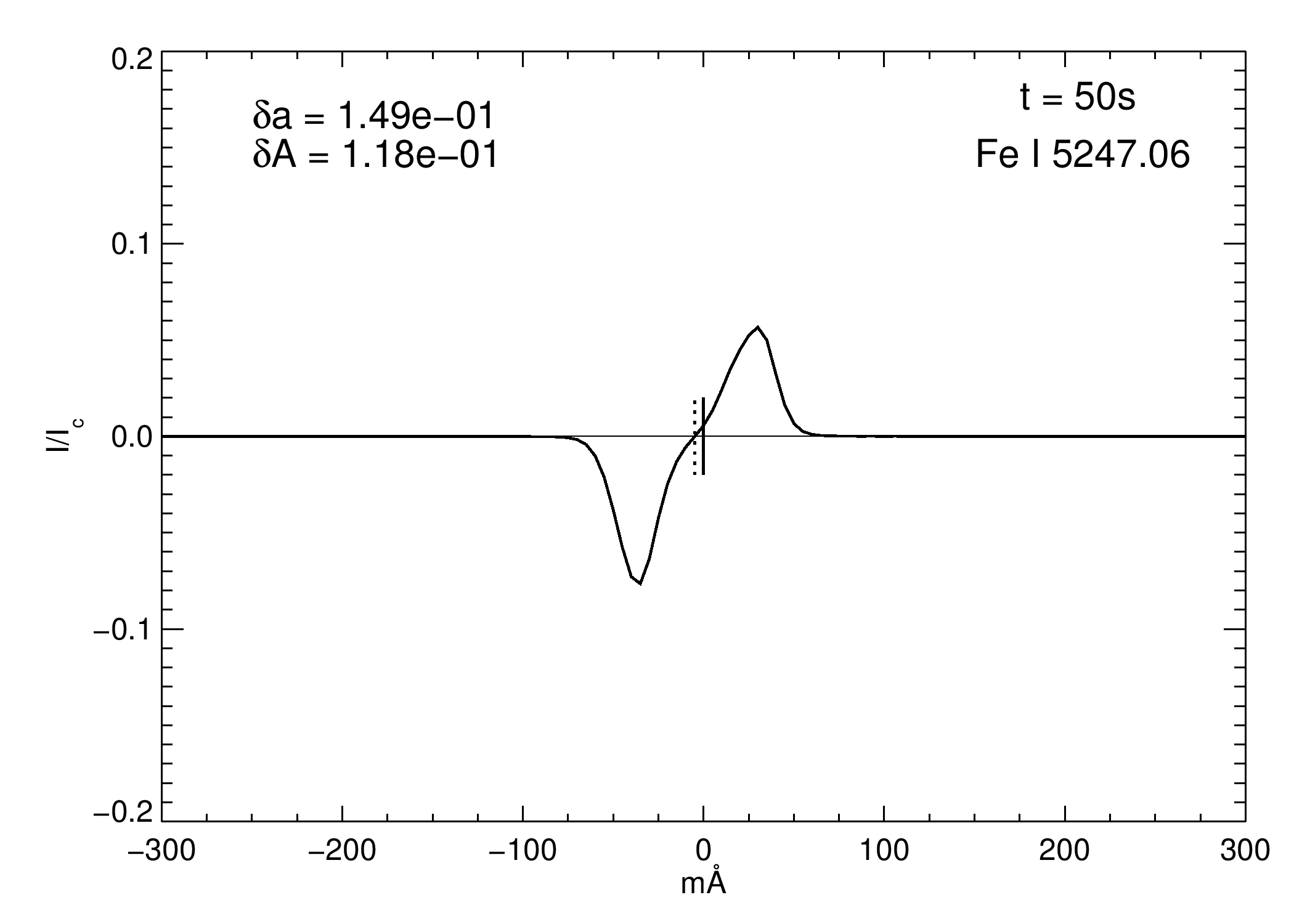}\\[-5pt]
{\small a)}& {\small b)}\\
\includegraphics[width=0.49\textwidth]{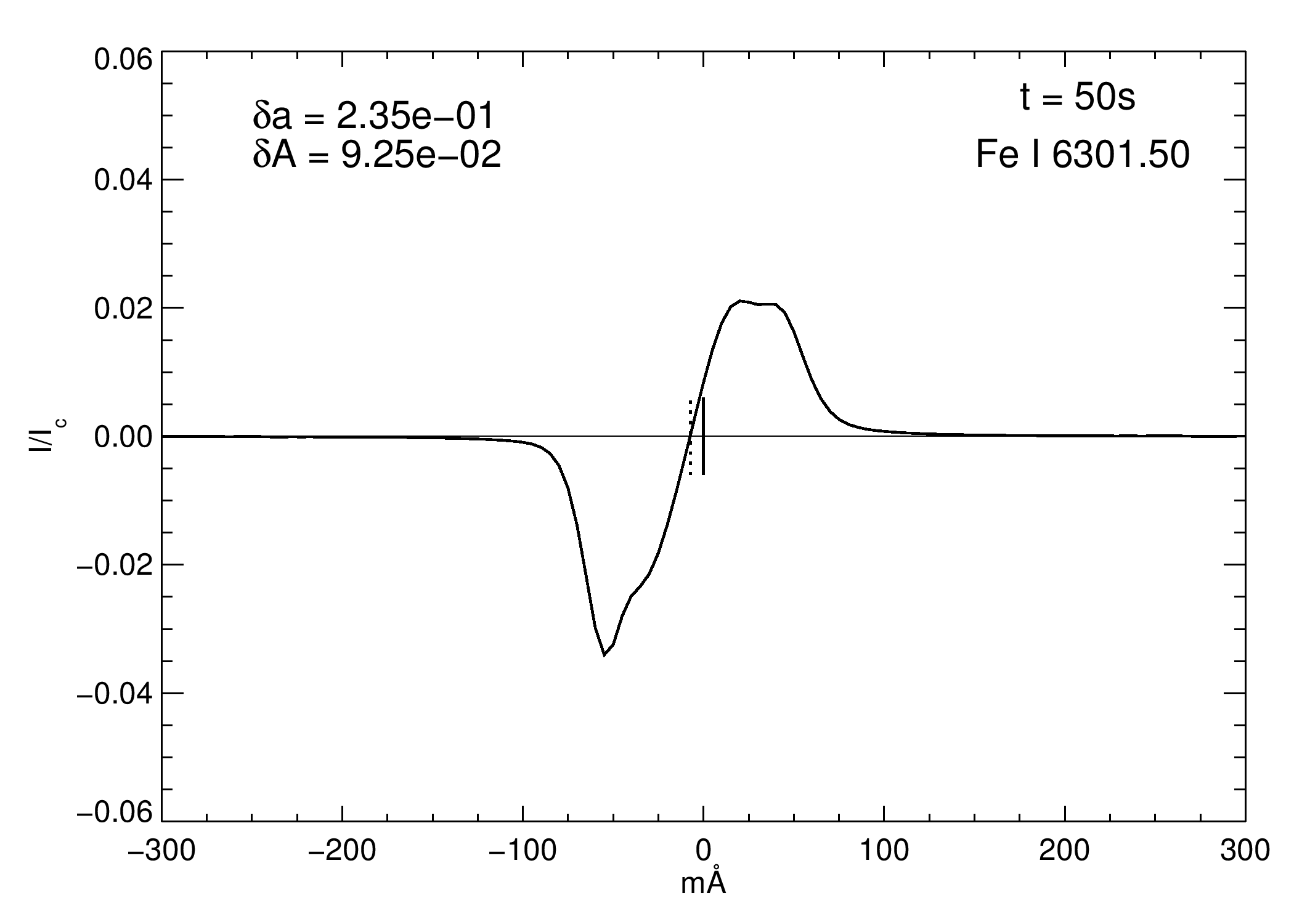} &
\includegraphics[width=0.49\textwidth]{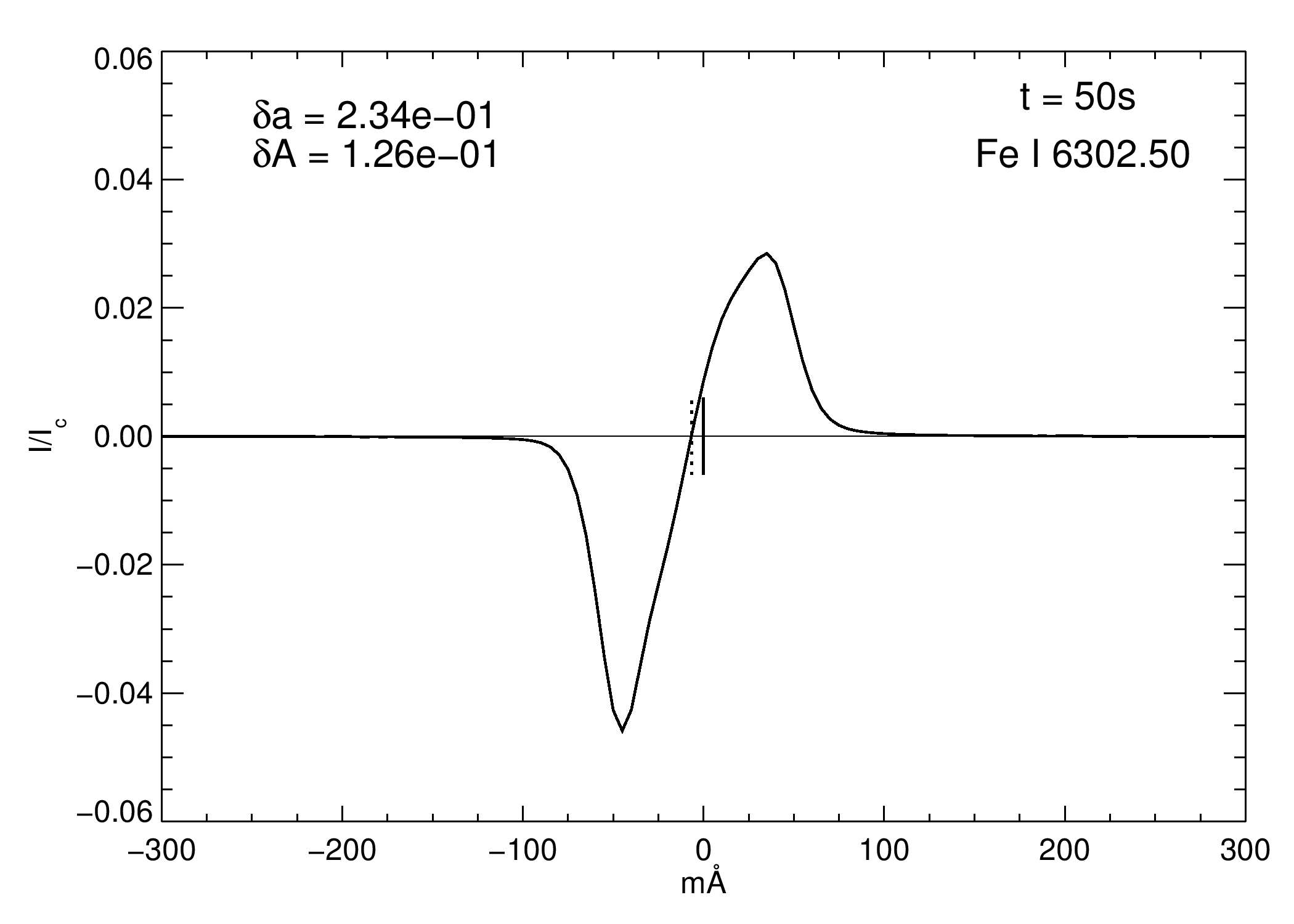}\\[-5pt]
{\small c)}& {\small d)}\\
\end{tabular}
\caption{Stokes-$V$ profiles of Fe~\textsc{i} (a) $\lambda$~5250.2~\AA, (b) $\lambda$~5247.06~\AA, (c) $\lambda$~6301.5~\AA, and (d) $\lambda$~6302.5~\AA\ at an elapsed time of 50~s from the vertical lines of sight in a slice ranging from $x=960$~km to $x=1160$~km (far right of the axis). The magnetic flux sheet has a field strength of 1600~G on the axis at $z=0$. The solid and dotted vertical lines are as in Figure~\ref{fig:Stokes_1000_left}.}
\label{fig:Stokes_1600_right}
\end{figure}

%
\subsection{Strong Field Case}

The emergent Stokes-$V$ profiles were also computed for a flux sheet with a magnetic field strength of 1600~G on the axis at $z=0$~km. Of the two cases shown in Figure~\ref{fig:1600_delta_t}, we consider in the following only the case with the wide excitation region. When spatially averaging these profiles over the entire width of the box, they do not show significant variation with time, revealing no sign of wave propagation inside the box, similar to the case with a field strength of 1000~G. Profiles averaged over smaller slices on either side of the axis show signs of wave propagation, once again emphasizing that lines of sight that are placed away from the symmetry axis of the flux sheet yield more information about the wave activity. Here, we carry out a similar study of the evolution of the Stokes-$V$ asymmetries for the 1600~G case as was done for the moderate field case. Figures~\ref{fig:Stokes_1600_left} and \ref{fig:Stokes_1600_right} show the Stokes-$V$ profiles at time $t=50$~s of the four Fe~\textsc{i} lines under study.
Figure~\ref{fig:Stokes_1600_left} shows the Stokes-$V$ profiles averaged over a horizontal range from $x=120$~km to $x=320$~km  (far left bundle of lines of sight) and Figure~\ref{fig:Stokes_1600_right} shows the profiles averaged over a range from $x=960$~km to $x=1160$~km (far right bundle of lines of sight).

\begin{figure}[]
\centering
\begin{tabular}{cc}
\includegraphics[scale=0.3]{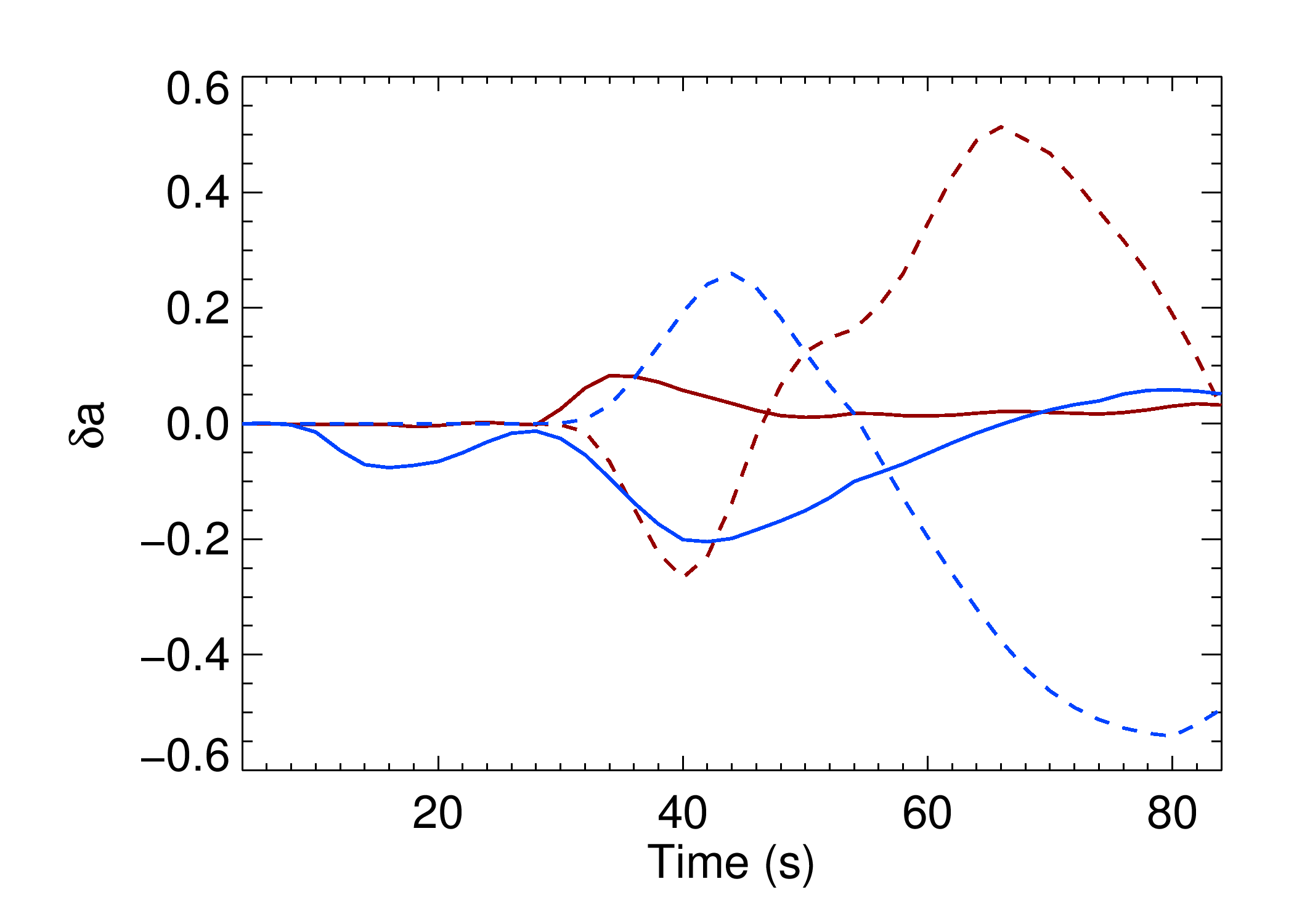}  &
\includegraphics[scale=0.3]{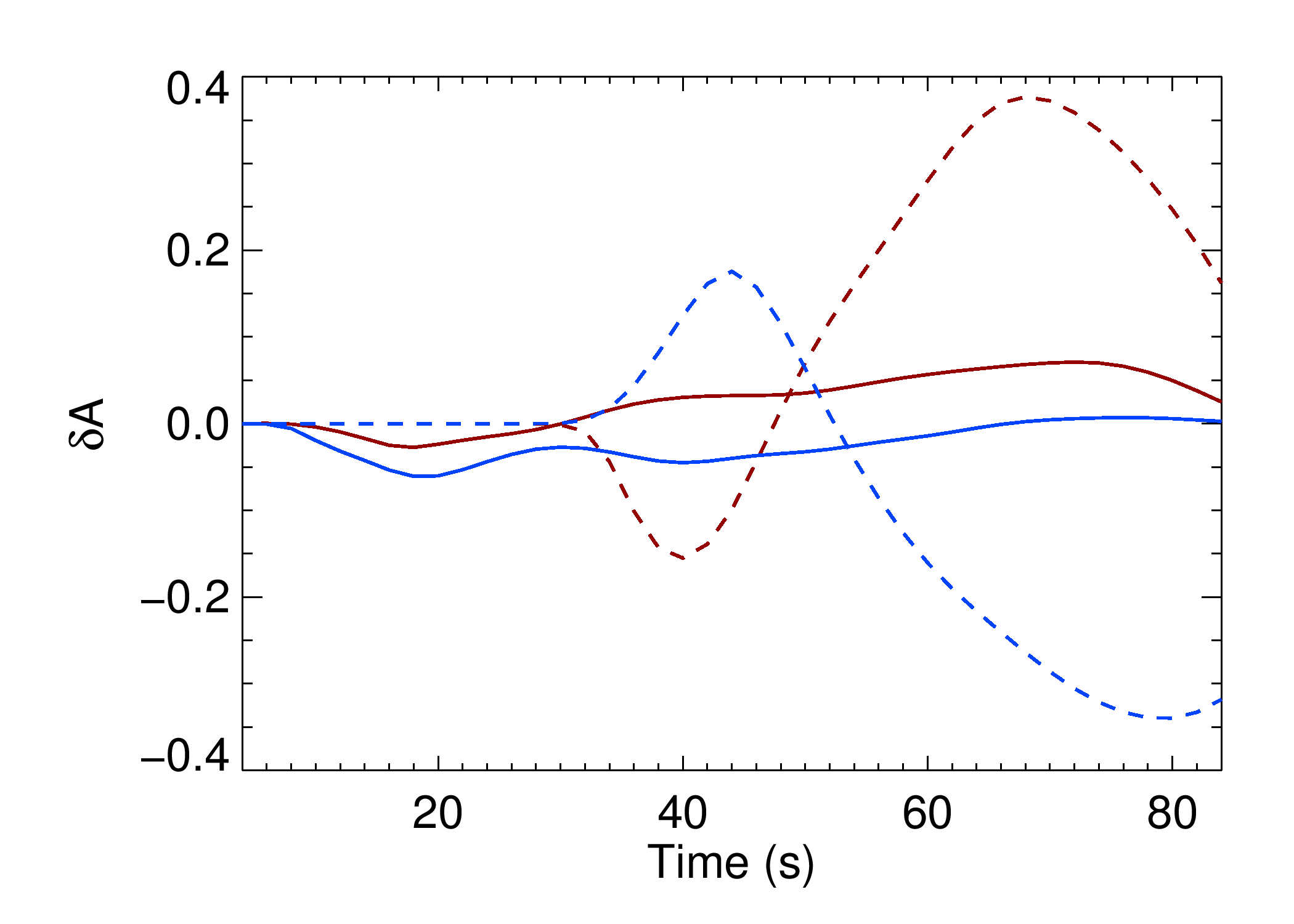} \\[-10pt]
\multicolumn{2}{c}{\footnotesize i) Fe~\textsc{i} $\lambda$ 5247.06~\AA}\\
\includegraphics[scale=0.3]{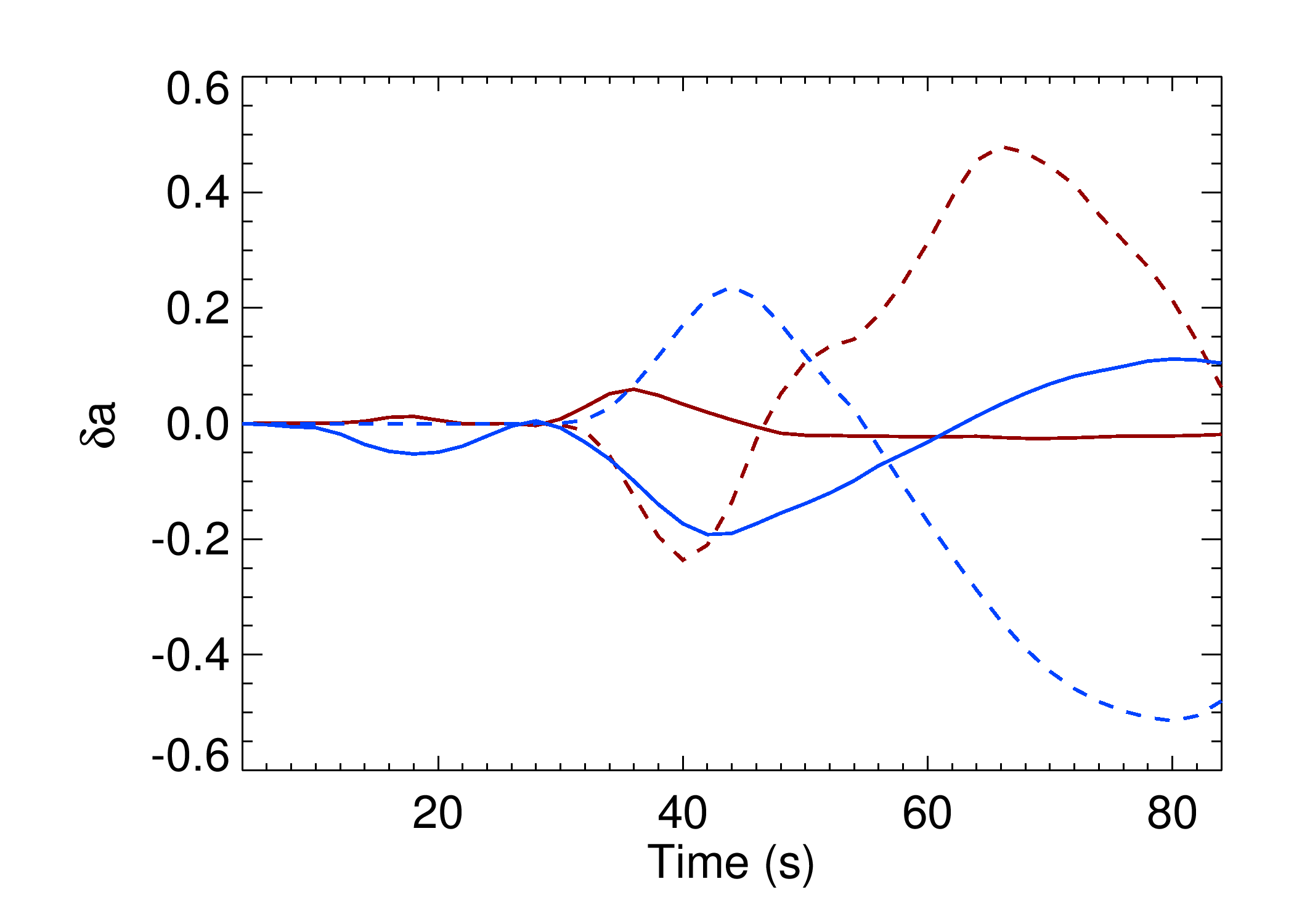}  &
\includegraphics[scale=0.3]{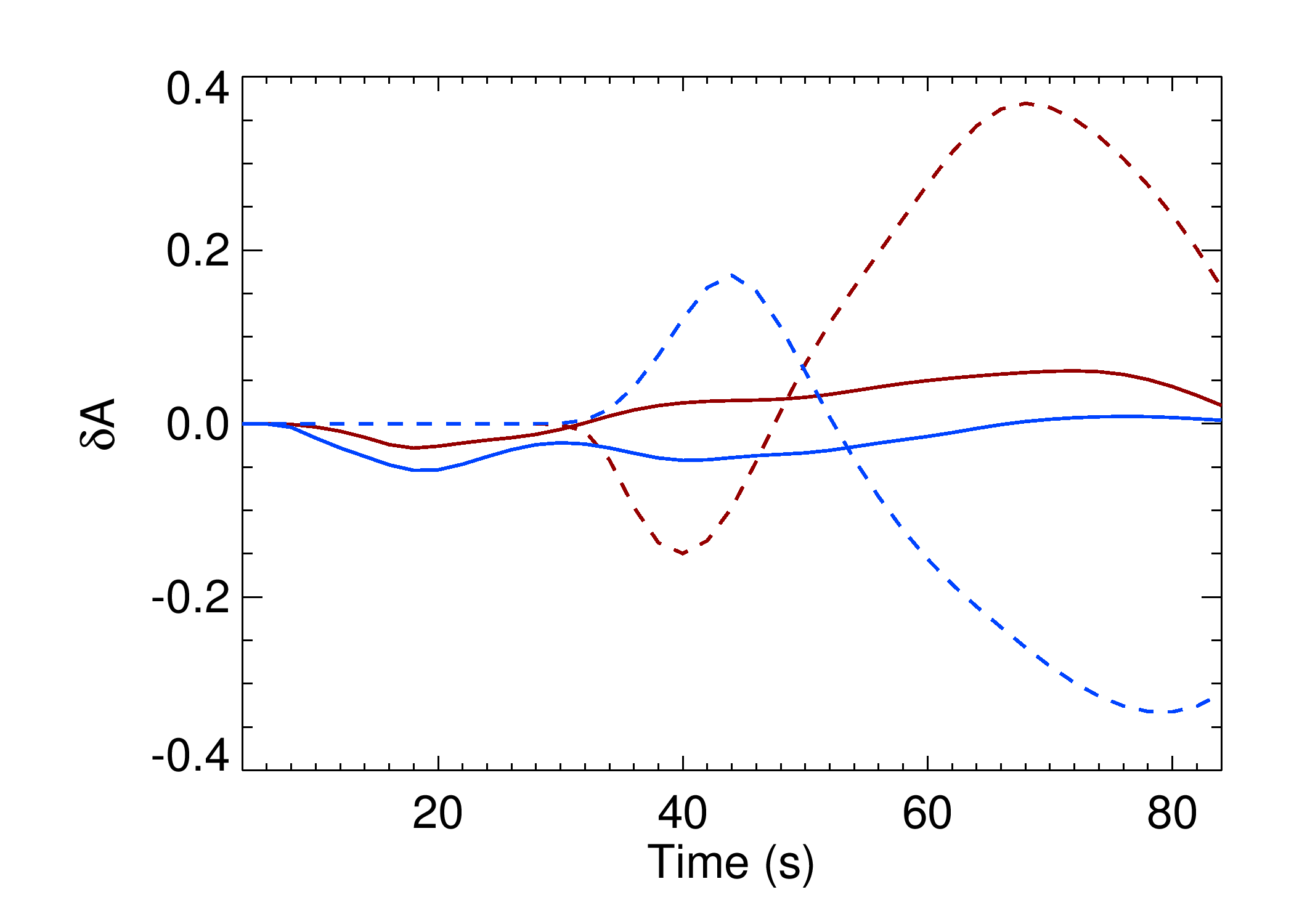} \\[-10pt]
\multicolumn{2}{c}{\footnotesize ii) Fe~\textsc{i} $\lambda$ 5250.2~\AA}\\
\includegraphics[scale=0.3]{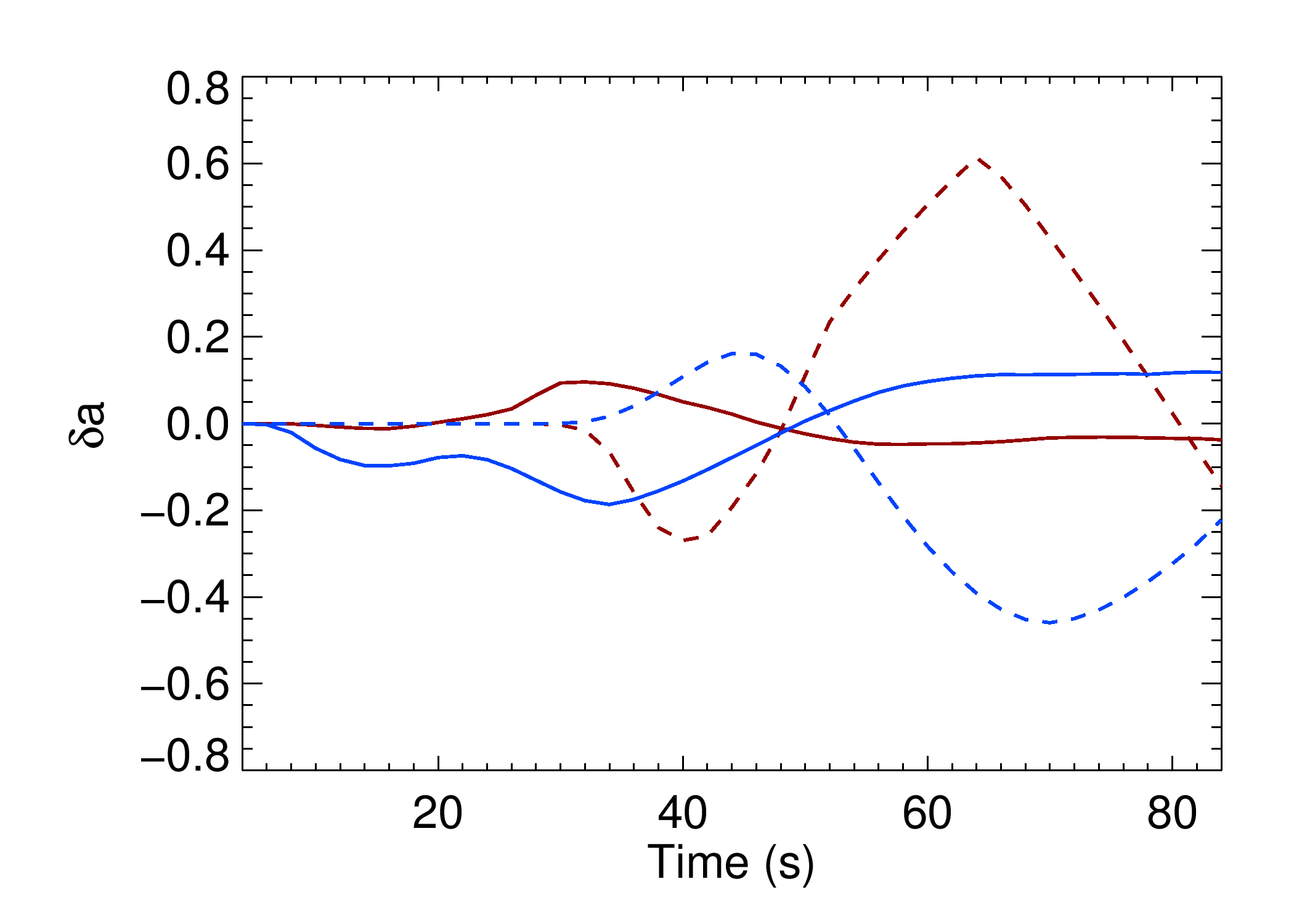}  &
\includegraphics[scale=0.3]{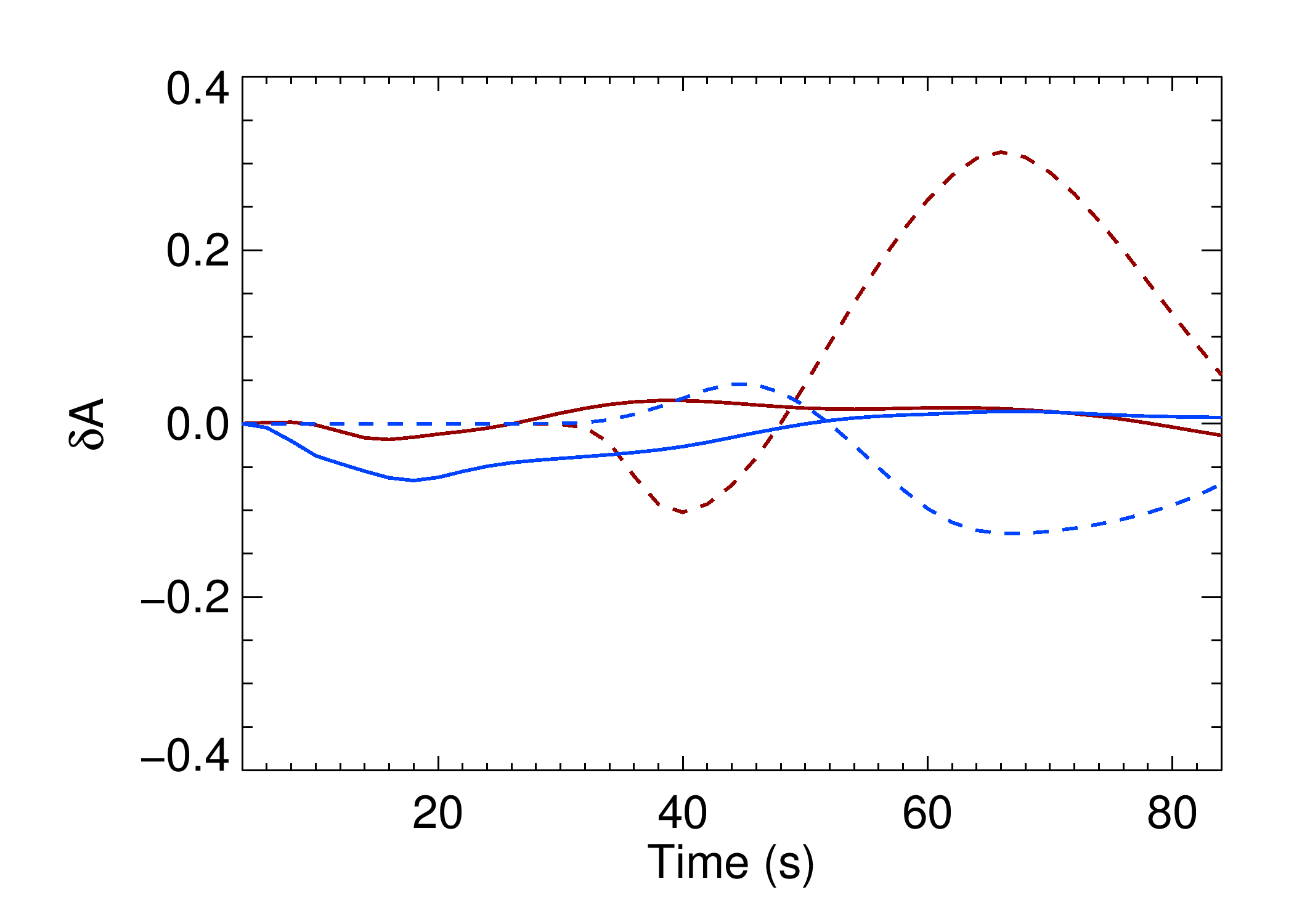} \\[-10pt]
\multicolumn{2}{c}{\footnotesize iii) Fe~\textsc{i} $\lambda$ 6301.5~\AA}\\
\includegraphics[scale=0.3]{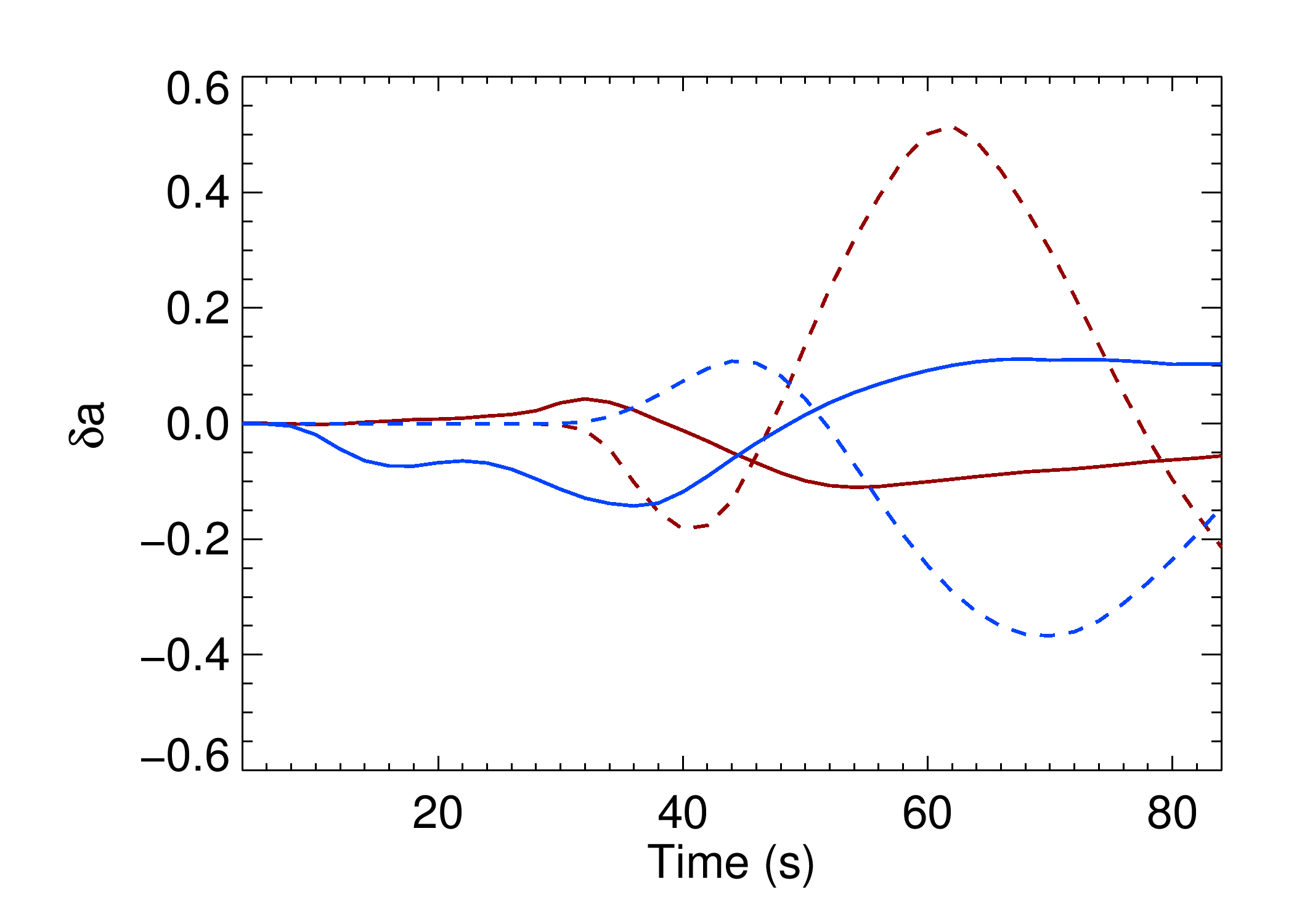}  &
\includegraphics[scale=0.3]{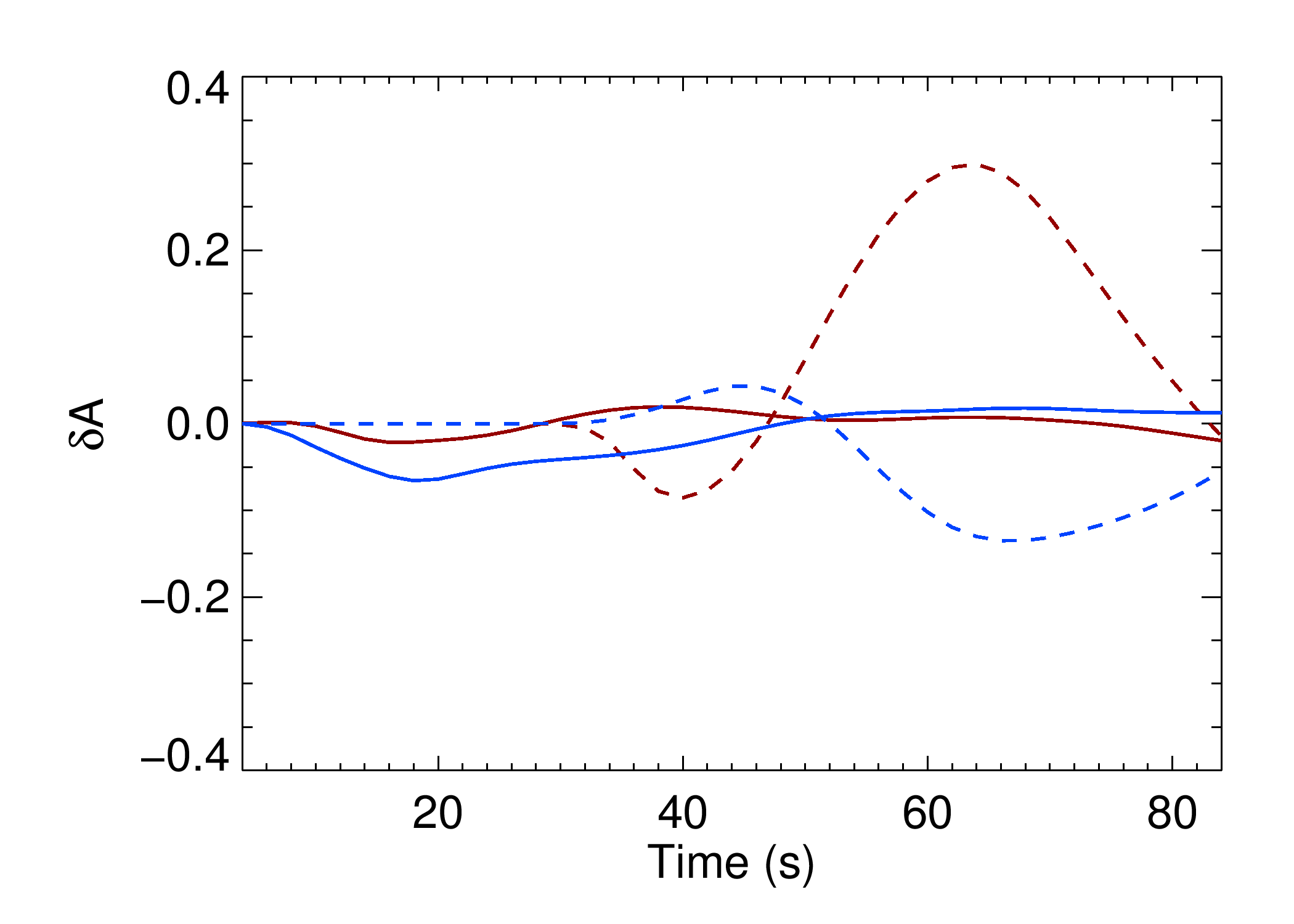} \\[-10pt]
\multicolumn{2}{c}{\footnotesize iv) Fe~\textsc{i} $\lambda$ 6302.5~\AA}\\
{\small a) Amplitude Asymmetry} & b) {\small Area Asymmetry} \\
\end{tabular}
\caption{The Stokes-$V$ (a) amplitude asymmetry and (b) area asymmetry for the four {Fe~\textsc{i}} lines as functions of time for the strong field case with 1600~G. The red solid curves represent the right side of the flux-sheet axis. The red dashed curves represent the bundle on the far right. Blue solid curves are for the left bundle and the blue dashed curves are for the far left bundle.}
\label{fig:Stokes_1600_evolution}
\end{figure}

The Stokes-$V$ amplitude asymmetry and  area asymmetry as functions of time for these lines are shown in Figure~\ref{fig:Stokes_1600_evolution}. The colour coding is the same as in Figure~\ref{fig:Stokes_1000_evolution}. The temperature perturbation and the velocity field for different time instances are shown in Figure~\ref{fig:1600_delta_t}b.
There, the colours represent the value of $\Delta T$ and the arrows show the velocity vectors at times $t=20$, 30, 40, and 50~s.

Unlike in the case with 1000~G, here we have significant fast, predominantly magnetic waves, which get
refracted within the flux sheet and convert to fast acoustic waves when they encounter the flux-sheet
boundary, where they leave the flux sheet and enter the field-free domain. This causes the wing-like feature in the temperature perturbations of Figure~\ref{fig:1600_delta_t}b that extends from $z=300$~km to $z=700$~km on both sides of the flux sheet (approximately along the $\beta=1$ contour) at time $t=40$~s. The velocities associated with both the fast and the slow waves result in the shift and the asymmetries of the Stokes-$V$ profiles.

In the following, we consider a bundle of lines of sight on the far right side of the flux-sheet axis. Here, the front of the fast, predominantly magnetic wave starts to become effective at a $t\approx 40$~s. The velocities are directed downwards in the magnetic region, resulting in a red-shifted Stokes-$V$ contribution illuminated by light from the unshifted absorption formed in the static layer below, which suppresses the blue lobe, leading to a negative asymmetry (dashed red curve in Figure~\ref{fig:Stokes_1600_evolution}a). At $t=60$~s, this downflow has moved out of the line-of-sight bundle into the field-free region while within the magnetic region an upflow evolves. This upflow, which is due to the following slow mode, gives rise to a positive asymmetry leading to a strong positive bump of the red dashed curve around $t=70$~s. This behaviour is different from the 1000~G case shown in Figure~\ref{fig:Stokes_1000_evolution}a wherein the first negative bump due to the fast converting mode is less pronounced or missing because of the relatively weak magnetic field. The line of sight on the left  hand side of the flux-sheet axis shows a similar behaviour but of opposite sign. This behaviour can also be seen in the area asymmetry of the outer bundles of lines of sight. Thus, $\delta a$ and $\delta A$ show a clear signature of both the fast and the slow mode.

\begin{figure}[]
\centering
\begin{tabular}{cc}
\includegraphics[scale=0.3]{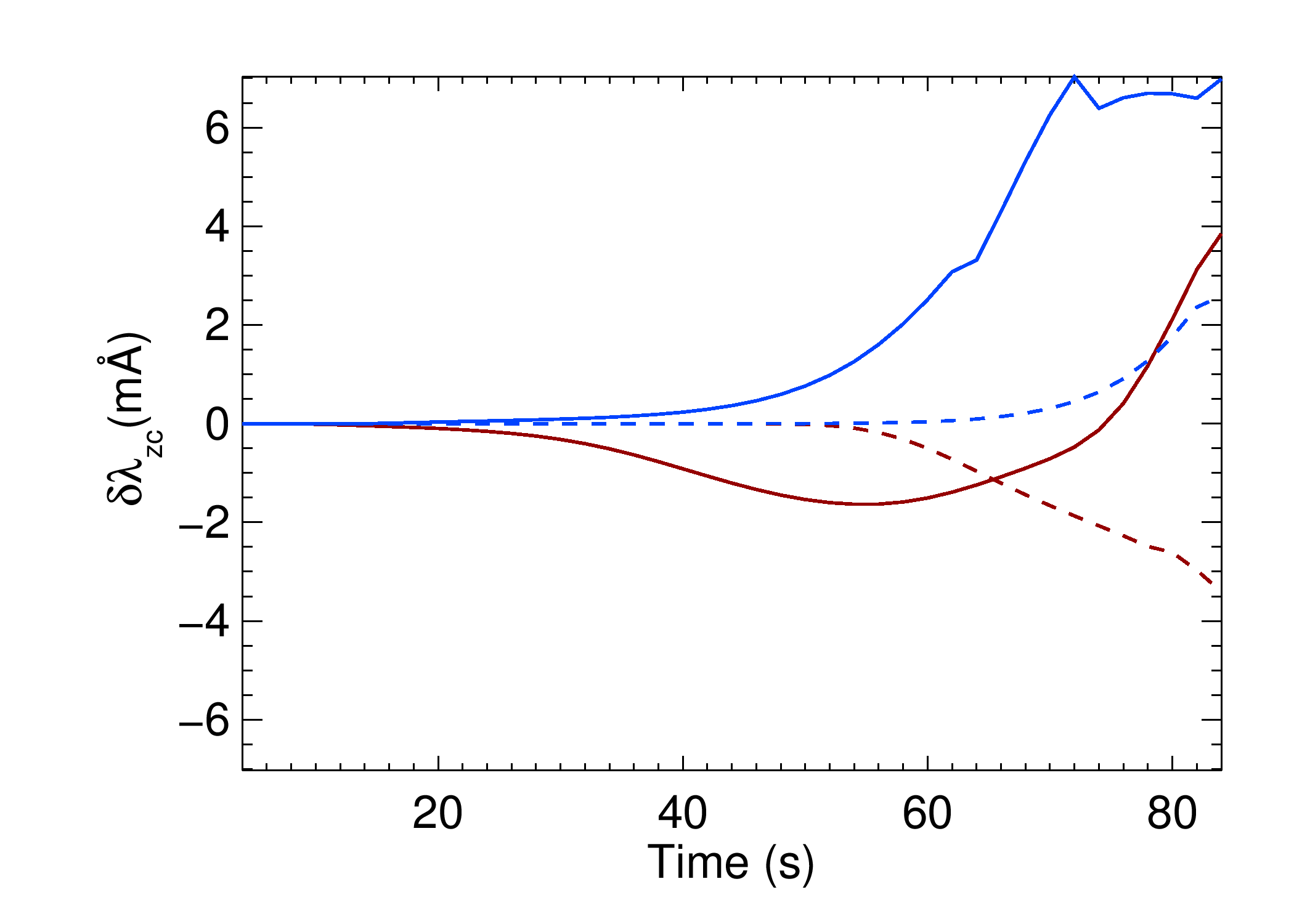}  &
\includegraphics[scale=0.3]{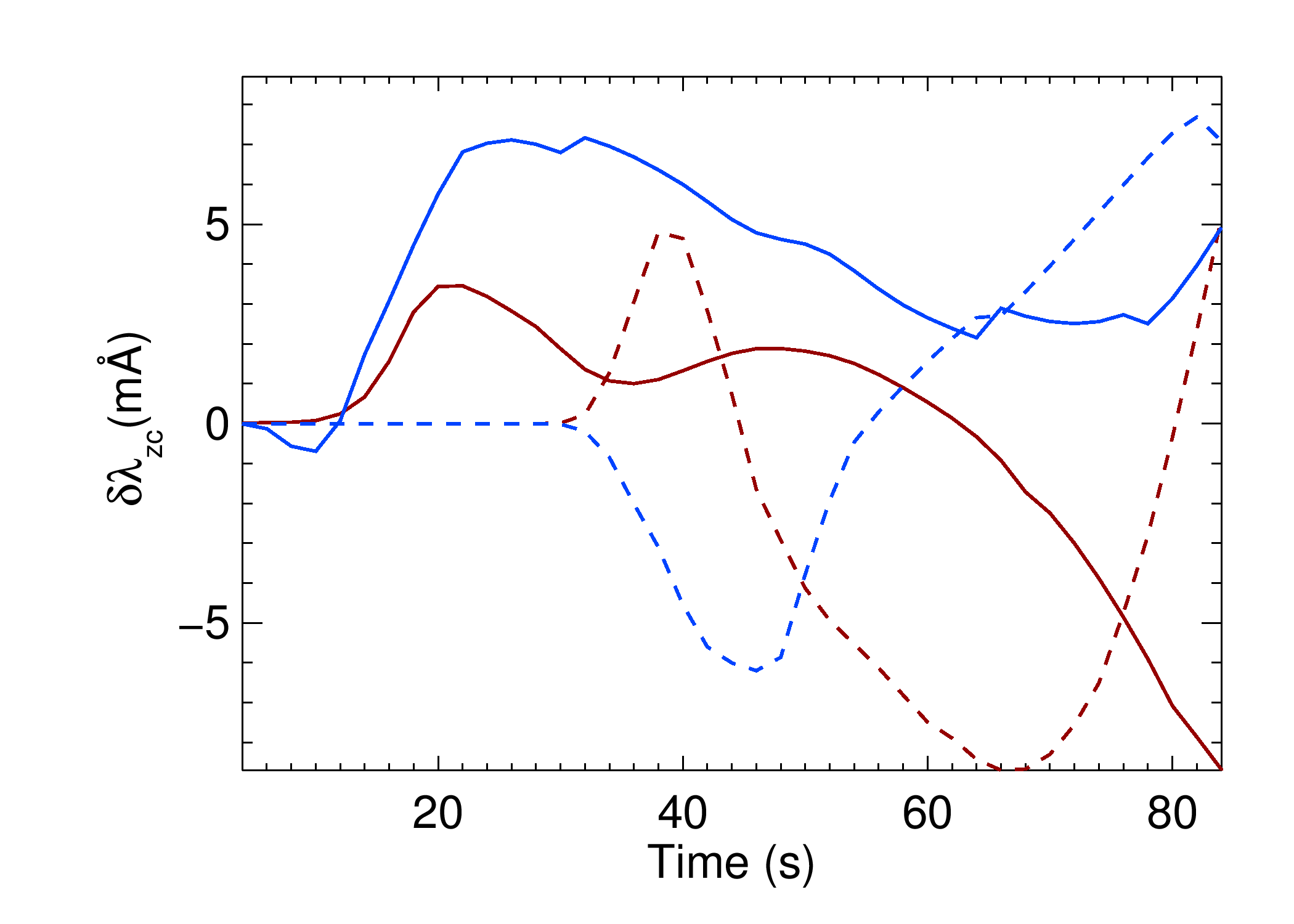} \\[-10pt]
\multicolumn{2}{c}{\footnotesize i) Fe~\textsc{i} $\lambda$ 5247.06~\AA}\\
\includegraphics[scale=0.3]{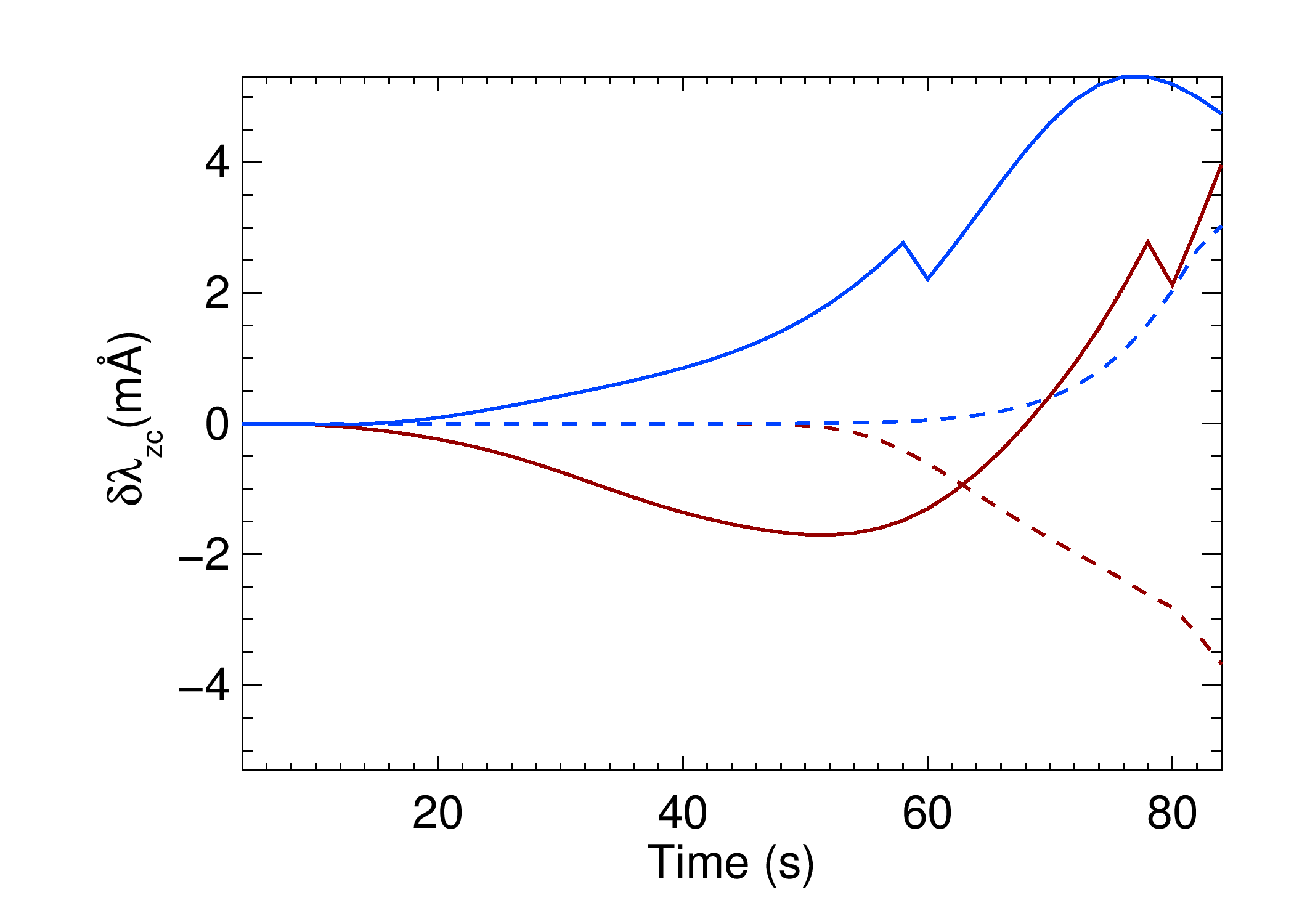}  &
\includegraphics[scale=0.3]{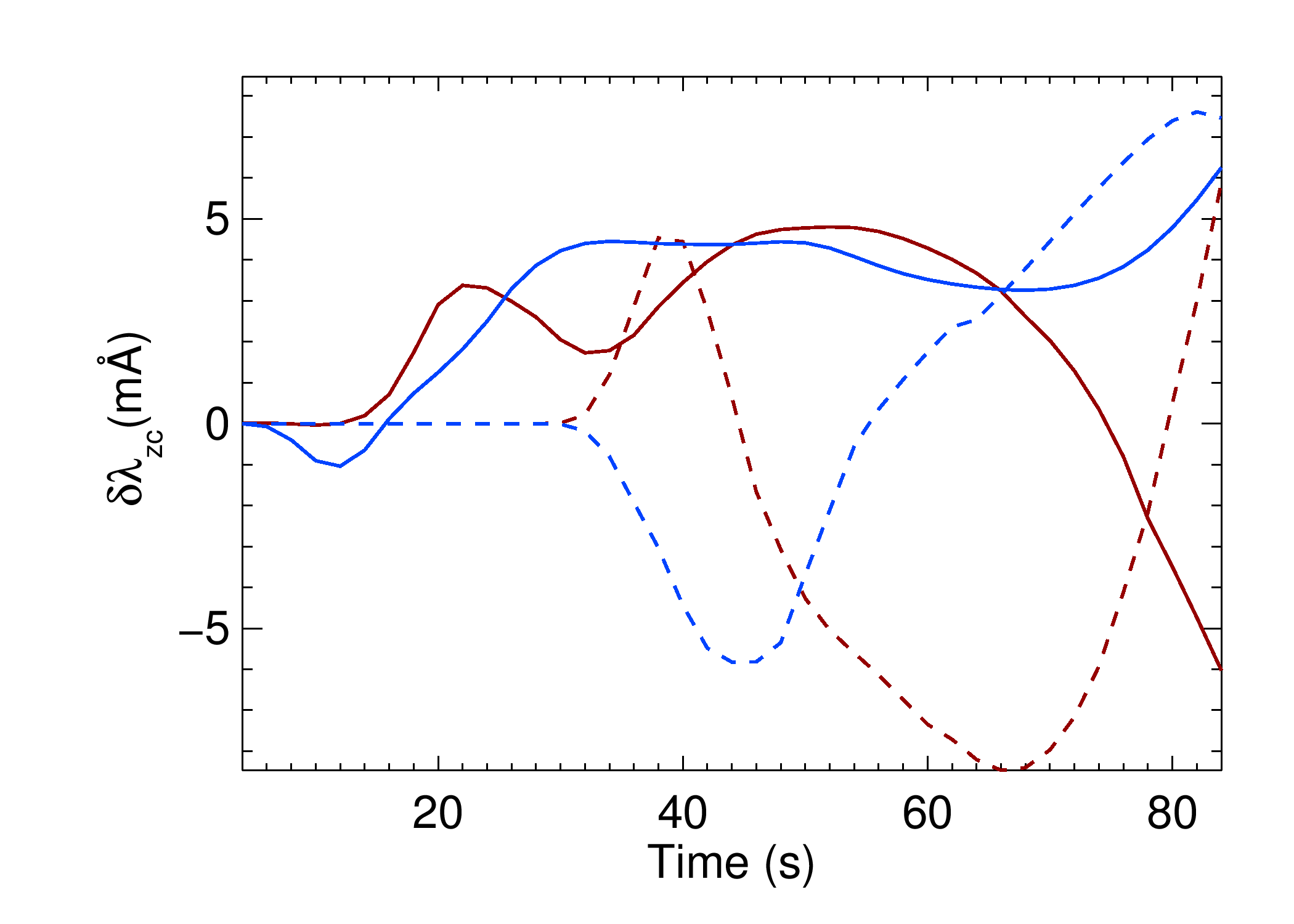} \\[-10pt]
\multicolumn{2}{c}{\footnotesize ii) Fe~\textsc{i} $\lambda$ 5250.2~\AA}\\
\includegraphics[scale=0.3]{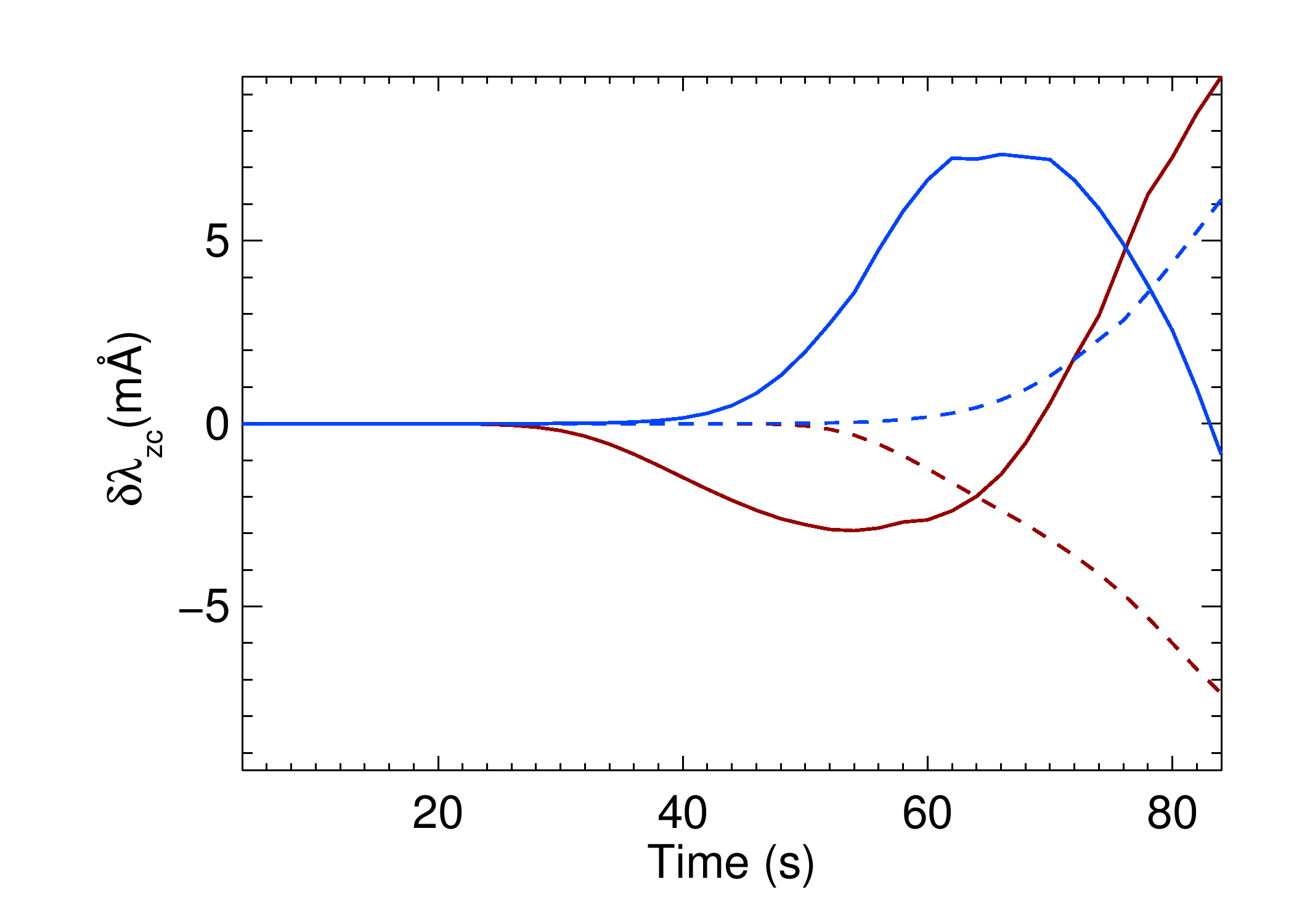}  &
\includegraphics[scale=0.3]{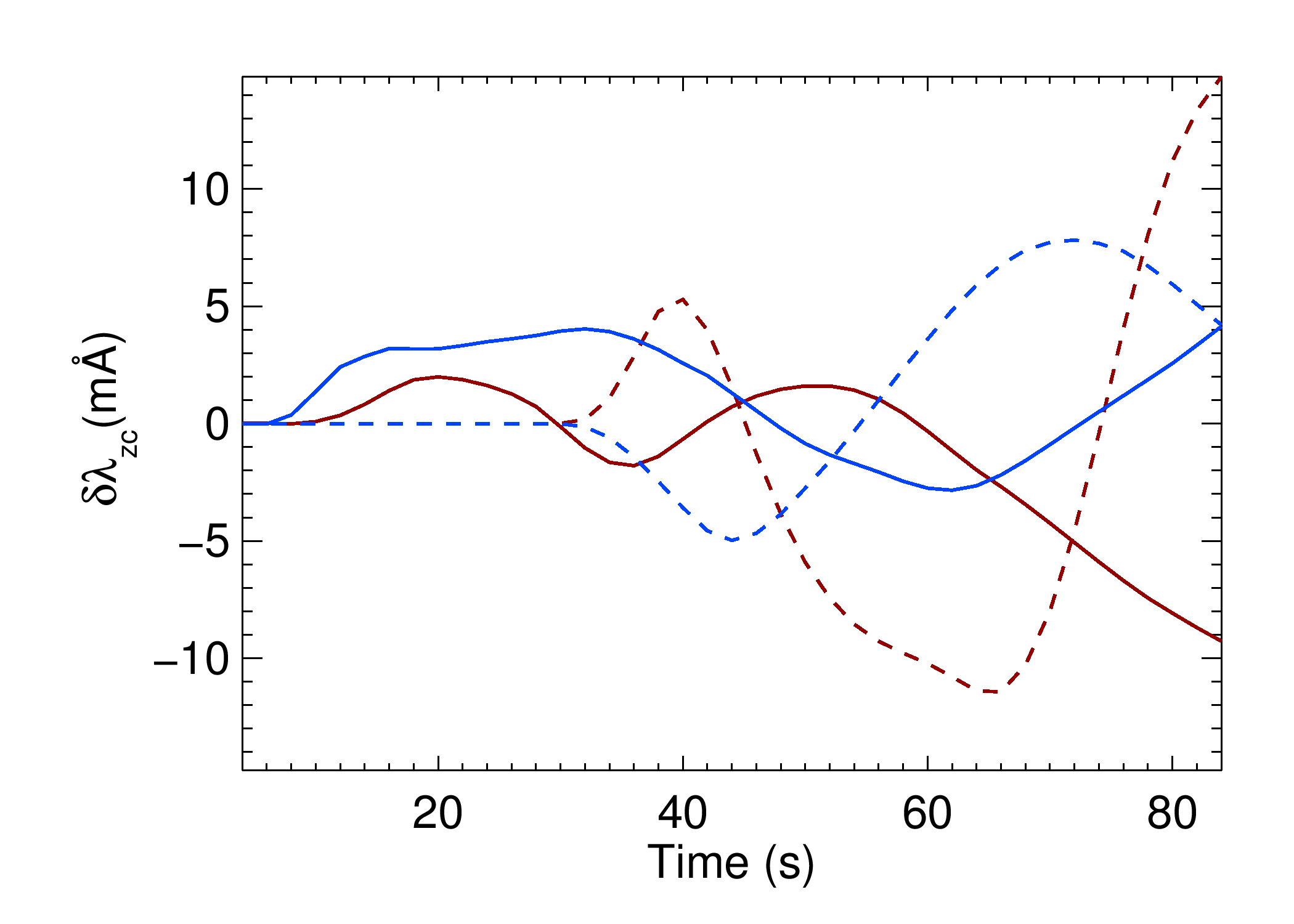} \\[-10pt]
\multicolumn{2}{c}{\footnotesize iii) Fe~\textsc{i} $\lambda$ 6301.5~\AA}\\
\includegraphics[scale=0.3]{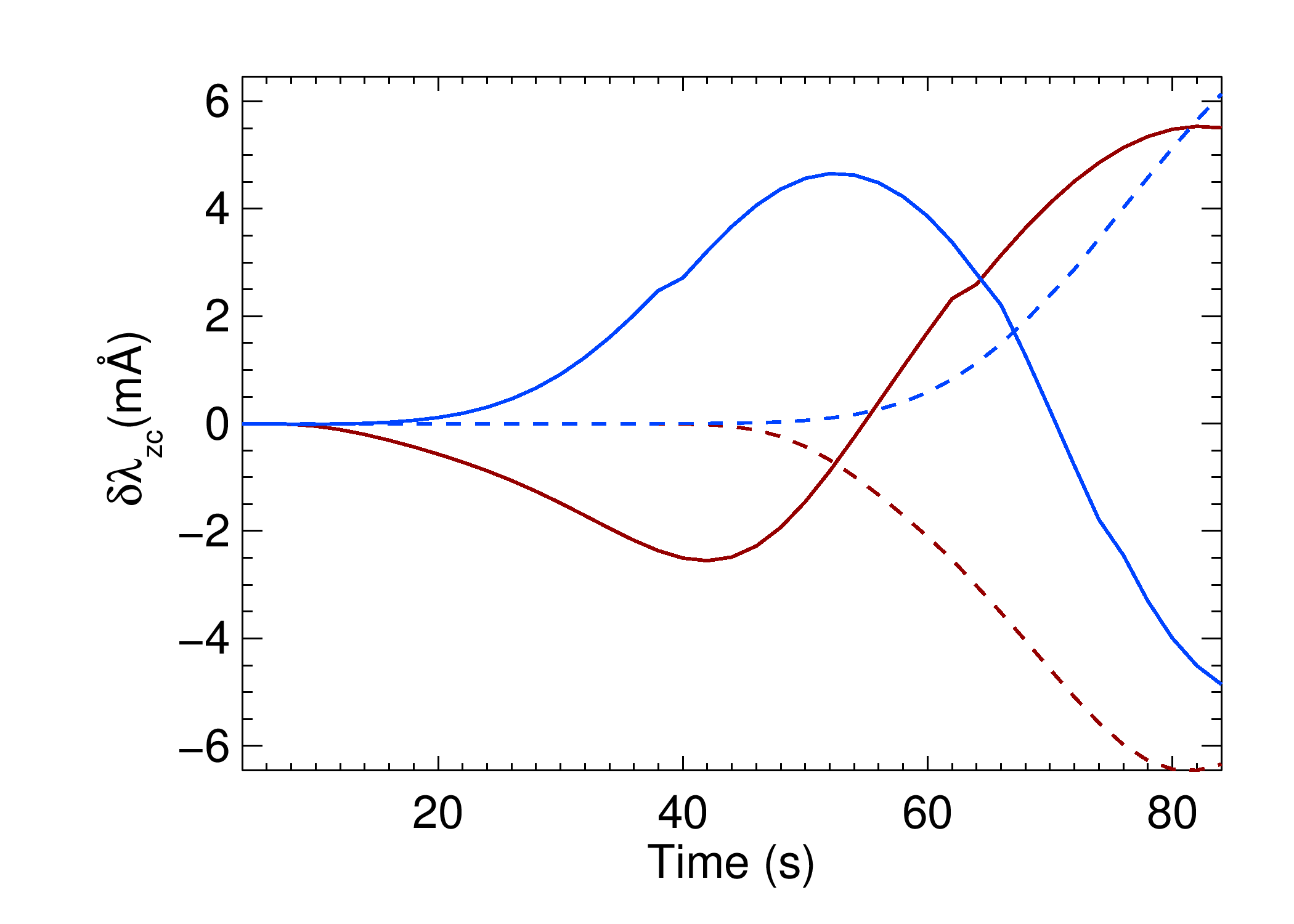}  &
\includegraphics[scale=0.3]{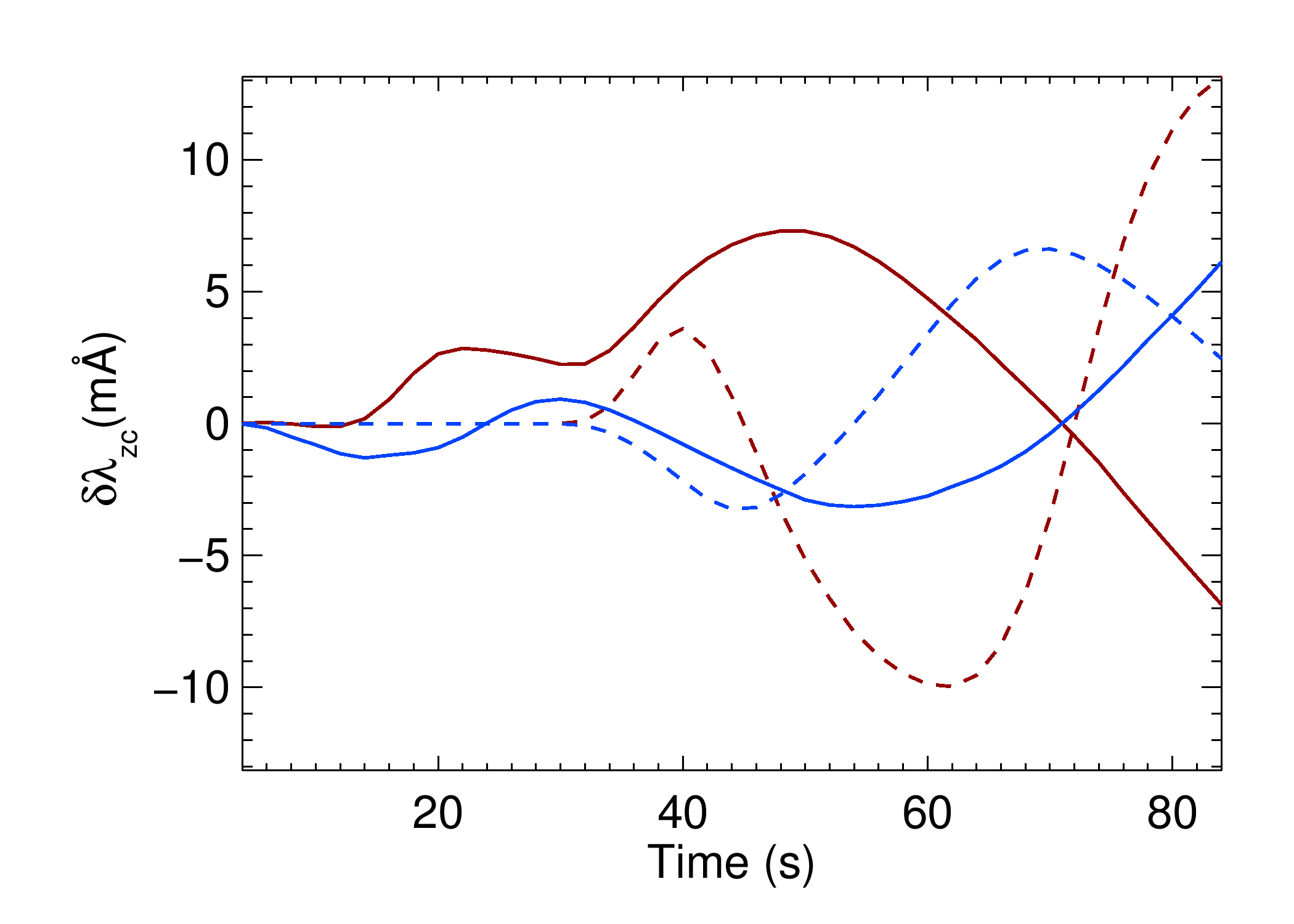} \\[-10pt]
\multicolumn{2}{c}{\footnotesize iv) Fe~\textsc{i} $\lambda$ 6302.5~\AA}\\
{\small a) 1000~G} & b) {\small 1600~G} \\
\end{tabular}
\caption{Stokes-$V$ zero-crossing shifts for the four Fe~\textsc{i} lines as functions of time for (a) the moderate field case with 1000~G and (b) the strong field case with 1600~G. The red solid curves represent the slice on the right side of the axis. The red dashed curves represent the slice on the far right. Blue solid curves are for the left slice and the blue dashed curves are for the far left slice.
}
\label{fig:Stokes_zero_cross_evolution}
\end{figure}

%
\subsubsection{Zero-Crossing Shift}
The flows within the magnetic elements can be estimated by the Stokes-$V$ zero-crossing shift, given by Equation~(\ref{eq:zero_crossing}). The 180$^{\circ}$ out-of-phase flow pattern formed on the two sides of the flux-sheet axis creates phase shifted, opposite zero-crossing shifts on the two sides. Figure~\ref{fig:Stokes_zero_cross_evolution} shows the Stokes-$V$ zero-crossing shift as a function of time for the moderate and the strong field cases. In case of the moderate field and line-of-sight bundles close to the sheet axis, the upflow on the right side and the downflow on the left side of the axis, which stem from the respective compressional and rarefactional fronts of the slow (acoustic) wave, result in a zero-crossing shift of opposite signs. The curves are shifted in time for the four lines and the two line-of-sight
bundles due to the difference in the formation heights of the lines and due to the time lag of the waves to the left and to the right of the flux-sheet axis, respectively. A similar trend in the zero-crossing shift can be seen in the outer line-of-sight bundles, where the wave fronts arrive later. The flow pattern associated with the fast (magnetic) wave is a prominent feature in the case with the strong field. It creates a significant bump of opposite sign on either side of the flux sheet for the outer line-of-sight bundles around $t\approx40$~s, visible in the
dashed curves of Figure~\ref{fig:Stokes_zero_cross_evolution}b.

%
\section{Summary and Conclusions}\label{s:summary_conclusion}

This work is an extension of a previous work by
\inlinecite{vigeesh2009},
which focused on the dynamics and the energy transport that occur
in intense flux tubes as a consequence of an impulsive transversal footpoint motion. In the present work, we have constructed flux tubes embedded in the photosphere and used the results of our simulation to compute the Stokes profiles that emerge from the top of the simulation box in order to study observational signatures of wave propagation inside the tubes.

The nature of the excited modes depends on the value of plasma $\beta$ at the place where the driving motion occurs. Depending upon the extent of the region of excitation, the energies imparted to the different modes vary. When the excitation occurs in a high-$\beta$ plasma, we observe that the excited modes are a slow acoustic wave and a fast magnetic wave that undergo mode conversion and transmission across the $\beta =1$ layer. In the case of excitation of a low-$\beta$ flux sheet over a large enough impact area, most of the energy will go to the fast (magnetic) mode. If the area of impact is smaller, then most of the energy goes into the slow (acoustic) wave, which is channelled up along the flux tube and eventually dissipates by shock formation. Hence, the impact of a large granule may impart more energy to the fast (magnetic) mode, but the non-magnetic atmosphere gains back this energy in the form of a fast (acoustic) wave due to refraction of the fast mode and mode conversion.

The anti-symmetry in the wave pattern with respect to the flux-sheet axis gives rise to distinct observational signatures in Stokes $V$. While the average Stokes-$V$ profile over the whole domain does not show any significant variation with time, clear evidence of the wave phenomena can be detected when looking at higher resolved lines of sight on either side of the flux sheet. Stokes-$V$ profiles become asymmetric, showing opposite temporal behaviour on the two sides of the flux-sheet axis. Furthermore, effects of refraction of the fast,
predominantly magnetic wave in the case of a strong magnetic field are clearly visible in the Stokes asymmetry and zero-crossing shift as a function of time. Our results show a clear signature of the fast, predominantly magnetic wave in these profiles. We come to the conclusion that polarimetric signatures of wave propagation in magnetic elements can be observed, provided that the spatial resolution is high enough so that magnetic concentrations can be resolved into different regions within the flux concentration.
Although the simulated Stokes asymmetries would be detectable with current polarimetric instruments, observations of considerably higher spatial resolution, capable of resolving individual flux concentrations, are needed in order to reveal the propagation of waves in individual flux concentrations and to detect the different modes  of the MHD waves. We have highlighted the importance of using the Stokes-$V$ asymmetries as a possible diagnostic tool to study wave propagation in magnetic elements at disk centre. Observations off disk centre would possibly slightly relax the high requirements of spatial resolution because the wave pattern can be expected to be less anti-symmetric in this case. However, the interpretation of the polarimetric signals as functions of time would become more intricate. The analysis in this work is based on photospheric lines under the assumption of local thermodynamic equilibrium (LTE). Hence our conclusions are not valid for lines formed in the chromosphere, since the LTE approximation is no longer valid in this region. A more realistic modeling should be carried out in three spatial dimensions and include NLTE effects.

%

%

%

%
\begin{acks}
This work was supported by the German Academic Exchange
Service (DAAD), grant D/05/57687, and the Indian Department of
Science \& Technology (DST), grant DST/INT/DAAD/P146/2006. We thank the anonymous referee for his/her useful comments which helped us to substantially improve the manuscript.
\end{acks}

%
%
\bibliographystyle{spr-mp-sola}
\bibliography{vigeesh_ms}

\end{article}
\end{document}